\documentclass[12pt]{article}
\usepackage{amsmath,amssymb,amsfonts}
\usepackage[dvips]{graphicx}
\usepackage{epsfig}

\usepackage[vcentermath]{youngtab}

\makeatletter \@addtoreset{equation}{section} \makeatother

\renewcommand{\baselinestretch}{1.2}
\begin{document}

\renewcommand{\thefootnote}{\alph{footnote}}

\begin{titlepage}

\begin{center}
\hfill {\tt Imperial/TP/09/SK/01}\\
\hfill {\tt arXiv:0903.4172}

\vspace{2cm}

{\large\bf \hspace*{-0.4cm}
\mbox{The Complete Superconformal Index
for $\mathcal{N}=6$ Chern-Simons Theory}}

\vspace{2cm}

\renewcommand{\thefootnote}{\alph{footnote}}

{\large Seok Kim}

\vspace{1cm}

\textit{
Theoretical Physics Group, Blackett Laboratory,\\
Imperial College, London SW7 2AZ, U.K.}\\

\vspace{0.3cm}

\textit{\&}

\textit{
Institute for Mathematical Sciences,\\
Imperial College, London SW7 2PG, U.K.}\\

\vspace{0.7cm}

E-mail
: {\tt 
s.kim@imperial.ac.uk}

\end{center}

\vspace{1cm}

\begin{abstract}

We calculate the superconformal index for $\mathcal{N}\!=\!6$
Chern-Simons-matter theory with gauge group $U(N)_k\times
U(N)_{\!-\!k}$ at arbitrary allowed value of the Chern-Simons level
$k$. The calculation is based on localization of the path integral
for the index. Our index counts supersymmetric gauge invariant
operators containing inclusions of magnetic monopole operators,
where latter operators create magnetic fluxes on 2-sphere. Through
analytic and numerical calculations in various sectors, we show that
our result perfectly agrees with the index over supersymmetric
gravitons in $AdS_4\times S^7/\mathbb{Z}_k$ in the large $N$ limit.
Monopole operators in nontrivial representations of $U(N)\times
U(N)$ play important roles. We also comment on possible applications
of our methods to other superconformal Chern-Simons theories.

\end{abstract}

\end{titlepage}

\renewcommand{\thefootnote}{\arabic{footnote}}

\setcounter{footnote}{0}

\renewcommand{\baselinestretch}{0}

\tableofcontents

\renewcommand{\baselinestretch}{1.2}

\section{Introduction}

An important problem in AdS/CFT \cite{Maldacena:1997re} is to
understand the Hilbert spaces of both sides. The string or M-theory
is put on global AdS, and the dual conformal field theory (CFT) is
radially quantized. Partition function encodes the information on
Hilbert space. In particular, one has to understand the spectrum of
strongly interacting CFT, which is in general difficult, to use the
dual string/M-theory to study various phenomena in conventional
gravity.

With supersymmetry, one can try to circumvent this difficulty by
considering quantities which contain possibly less information than
the partition function but do not depend on (or depend much more
mildly on) the coupling constants controlling the interaction.

This has been considered in the context of AdS/CFT. If a
superconformal theory has \textit{continuous} parameters, one can
construct a function called the superconformal index which does not
depend on changes of them \cite{Kinney:2005ej,Bhattacharya:2008zy}.
The general structure of the superconformal index was investigated
in 4 dimension \cite{Kinney:2005ej}, and then in 3, 5, 6 dimensions
\cite{Bhattacharya:2008zy}. See also \cite{Romelsberger:2005eg}. In
all these cases, the superconformal index is essentially the Witten
index \cite{Witten:1982df} and acquires nonzero contribution only
from states preserving supersymmetry. The superconformal index was
computed for a class of SCFT$_4$ in
\cite{Kinney:2005ej,Nakayama:2005mf}, including the
$\mathcal{N}\!=\!4$ Yang-Mills theory. Similar quantity called the
elliptic genus was also studied in 2 dimensional SCFT
\cite{Schellekens:1986yi}. The latter index played a major role in
understanding supersymmetric black holes in AdS$_3$/CFT$_2$
\cite{Strominger:1996sh}.

In this paper, we study the superconformal index in AdS$_4$/CFT$_3$.

Recently, based on the idea of using superconformal Chern-Simons
theory \cite{Schwarz:2004yj} to describe low energy dynamics of
M2-branes, and after the first discovery of a class of
$\mathcal{N}\!=\!8$ superconformal Chern-Simons theories
\cite{Bagger:2006sk,Gustavsson:2007vu}, $\mathcal{N}\!=\!6$
superconformal Chern-Simons theory with gauge group $U(N)_k\times
U(N)_{-k}$ has been found and studied, where the integers $k$ and
$-k$ denote the Chern-Simons levels associated with two gauge groups
\cite{Aharony:2008ug}. This theory describes the low energy dynamics
of $N$ parallel M2-branes placed at the tip of
$\mathbb{C}^4/\mathbb{Z}_k$, and is proposed to be dual to M-theory
on $AdS_4\times S^7/\mathbb{Z}_k$. See
\cite{Benna:2008zy,Hosomichi:2008jb,Bandres:2008ry,Aharony:2008gk}
for further studies of this theory.

Various tests have been made for this proposal. For instance,
studies of the moduli space \cite{Aharony:2008ug} and $D2$ branes
\cite{Mukhi:2008ux,Aharony:2008ug}, chiral operators
\cite{Aharony:2008ug,Hanany:2008qc}, higher derivative correction in
the broken phase \cite{Hosomichi:2008ip,Baek:2008ws} (see also
\cite{Alishahiha:2008rs}), integrability and nonsupersymmetric
spectrum \cite{integ} have been made. Many of these tests, perhaps
except the last example above, rely on supersymmetry in some form.

One of the most refined application of supersymmetry to test this
duality so far would be the calculation of the superconformal index
in the type IIA limit \cite{Bhattacharya:2008bja}. See also
\cite{Choi:2008za,Dolan:2008vc}.
The superconformal index was calculated in the
't Hooft limit, in which $N$ and $k$ are taken to be large keeping
$\lambda=\frac{N}{k}$ finite. The authors of
\cite{Bhattacharya:2008bja} argue that $\lambda$ can be regarded as
a continuous parameter in the 't Hooft limit. The index calculated in
the free Chern-Simons theory, $\lambda\rightarrow 0$, is expected to
be the same as the index in the opposite limit $\lambda\gg 1$ from
the continuity argument. M-theory can be approximated by type
IIA supergravity on $AdS_4\times\mathbb{CP}^3$ in the latter limit,
where $\mathbb{CP}^3$ appears as the base space in the Hopf
fibration of $S^7/\mathbb{Z}_k$ \cite{Aharony:2008ug}. The index
over multiple type IIA gravitons perfectly agreed with the gauge
theory result \cite{Bhattacharya:2008bja}.

It is tempting to go beyond the 't Hooft limit and calculate the
full superconformal index for any value of $k$ (and $N$) for this
theory. This is the main goal of this paper. The general index is
expected to capture the contribution from states carrying
Kaluza-Klein (KK) momenta along the fiber circle of
$S^7/\mathbb{Z}_k$, or $D0$ brane charges from type IIA point of
view. As $k$ increases, the radius of the circle decreases as
$\frac{1}{k}$, and the energy of the KK states would grow. So even
when $k$ is very large that weakly coupled type IIA string theory is
reliable, our index captures nonperturbative correction to
\cite{Bhattacharya:2008bja} from heavy $D0$ branes.

The gauge theory dual to the KK-momentum is argued to be appropriate
magnetic flux on $S^2$ \cite{Aharony:2008ug}. The gauge theory
operators creating magnetic fluxes are called the monopole
operators, or 't Hooft operators
\cite{Aharony:2008ug,Berenstein:2008dc,Klebanov:2008vq,Borokhov:2002ib}.
These operators are not completely understood to date.

Just like the ordinary partition function at finite temperature, the
superconformal index for a radially quantized SCFT admits a path
integral representation. In this paper we calculate the index from
this path integral for the $\mathcal{N}\!=\!6$ Chern-Simons theory
on Euclidean $S^2\times S^1$. The index in the sector with
nonzero KK-momentum is given by integrating over configurations
carrying nonzero magnetic fluxes, which will turn out to be fairly
straightforward. Therefore, lack of our understanding on 't Hooft
operators will not cause a problem for us. In fact, monopole operators
have been most conveniently studied in radially quantized theories
\cite{Borokhov:2002ib}.

Our computation is based on the fact that this path integral is
`supersymmetric,' or has a fermionic symmetry. As a Witten index,
the superconformal index acquires contribution from states
preserving a particular pair of supercharges which are mutually
Hermitian conjugates. Calling one of them $Q$, which is nilpotent,
$Q^2=0$, the fermionic symmetry of the integral is associated with
$Q$. An integral of this kind can be computed by localization. See
\cite{Witten:1992xu} and related references therein for a
comprehensive discussion. A simple way of stating the idea is that
one can deform the integrand by adding a $Q$-exact term $QV$ in the
measure, for any gauge invariant expression $V$, without changing
the integral. For a given $V$, one can add $tQV$ to the action
$S\!\rightarrow\!S+tQV$ where $t$ is a continuous parameter. With a
favorable choice of $V$ as will be explained later, $t$ can be
regarded as a continuous coupling constant of the deformed action
admitting a `free' theory limit as $t\rightarrow\infty$. As already
mentioned in \cite{Kinney:2005ej}, it suffices for the deformed
action to preserve only a subset of the full superconformal
symmetry, involving $Q$ and symmetries associated with charges with
which we grade the states in the index. We only take advantage of a
nilpotent symmetry rather than full superconformal symmetry.

As in \cite{Kinney:2005ej,Sundborg:1999ue,Aharony:2003sx,Aharony:2005bq},
our result is given by an integral of appropriate unitary matrices.

We use our superconformal index to provide a nontrivial test of the
$\mathcal{N}\!=\!6$ AdS/CFT proposal. The readers may also regard it
as subjecting our calculation to a test against known results from
gravity, if they prefer to. In the large $N$ limit, still keeping
$k$ finite, our index is expected to be the index over
supersymmetric gravitons of M-theory at low energy. In the sectors
with one, two and three units of magnetic fluxes (KK-momenta), we
provide analytic calculations or evaluate the unitary matrix
integral numerically, up to a fairly nontrivial order, to see that
the two indices perfectly agree. Similar comparisons with gravitons
in dual string theories are made in other dimensions, e.g. for the 2
dimensional elliptic genus index \cite{de Boer:1998ip} and also for
the 4 dimensional index in $\mathcal{N}\!=\!4$ Yang-Mills theory
\cite{Kinney:2005ej}. We will find that monopole operators in
nontrivial representations of $U(N)\times U(N)$, beyond those studied
in \cite{Aharony:2008ug,Berenstein:2008dc,Klebanov:2008vq},
play crucial roles for the two indices to agree.

An interesting question would be whether our index captures
contribution from supersymmetric black holes beyond the low energy
limit. In AdS$_3$/CFT$_2$, the contribution to the elliptic genus
index from BTZ black holes is calculated and further discussed
\cite{Strominger:1996sh}. See also \cite{Mandal:2007ug} for a recent study
of elliptic genus beyond gravitons.
However, in 4 dimension, it has been found
that the index does not capture the contribution from supersymmetric
black holes \cite{Kinney:2005ej} in the large $N$ limit, possibly
due to a cancelation between bosonic and fermionic states. In this
paper, following \cite{Kinney:2005ej}, we consider the large $N$
limit in which chemical potentials are set to order 1 (in the unit given
by the radius of $S^2$). The situation here is somewhat similar to
$d\!=\!4$ in that a deconfinement phase transition at order $1$
temperature like
\cite{Sundborg:1999ue,Aharony:2003sx,Aharony:2005bq} is not found.
However, more comment is given in the conclusion.

The methods developed in this paper can be applied to other
superconformal Chern-Simons theories. $\mathcal{N}\!=\!5$,
$\mathcal{N}\!=\!4$ Chern-Simons-matter theories and the gravity
duals of some of them are studied in
\cite{Hosomichi:2008jb,Aharony:2008gk} and \cite{Gaiotto:2008sd},
respectively. There is an abundance of interesting superconformal
Chern-Simons theories with $\mathcal{N}\!\leq\!3$ supersymmetry.
\cite{Gaiotto:2007qi} provides a basic framework. For example, some
$\mathcal{N}\!=\!3,2$ theories are presented in
\cite{Jafferis:2008qz} with hyper-K\"{a}hler and Calabi-Yau moduli
spaces. Comments on possible applications of our index to these
theories are given in conclusion.

The rest of this paper is organized as follows. In section 2 we
summarize some aspects of $\mathcal{N}\!=\!6$ Chern-Simons theory
and the superconformal index. We also set up the index calculation
and explain our results. In section 3 we consider a large $N$ limit
and compare our result with the index of M-theory gravitons. Section
4 concludes with comments and further directions. Most of the
detailed calculation is relegated to appendices A and B. Appendix C
summarizes the index over M-theory gravitons.

\section{Superconformal index for $\mathcal{N}=6$ Chern-Simons theory}

\subsection{The theory}

The action and supersymmetry of $\mathcal{N}=6$ Chern-Simons-matter
theory are presented and further studied in
\cite{Aharony:2008ug,Benna:2008zy,Hosomichi:2008jb,Bandres:2008ry}.
The Poincare and special supercharges form vector representations of
$SO(6)$ R-symmetry, or equivalently rank 2 antisymmetric
representations of $SU(4)$ with reality conditions:
\begin{equation}
  Q_{IJ\alpha}=\frac{1}{2}\epsilon_{IJKL}\bar{Q}^{KL}_\alpha\ ,\ \
  S^{IJ}_\alpha=\frac{1}{2}\epsilon^{IJKL}\bar{S}_{KL\alpha}\ ,
\end{equation}
where $I,J,K,L\!=\!1,2,3,4$ and $\alpha\!=\!\pm$. Under radial
quantization, the special supercharges are Hermitian conjugate to
the Poincare supercharges: $S^{IJ\alpha}=(Q_{IJ\alpha})^\dag$,
$\bar{S}_{IJ}^{\alpha}=(\bar{Q}^{IJ}_\alpha)^\dag$. There are two
(Hermitian) gauge fields $A_\mu$, $\tilde{A}_\mu$ for $U(N)\times
U(N)$. The matter fields are complex scalars and fermions in ${\bf
4}$ and ${\bf \bar{4}}$ of $SU(4)$, respectively. We write them as
$C_I$ and $\Psi^I_\alpha$. They are all in the bifundamental
representation $({\bf N},{\bf \bar{N}})$ of $U(N)\times U(N)$.

In this paper we are interested in the superconformal index
associated with a special pair of supercharges. We pick $Q\equiv
Q_{34-}$ and $S\equiv S^{34-}$ without losing generality. For our
purpose, it is convenient to decompose the fields in
super-multiplets of $d\!=\!3$, $\mathcal{N}\!=\!2$ supersymmetry
generated by $Q_\alpha\equiv Q_{34\alpha}$. Writing the matter
fields as $C_I=(A_1,A_2,\bar{B}^{\dot{1}},\bar{B}^{\dot{2}})$ and
$\Psi^I\sim(-\psi_2,\psi_1,-\bar{\chi}^{\dot{2}},\bar{\chi}^{\dot{1}})$,
they group into 4 chiral multiplets as
\begin{equation}
  (A_a,\psi_{a\alpha})\ \ {\rm in}\ ({\bf N}, {\bf \bar{N}})\ ,\ \
  (B_{\dot{a}},\chi_{\dot{a}\alpha})\ \ {\rm in}
  \ ({\bf \bar{N}}, {\bf N})\ ,
\end{equation}
where $a=1,2$ and $\dot{a}=\dot{1},\dot{2}$ are doublet indices for
$SU(2)\times SU(2)\subset SU(4)$ commuting with $Q_\alpha$. The
global charges of the fields and supercharges are presented in Table
1. $h_1,h_2,h_3$ are three Cartans of $SO(6)$ in the `orthogonal
2-planes' basis, $\frac{1}{2}(h_1\!\pm\!h_2)$ being the Cartans of the above
$SU(2)\times SU(2)$. $j_3$ is the Cartan of $SO(3)\subset SO(3,2)$.
$\epsilon$ is the energy in radial quantization, or the scale
dimension of operators. $h_4$ is the baryon-like charge commuting
with the $\mathcal{N}\!=\!6$ superconformal group $Osp(6|4)$.
\begin{table}[t]\label{charges}
$$
\begin{array}{c|ccc|cc|c}
  \hline{\rm fields}&h_1&h_2&h_3&j_3&\epsilon& h_4\\
  \hline(A_1,A_2)&(\frac{1}{2},-\frac{1}{2})&(\frac{1}{2},-\frac{1}{2})&
  (-\frac{1}{2},-\frac{1}{2})&0&\frac{1}{2}&\frac{1}{2}\\
  (B_{\dot{1}},B_{\dot{2}})&(\frac{1}{2},-\frac{1}{2})&(-\frac{1}{2},
  \frac{1}{2})&(-\frac{1}{2},-\frac{1}{2})&0&\frac{1}{2}&-\frac{1}{2}\\
  (\psi_{1\pm},\psi_{2\pm})&(\frac{1}{2},-\frac{1}{2})&(\frac{1}{2},
  -\frac{1}{2})&(\frac{1}{2},\frac{1}{2})&\pm\frac{1}{2}&1&\frac{1}{2}\\
  (\chi_{\dot{1}\pm},\chi_{\dot{2}\pm})&(\frac{1}{2},-\frac{1}{2})&
  (-\frac{1}{2},\frac{1}{2})&(\frac{1}{2},\frac{1}{2})&\pm\frac{1}{2}&1
  &-\frac{1}{2}\\
  \hline A_\mu,\tilde{A}_\mu&0&0&0&(1,0,-1)&1&0\\
  \lambda_{\pm},\tilde\lambda_\pm&0&0&-1&\pm\frac{1}{2}&\frac{3}{2}&0\\
  \sigma,\tilde\sigma&0&0&0&0&1&0\\
  \hline Q_\pm&0&0&1&\pm\frac{1}{2}&\frac{1}{2}&0\\
  S^\pm&0&0&-1&\mp\frac{1}{2}&-\frac{1}{2}&0\\
  \hline
\end{array}
$$
\caption{charges of fields and supercharges}
\end{table}

The Lagrangian is presented, among others, in
\cite{Aharony:2008ug,Benna:2008zy}. It is convenient to introduce
auxiliary fields $\lambda_\alpha,\sigma$ and
$\tilde\lambda_\alpha,\tilde\sigma$ which form vector multiplets
together with gauge fields. We closely follow the notation of
\cite{Benna:2008zy}. The action is given by
\begin{equation}
  \mathcal{L}=\mathcal{L}_{CS}+\mathcal{L}_{m}\ ,
\end{equation}
where the Chern-Simons term is given by\footnote{Fields are related
as $(A_\mu,\tilde{A}_\mu)_{{\rm ours}}=-(A_\mu,\hat{A}_\mu)_{\rm
theirs}$, $(\sigma,D,\tilde\sigma,\tilde{D})_{\rm ours}
=(\sigma,D,\hat\sigma,\hat{D})_{\rm theirs}$,
$(\lambda_\alpha,\tilde\lambda_\alpha)_{\rm ours}=(\chi_\alpha,
\hat\chi_\alpha)_{\rm theirs}$, $(A_a,B_{\dot{a}})=(Z^A,W_A)$,
$\frac{k}{4\pi}=2K$.}
\begin{equation}\label{cs-action}
  \mathcal{L}_{CS}=\frac{k}{4\pi}
  {\rm tr}\left(A\wedge dA-\frac{2i}{3}A^3
  +i\bar\lambda\lambda-2D\sigma\right)-\frac{k}{4\pi}
  {\rm tr}\left(\tilde{A}\wedge d\tilde{A}
  -\frac{2i}{3}\tilde{A}^3
  +i\bar{\tilde\lambda}\tilde\lambda-2\tilde{D}\tilde\sigma\right)
\end{equation}
and
\begin{eqnarray}\label{matter-action}
  \hspace*{-0.5cm}\mathcal{L}_{m}&=&
  {\rm tr}\left[\frac{}{}\right.\!\!
  -D_\mu\bar{A}^aD^\mu A_a-D_\mu\bar{B}^{\dot{a}}
  D_\mu B_{\dot{a}}-i\bar\psi^a\gamma^\mu D_\mu\psi_a-i\bar\chi^{\dot{a}}
  \gamma^\mu D_\mu\chi_{\dot{a}}\nonumber\\
  &&\hspace{0.5cm}
  -\left(\sigma A_a-A_a\tilde\sigma\right)
  \left(\bar{A}^a\sigma-\tilde\sigma\bar{A}^a\right)
  -\left(\tilde\sigma B_{\dot{a}}-B_{\dot{a}}\sigma\right)
  \left(\bar{B}^{\dot{a}}\tilde\sigma-\sigma\bar{B}^{\dot{a}}
  \right)\nonumber\\
  &&\hspace{0.5cm}
  +\bar{A}^aDA_a-A_a\tilde{D}\bar{A}^a-B_{\dot{a}}D\bar{B}^{\dot{a}}
  +\bar{B}^{\dot{a}}\tilde{D}B_{\dot{a}}\nonumber\\
  &&\hspace{0.5cm}-i\bar\psi^a\sigma\psi_a+i\psi_a\tilde\sigma\bar\psi^a
  +i\bar{A}^a\lambda\psi_a+i\bar\psi^a\bar\lambda A_a
  -i\psi_a\tilde\lambda\bar{A}^a-iA_a\bar{\tilde\lambda}\bar\psi^a\nonumber\\
  &&\hspace{0.5cm}+i\chi_{\dot{a}}\sigma\bar\chi^{\dot{a}}
  -i\bar\chi^{\dot{a}}\tilde\sigma\chi_{\dot{a}}
  -i\chi_{\dot{a}}\lambda\bar{B}^{\dot{a}}
  -iB_{\dot{a}}\bar\lambda\bar\chi^{\dot{a}}
  +i\bar{B}^{\dot{a}}\tilde\lambda\chi_{\dot{a}}
  +i\bar\chi^{\dot{a}}\bar{\tilde\lambda}B_{\dot{a}}\!\left.\frac{}{}\right]
  +\mathcal{L}_{\rm sup}\ .
\end{eqnarray}
$\mathcal{L}_{\rm sup}$ contains scalar potential and Yukawa
interaction obtained from a superpotential
\begin{equation}
  W=-\frac{2\pi}{k}\epsilon^{ab}
  \epsilon^{\dot{a}\dot{b}}{\rm tr}(A_aB_{\dot{a}}A_bB_{\dot{b}})
\end{equation}
where the fields $A_a,B_{\dot{a}}$ in the superpotential are
understood as chiral superfields
$A_a+\sqrt{2}\theta\psi_a+\theta^2F_{A_a}$ and
$B_{\dot{a}}+\sqrt{2}\theta\chi_{\dot{a}}+\theta^2F_{B_{\dot{a}}}$.
Integrating out the auxiliary fields, one can easily obtain the
expressions for $\sigma,\tilde\sigma,\lambda,\tilde\lambda$ in terms
of the matter fields.

The $\mathcal{N}\!=\!2$ supersymmetry transformation under
$Q_\alpha\!=\!Q_{34\alpha}$ can be obtained from the superfields.
Most importantly,
\begin{eqnarray}
  Q_\alpha\psi_{a\beta}\!&\!=\!&\!\sqrt{2}\epsilon_{\alpha\beta}
  \partial_{\bar{A}^a}\bar{W}\ ,\ \
  Q_\alpha\chi_{\dot{a}\beta}=\sqrt{2}\epsilon_{\alpha\beta}
  \partial_{\bar{B}^{\dot{a}}}\bar{W}\nonumber\\
  Q_\alpha\bar\psi^a_\beta\!&\!=\!&\!-\sqrt{2}i
  (\gamma^\mu)_{\alpha\beta}D_\mu\bar{A}^a
  +\sqrt{2}i\epsilon_{\alpha\beta}\left(\tilde\sigma\bar{A}^a-
  \bar{A}^a\sigma\right)\nonumber\\
  Q_\alpha\bar\chi^{\dot{a}}_\beta\!&\!=\!&\!-\sqrt{2}i
  (\gamma^\mu)_{\alpha\beta}D_\mu\bar{B}^{\dot{a}}
  +\sqrt{2}i\epsilon_{\alpha\beta}\left(\sigma\bar{B}^{\dot{a}}
  -\bar{B}^{\dot{a}}\tilde\sigma\right)\nonumber\\
  Q_\alpha \lambda_\beta\!&\!=\!&\!-\sqrt{2}i\left[\frac{}{}\!
  (\gamma^\mu)_{\alpha\beta}
  (D_\mu\sigma+i\star F_\mu)+\epsilon_{\alpha\beta}D\right]
  \nonumber\label{gaugino-susy}\\
  Q_\alpha\bar\lambda_\beta\!&\!=\!&\!0\label{gaugino-susy}
\end{eqnarray}
where $\epsilon^{012}=1$. Following \cite{Benna:2008zy}, we choose
$(\gamma^\mu)_\alpha^{\ \beta}=(i\sigma^2,\sigma^1,\sigma^3)$ so
that $(\gamma^\mu)_{\alpha\beta}=(-1,-\sigma^3,\sigma^1)$.

We will be interested in the Euclidean version of this theory. The
action and supersymmetry transformation can easily be changed to the
Euclidean one by Wick rotation, i.e. by replacements $x^0=-ix^0_E$,
$A_0=i(A_E)_0$, etc. Note that $D$, playing the role of Lagrange
multiplier in (\ref{cs-action}), (\ref{matter-action}), should be
regarded as an imaginary field. After Wick rotation one obtains
$\gamma_E^\mu=(-\sigma^2,\sigma^1,\sigma^3)$. The spinors, say,
$\psi_a$ and $\bar\psi^a$ are no longer complex conjugates to each
other. The notation of \cite{Benna:2008zy} naturally lets us regard
them as independent chiral and anti-chiral spinors $\psi_\alpha$ and
$\bar\psi_{\dot\alpha}$ in Euclidean 4 dimension, reduced down to
$d\!=\!3$. Indeed, upon identifying $\sigma=A_3$ etc., the kinetic
term plus the coupling to $\sigma,\tilde\sigma$ can be written as
\begin{equation}
  i\bar\psi^{a\alpha}(\gamma^\mu_E)_\alpha^{\ \beta}
  D_\mu\psi_{a\beta}+\bar\psi^{a\alpha}D_3\psi_{a\alpha}\equiv
  -\bar\psi^a_\alpha(\bar\sigma^\mu)^{\alpha\beta}D_\mu\psi_{a\beta}
\end{equation}
in 4 dimensional notation, where
$D_3\psi_a\equiv-i\sigma\psi_a+i\psi_a\tilde\sigma$ and
$(\bar\sigma^\mu)^{\alpha\beta}\equiv i\epsilon^{\alpha\gamma}
(\gamma_E^\mu,-i)_\gamma^{\ \beta}=(1,i\sigma^3,-i\sigma^1,i\sigma^2
)^{\alpha\beta}$. In our computation in appendices, it will be more
convenient to choose a new $SO(3)$ frame for spinors so that
\begin{equation}
  \bar\sigma^\mu=(1,-i\vec\sigma)=(1,-i\sigma^1,-i\sigma^2,-i\sigma^3)\ .
\end{equation}
To avoid formal manipulations in the main text being a bit nasty,
this change of frame will be assumed only in appendices. Similar
rearrangement can be made for $\chi_{\dot{a}},\bar\chi^{\dot{a}}$.

In Euclidean theory, $Q_\alpha\lambda_\beta$ in (\ref{gaugino-susy})
is given by
\begin{equation}
  Q_\alpha \lambda_\beta=-\sqrt{2}i\left[\frac{}{}\!
  (\gamma^\mu_E)_{\alpha\beta}
  (D_\mu\sigma-\star F_\mu)+\epsilon_{\alpha\beta}D\right]\ .
\end{equation}
Configurations preserving two supercharges $Q_\alpha$ are described
by the Bogomolnyi equations $(\star F)_\mu=D_\mu\sigma$ and $D=0$.
The first one is the BPS equation for magnetic monopoles in
Yang-Mills theory, with a difference that $\sigma$ is a composite
field here. See \cite{Hosomichi:2008ip} for related discussions. We
shall shortly deform the theory with a $Q$-exact term. $\sigma$ will
not be a composite field then.

A conformal field theory defined on $\mathbb{R}^{d+1}$ can be
radially quantized to a theory living on $S^d\times\mathbb{R}$,
where $\mathbb{R}$ denotes time. The procedures of radial
quantization are summarized in appendix A. See also
\cite{Bhattacharyya:2007sa,Grant:2008sk} for related discussions.

In the radially quantized theory, one can consider configurations in
which nonzero magnetic flux is applied on spatial $S^2$. From the
representations of matter fields under $U(N)\times U(N)$, one finds
that ${\rm tr}F={\rm tr}\tilde{F}$ should be satisfied. The
Kaluza-Klein momentum in the dual M-theory along the fiber circle of
$S^7/\mathbb{Z}_k$ is given by
\begin{equation}\label{kk-momentum}
  P=\frac{k}{4\pi}\int_{S^2}{\rm tr}F=
  \frac{k}{4\pi}\int_{S^2}{\rm tr}\tilde{F}\ \in~\frac{k}{2}~\mathbb{Z}
\end{equation}
in the gauge theory \cite{Aharony:2008ug}. This, via Gauss' law constraint,
turns out to be proportional to $h_4$ in Table 1 \cite{Aharony:2008ug}.

\subsection{The superconformal index and localization}

The superconformal symmetry of this theory is $Osp(6|4)$, whose
bosonic generators form $SO(6)\times SO(3,2)$. Its Cartans are given
by five charges: $h_1,h_2,h_3$ and $\epsilon,j_3$ in $SO(2)\times
SO(3)\subset SO(3,2)$. Some important algebra involving our special
supercharges is
\begin{equation}
  Q^2=S^2=0\ ,\ \ \{Q,S\}=\epsilon-h_3-j_3\ .
\end{equation}
The first equation says $Q,S$ are nilpotent, while the second one
implies the BPS energy bound $\epsilon\!\geq\!h_3\!+\!j_3$. The
special supercharges $Q,S$ are charged under some Cartans. From
\mbox{Table 1}, four combinations $h_1,h_2$, $\epsilon\!+\!j_3$,
$\epsilon\!-\!h_3\!-\!j_3$ commute with $Q,S$. The last is nothing
but $\{Q,S\}$. The superconformal index for a pair of supercharges
$Q,S$ is given by \cite{Bhattacharya:2008bja,Bhattacharya:2008zy}
\begin{equation}\label{superconf-index}
  I(x,y_1,y_2)={\rm Tr}\left[(-1)^Fe^{-\beta^\prime\{Q,S\}}
  e^{-\beta(\epsilon+j_3)}e^{-\gamma_1h_1-\gamma_2h_2}\right]
\end{equation}
where $x\equiv e^{-\beta}$, $y_1\equiv e^{-\gamma_1}$, $y_2\equiv
e^{-\gamma_2}$. $F$ is the fermion number. The charges we use to
grade the states in the index should commute with $Q$ and $S$
\cite{Kinney:2005ej}. As a Witten index, this function does not
depend on $\beta^\prime$ since it gets contribution only from states
annihilated by $Q$ and $S$.

The above index admits a path integral representation on $S^2\times
S^1$, where the radius of the last circle is given by the inverse
temperature $\beta+\beta^\prime$. Had the operator inserted inside
the trace been $e^{-(\beta+\beta^\prime)\epsilon}$, the measure of
the integral would have been given by the Euclidean action with the
identification $\epsilon=-\frac{\partial}{\partial\tau}$ with
Euclidean time $\tau$. The insertion of $(-1)^F$ would also have
made all fields periodic in
$\tau\sim\tau\!+\!(\beta\!+\!\beta^\prime)$. The actual insertion
$(-1)^Fe^{-(\beta+\beta^\prime)\epsilon-(\beta-
\beta^\prime)j_3+\beta^\prime h_3-\gamma_1 h_1-\gamma_2h_2}$ twists
the boundary condition: alternatively, this twist can be undone by
replacing all time derivatives in the action by
\begin{equation}
  \partial_\tau\rightarrow\partial_\tau-
  \frac{\beta-\beta^\prime}{\beta+\beta^\prime}j_3
  +\frac{\beta^\prime}{\beta+\beta^\prime}h_3
  -\frac{\gamma_1}{\beta+\beta^\prime}h_1-\frac{\gamma_2}{\beta+
  \beta^\prime}h_2\ ,
\end{equation}
leaving all fields periodic. The generators of Cartans assume
appropriate representations depending on the field they act on. The
angular momentum $j_3$ is given for each mode after expanding fields
with spherical harmonics, or with the so called monopole spherical
harmonics \cite{Wu:1976ge} if nontrivial magnetic field is applied
on $S^2$. In actual computation we will often formulate the theory
on $\mathbb{R}^3$ in Cartesian coordinates, with $r=e^\tau$ (see
appendix A and also \cite{Bhattacharyya:2007sa,Grant:2008sk}).
Change in derivatives on $\mathbb{R}^3$ due to the above twist is
\begin{equation}
  \vec\nabla\rightarrow\vec\nabla+\frac{\vec{r}}{r^2}
  \left(-\frac{\beta-\beta^\prime}{\beta+\beta^\prime}j_3
  +\frac{\beta^\prime}{\beta+\beta^\prime}h_3
  -\frac{\gamma_1}{\beta+\beta^\prime}h_1-\frac{\gamma_2}
  {\beta+\beta^\prime}h_2\right)\ .
\end{equation}
From now on our derivatives are understood with this shift, hoping
it will not cause confusion.

Insertion of a $Q$-exact operator $\{Q,V\}$, for any gauge-invariant
operator $V$, to the superconformal index (\ref{superconf-index})
becomes zero due to the $Q$-invariance of the Cartans appearing in
(\ref{superconf-index}) and the periodic boundary condition for the
fields due to $(-1)^F$ \cite{Cecotti:1981fu}.\footnote{ In
localization calculations in different contexts, $Q^2$ is often zero
up to a gauge transformation. In this case $Q$ defines the so called
\textit{equivariant} cohomology, in which case $V$ should be gauge
invariant. Although $Q^2\!=\!0$ in our case, we simply choose $V$ to
be gauge-invariant even if we do not seem to be forced to.} From the
nilpotency of $Q$, the operator $e^{-t\{Q,V\}}$ takes the form
$1+Q(\cdots)$ for a given $V$ and a continuous parameter $t$.
Therefore, in the path integral representation, we may add the
$Q$-exact term to the action $S\rightarrow S+t\{Q,V\}$ without
changing the integral. The parameter $t$ can be set to a value with
which the calculation is easiest.\footnote{There is a subtle caveat
in this argument when the integration domain is non-compact. Since
we integrate over the non-compact space of fields, irrelevant saddle
points may `flow in from infinity' as we change $t$. See
\cite{Witten:1992xu} for more explanation. Fortunately, this problem
appears to be absent with the saddle points we find below.} In
particular, by suitably choosing $V$, setting $t\rightarrow+\infty$
may be regarded as a semi-classical limit with $t$ being
$\hbar^{-1}$. This semi-classical or Gaussian `approximation' then
provides the exact result since the integral is $t$-independent.

We choose to deform the action of the $\mathcal{N}\!=\!6$
Chern-Simons theory by a $Q$-exact term which looks similar to the
$d\!=\!3$ $\mathcal{N}\!=\!2$ `Yang-Mills' action as follows. Using
the gaugino superfield (Euclidean)
\begin{equation}
  \mathcal{W}_\alpha(y)\sim-\sqrt{2}i\lambda_\alpha(y)+2D(y)\theta_\alpha+
  \left(\gamma^\mu\theta\right)_{\alpha}\left(D_\mu\sigma-\star
  F_\mu\right)(y)+\sqrt{2}\theta^2(\gamma^\mu D_\mu\bar\lambda(y))_\alpha
\end{equation}
with $y^\mu=x^\mu+i\theta\gamma^\mu\bar\theta$, we add
\begin{equation}
  t\{Q,V\}=\left.\frac{1}{g^2}\int d^3x\ r
  \mathcal{W}^\alpha\mathcal{W}_\alpha\right|_{\theta^2\bar\theta^0}
  \ \ \ \ \ \ ({\rm taking}\ t=\frac{1}{g^2}~)
\end{equation}
to the original action. Let us provide supplementary explanations.
The multiplication of $r$ in the integrand makes this term scale
invariant: it is crucial to have this symmetry since it will be our
time translation symmetry after radial quantization. Of course
translation symmetry on $\mathbb{R}^3$ is broken, which does not
matter to us. It is also easy to show that the above deformation is
$Q$-exact. Taking the coefficient of $\theta^2$ is equivalent to
$\partial_{\theta^-}\partial_{\theta^+}$, which in turn is related
to $Q_\alpha$ by
$Q_\alpha=\partial_\alpha-i(\gamma^\mu\bar\theta)_\alpha\partial_\mu$.
However, since we are keeping terms with $\bar\theta^0$,
$\partial_\alpha$ is effectively $Q_\alpha$. Furthermore, $y^\mu$
can be replaced by $x^\mu$ for the same reason. Therefore the added
term indeed takes the form $Q_-Q_+(\cdots)$ with $y^\mu\rightarrow
x^\mu$ everywhere. Finally, note that we do not add a term of the
form $\int d^2\bar\theta\
\bar{\mathcal{W}}_\alpha\bar{\mathcal{W}}^\alpha$. Expanding in
components, one finds
\begin{equation}\label{exact-deform}
  \Delta S=t\{Q,V\}=\frac{1}{2g^2}\int_{1\leq r\leq e^{\beta\!+\!
  \beta^\prime}}\!\!\!d^3x\ r\left[\frac{}{}\!\!\left(\star F_\mu-D_\mu\sigma
  \right)^2-D^2+\lambda^\alpha(\sigma^\mu)_{\alpha\beta}
  D_\mu\bar\lambda^\beta\right]
\end{equation}
where $\sigma^\mu=(1,-i\sigma^3,i\sigma^1,-i\sigma^2)$ in the basis
of \cite{Benna:2008zy}, or $(1,i\vec\sigma)$ in the basis we use in
the appendix, and $D_3\bar\lambda^\beta=-i\sigma\bar\lambda^\beta+
i\bar\lambda^\beta\sigma$. We already turned $D$ to an imaginary
field during Wick rotation, which makes $-D^2$ positive. Everything
goes similarly for the other vector multiplet
$\tilde{A}_\mu,\tilde\sigma,\tilde\lambda_\alpha$. Some 1-loop study
has been made for `ordinary' Yang-Mills Chern-Simons theories
\cite{Kao:1995gf}. Due to various differences in our construction,
we will obtain very different results.

\subsection{Calculation of the index}

Having set up the localization problem in the previous subsection,
we first find saddle points in the limit $g\rightarrow 0$, and then
compute the 1-loop determinants around them.

All fields are subject to the periodic boundary condition along
$S^1$, or the radial direction: working with fields in
$\mathbb{R}^3$, the equivalent boundary condition is
\begin{equation}\label{boundary}
  \Psi(r=e^{\beta})=e^{-\beta\Delta_\Psi}\Psi(r=1)\ ,
\end{equation}
where $\Delta_\Psi$ is the scale dimension of the field $\Psi$. See
appendix A and \cite{Bhattacharyya:2007sa,Grant:2008sk}.

The saddle point equations, which can be deduced either from
(\ref{exact-deform}) or from the supersymmetry transformation, are
given by
\begin{equation}\label{saddle-eqn}
  \star F_\mu=D_\mu\sigma\ ,\ \ D=0\ ,\ \
  \star \tilde{F}_\mu=D_\mu\tilde\sigma\ ,\ \ \tilde{D}=0\ .
\end{equation}
Note $\sigma,\tilde\sigma$ are no longer composite fields. From the
supersymmetry transformation of matter fermions, we find that only
$A_a=B_{\dot{a}}=0$ satisfies the above boundary condition. All
fermions are naturally set to zero. An obvious solution for the
fields $F_{\mu\nu},\sigma$ and $\tilde{F}_{\mu\nu},\tilde\sigma$ is
Dirac monopoles in the diagonals $U(1)^N\times U(1)^N\subset
U(N)\times U(N)$:
\begin{equation}
  \star F_\mu=\frac{x_\mu}{2r^3}
  {\rm diag}(n_1,n_2,\cdots,n_N)\ ,\ \
  \star\tilde{F}_\mu=\frac{x_\mu}{2r^3}{\rm diag}(\tilde{n}_1,\tilde{n}_2,
  \cdots,\tilde{n}_N)\ ,
\end{equation}
together with
\begin{equation}
  \sigma=-\frac{1}{2r}{\rm diag}(n_1,n_2,\cdots,n_N)\ ,\ \
  \tilde\sigma=-\frac{1}{2r}{\rm diag}(\tilde{n}_1,\tilde{n}_2,
  \cdots,\tilde{n}_N)\ .
\end{equation}
Since we are considering the region $1\leq r\leq e^\beta$ in
$\mathbb{R}^3$ excluding the origin $r=0$, this solution is regular.
The spherical symmetry of the solution implies that there is no
twisting in derivatives. Note that all fields satisfy the boundary
condition (\ref{boundary}) with
$\Delta_\sigma=\Delta_{\tilde\sigma}=1$ and
$\Delta_{F}=\Delta_{\tilde{F}}=2$. The coefficients $n_i$ and
$\tilde{n}_i$ ($i=1,2,\cdots,N$) have to be integers since the
diagonals of $\int_{S^2}F$ and $\int_{S^2}\tilde{F}$ are $2\pi$
times integers.\footnote{More generally, one can consider fractional
fluxes $n_i=m_i+\frac{B}{k}$, $\tilde{n}_i=\tilde{m}_i+\frac{B}{k}$
where $m_i,\tilde{m}_i,B$ are integers. See
\cite{Klebanov:2008vq,Berenstein:2009sa} for details. Since fields
couple to differences of fluxes, $B$ appears only through the phase
factor (\ref{phase}) in (\ref{exact-index}), and in particular
disappears in our large $N$ calculation in section 3.}

Apart from the above Abelian solutions, we could not find any other
solutions satisfying the boundary condition. For instance, although
the governing equation is the same, non-Abelian solutions like the
embedding of $SU(2)$ 't Hooft-Polyakov monopoles are forbidden since
the boundary condition is not met. There still is a possibility that
deformation of derivatives from twisting might play roles, which we
have not fully ruled out. Anyway, we shall only consider the above
saddle points and find agreement with the graviton index, which we
regard as a strong evidence that we found all relevant saddle
points.

To the above solution, one can also superpose holonomy zero modes
along $S^1$ as follows. Since the solution is diagonal, turning on
constant $A_\tau,\tilde{A}_\tau$ diagonal in the same basis
obviously satisfies (\ref{saddle-eqn}). In terms of fields
normalized in $\mathbb{R}^3$, this becomes
\begin{equation}
  A_r=\frac{1}{(\beta\!+\!\beta^\prime)r}
  {\rm diag}(\alpha_1,\alpha_2,\cdots,\alpha_N)
  \ ,\ \ \tilde{A}_r=\frac{1}{(\beta\!+\!\beta^\prime)r}{\rm diag}
  (\tilde\alpha_1,\tilde\alpha_2,\cdots,\tilde\alpha_N)
\end{equation}
where we insert factors of $\beta\!+\!\beta^\prime$ for later
convenience. Taking into account the large gauge transformation
along $S^1$, the coefficients $\alpha_i$, $\tilde\alpha_i$ are all
periodic:
\begin{equation}
  \alpha_i\sim\alpha_i+2\pi\ ,\ \ \tilde\alpha_i\sim\tilde\alpha_i+2\pi\
  \ \ (i=1,2,\cdots,N).
\end{equation}
This holonomy along the time circle is related to the Polyakov loop
\cite{Aharony:2003sx}. The full set of saddle points is parametrized
by integer fluxes $\{n_i,\tilde{n}_i\}$ and the holonomy
$\{\alpha_i,\tilde\alpha_i\}$.

The analysis around the saddle point in which all $n_i$,
$\tilde{n}_i$ vanish was done in \cite{Bhattacharya:2008bja}. We
provide the 1-loop analysis around all saddle points. We explain
various ingredients in turn: classical contribution, gauge-fixing
and Faddeev-Popov measure, 1-loop contribution, Casimir-like energy
shift and finally the full answer.

We first consider the `classical' action. Plugging in the saddle
point solution to the action, the classical action proportional to
$g^{-2}$ is always zero, as expected since our final result should
not depend on $g$. Since the classical action itself involves two
classes of terms, one of order $\mathcal{O}(g^{-2})$ from $Q$-exact
deformation and another $\mathcal{O}(g^0)$ from original action, one
should keep the latter part of the classical action to do the
correct 1-loop physics. This comes from the Chern-Simons term. To
correctly compute it, one has to extend $S^2\times S^1$ to a
4-manifold $\mathcal{M}_4$ bounding it, and use
$\frac{1}{4\pi}\int_{S^2\times S^1}{\rm tr} A\wedge F=
\frac{1}{4\pi}\int_{\mathcal{M}_4}{\rm tr}F\wedge F$. If one chooses
a disk $D_2$ bounded by $S^1=\partial D_2$ and take
$\mathcal{M}_4=S^2\times D_2$, one finds
\begin{equation}
  \frac{1}{4\pi}{\rm tr}\int_{S^2\times D_2}F\wedge F=
  \frac{1}{2\pi}{\rm tr}\int_{D_2}F\cdot\int_{S^2} F=
  \frac{1}{2\pi}\int_{S^1}A\cdot\int_{S^2}F=\sum_{i=1}^Nn_i\alpha_i\ .
\end{equation}
Taking into account the second gauge field $\tilde{A}_\mu$ at level
$-k$, the exponential of the $\mathcal{O}(g^0)$ part of the
classical action is a phase given by
\begin{equation}\label{phase}
  e^{ik\sum_{i=1}^N(n_i\alpha_i-\tilde{n}_i\tilde\alpha_i)}\ .
\end{equation}
The exponent is linear in $\alpha_i,\tilde\alpha_i$. Apart from these
linear terms, there are no quadratic terms in the holonomy
(around any given value) in the rest of the classical action. Thus Gaussian
approximation is not applicable and they should be treated exactly.
We shall integrate over them after applying Gaussian approximation to all
other degrees.

Before considering 1-loop fluctuations, we fix the gauge. Following
\cite{Aharony:2003sx} we choose the Coulomb gauge, or the background
Coulomb gauge for the components of fluctuations coupled to the
background. In the calculation around the saddle point where all
fields are zero, the Faddeev-Popov determinant computed following
\cite{Aharony:2003sx} is that for the $U(N)\times U(N)$ unitary
matrices with eigenvalues
$\{e^{i\alpha_i}\},\{e^{i\tilde\alpha_i}\}$
\begin{equation}
  \prod_{i<j}\left[2\sin\left(\frac{\alpha_i\!-\!\alpha_j}{2}\right)\right]^2
  \prod_{i<j}\left[2\sin\left(\frac{\tilde\alpha_i\!-\!\tilde\alpha_j}{2}
  \right)\right]^2\ .
\end{equation}
In the saddle point with nonzero fluxes, the flux effectively
`breaks' $U(N)\times U(N)$ to an appropriate subgroup. For instance,
with fluxes $\{3,2,2,0,0\}$ on $U(1)^5\subset U(5)$, $U(5)$ is
broken to $U(1)\times U(2)\times U(2)$. As explained in appendix B,
the Faddeev-Popov measure for the saddle point with flux is the
unitary matrix measure for the \textit{unbroken subgroup} of
$U(N)\times U(N)$:
\begin{equation}
  \prod_{i<j;n_i=n_j}
  \left[2\sin\left(\frac{\alpha_i\!-\!\alpha_j}{2}\right)\right]^2
  \prod_{i<j;\tilde{n}_i=\tilde{n}_j}
  \left[2\sin\left(\frac{\tilde\alpha_i\!-\!\tilde\alpha_j}{2}\right)\right]^2
\end{equation}
where the restricted products keep a sine factor for a pair of
eigenvalues in the same unbroken gauge group only.

The fluctuations of fields with nonzero quadratic terms in the
action can be treated by Gaussian approximation. The result consists
of several factors of determinants from matter scalars, fermions and
also from fields in vector multiplets, which is schematically
\begin{equation}\label{schematic-det}
  \frac{\det_{\psi_a,\chi_a}\det_{\lambda}\det_{\tilde\lambda}}
  {\det_{A_a,B_a}\det_{A_\mu,\sigma}\det_{\tilde{A}_\mu,\tilde\sigma}}\ .
\end{equation}
The fields in $U(N)\times U(N)$ vector multiplets will turn out to
provide nontrivial contribution.

We first consider the 1-loop determinant from matter fields, namely
$A_a,B_{\dot{a}},\psi_a,\chi_{\dot{a}}$ and their conjugates. In
this determinant, the index over the so-called `letters' play
important roles.\footnote{Letters are defined by single basic fields
with many derivatives acting on them
\cite{Aharony:2003sx,Kinney:2005ej,Bhattacharya:2008bja}. In path
integral calculation, they are simply (monopole) spherical harmonics
modes of the fields.} To start with, we consider the basic fields
$A_a,\bar{B}^{\dot{a}},\psi_a,\bar\chi^{\dot{a}}$ in ($N$,$\bar{N}$)
representation of the gauge group and pick up the $ij$'th component,
where $i$ ($j$) runs over $1,2,\cdots,N$ and refers to the
fundamental (anti-fundamental) index of first (second) $U(N)$. In
the quadratic Lagrangian, these modes couple to the background
magnetic field with charge $n_i\!-\!\tilde{n}_j$. The index over
letters in this component is given by
\begin{equation}\label{letter-bif}
  f^+_{ij}(x,y_1,y_2)=x^{|n_i\!-\!\tilde{n}_j|}\left[
  \frac{x^{1/2}}{1-x^2}
  \left(\!\sqrt{\frac{y_1}{y_2}}+\!\sqrt{\frac{y_2}{y_1}}\right)-
  \frac{x^{3/2}}{1-x^2}
  \left(\!\sqrt{y_1y_2}+\!\frac{1}{\sqrt{y_1y_2}}\right)\right]\equiv
  x^{|n_i\!-\!\tilde{n}_j|}f^+(x,y_1,y_2)\ .
\end{equation}
See appendix B.1 for the derivation. Similarly, the index over the
$ij$'th component of letters in ($\bar{N}$,$N$) representation is
given by
\begin{equation}\label{letter-antibif}
  f^-_{ij}(x,y_1,y_2)=x^{|n_i\!-\!\tilde{n}_j|}
  \left[\frac{x^{1/2}}{1-x^2}
  \left(\!\sqrt{y_1y_2}+\!\frac{1}{\sqrt{y_1y_2}}\right)-
  \frac{x^{3/2}}{1-x^2}
  \left(\!\sqrt{\frac{y_1}{y_2}}+\!\sqrt{\frac{y_2}{y_1}}\right)
  \right]\equiv x^{|n_i\!-\!\tilde{n}_j|}f^-(x,y_1,y_2)\ .
\end{equation}
There is no dependence on the regulator $\beta^\prime$, as expected.
Again see appendix B.1 for details. With these letter indices, the
1-loop determinant for given $\alpha_i,\tilde\alpha_i$ is given by
\begin{equation}
  \prod_{i,j=1}^N\exp\left[\sum_{n=1}^\infty\frac{1}{n}\left(
  f^+_{ij}(x^n,y_1^n,y_2^n)e^{in(\tilde\alpha_j\!-\!\alpha_i)}+
  f^-_{ij}(x^n,y_1^n,y_2^n)e^{in(\alpha_i\!-\!\tilde\alpha_j)}\right)
  \right]\ .
\end{equation}
The expression $\exp\left[\sum_{n=1}^\infty\frac{1}{n}f(x^n)\right]$
above, sometimes called the Plethystic exponential of a function
$f(x)$, appears since we count operators made of identical letters
\cite{Benvenuti:2006qr}. Note that the determinant around the saddle
point in which all fluxes are zero, $n_i=\tilde{n}_i=0$, reduces to
the result obtained in \cite{Bhattacharya:2008bja} using
combinatoric methods in the free theory. When all $n_i,\tilde{n}_i$
are zero, our letter indices $f^\pm_{ij}$ all reduces to $f^\pm$,
which are exactly the letter indices obtained in \cite{Bhattacharya:2008bja}.

We also consider the determinant from fields in vector multiplets.
Here, the letter index over the $ij$'th component of the adjoint
fields $A_\mu,\lambda_\alpha,\sigma$ is given by
\begin{equation}\label{letter-adj1}
  f^{{\rm adj}}_{ij}(x)=-(1-\delta_{n_in_j})x^{|n_i\!-\!n_j|}
  =\left\{\begin{array}{ll}
  0&{\rm if}\ n_i=n_j\\-x^{|n_i\!-\!n_j|}&{\rm if}\ n_i\neq n_j\end{array}
  \right.\ ,
\end{equation}
and similarly for the $ij$'th component of the fields
$\tilde{A}_\mu,\tilde\lambda_\alpha,\tilde\sigma$ one finds
\begin{equation}\label{letter-adj2}
  \tilde{f}^{{\rm adj}}_{ij}(x)=-(1-\delta_{\tilde{n}_i\tilde{n}_j})
  x^{|\tilde{n}_i\!-\!\tilde{n}_j|}
  =\left\{\begin{array}{ll}0&{\rm if}\ \tilde{n}_i=\tilde{n}_j\\
  -x^{|\tilde{n}_i-\tilde{n}_j|}&{\rm if}\ \tilde{n}_i\neq\tilde{n}_j
  \end{array}\right.\ .
\end{equation}
The full 1-loop determinant from adjoint fields is again given by
the same exponential:
\begin{equation}
  \prod_{i,j=1}^N\exp\left[\sum_{n=1}^\infty\frac{1}{n}\left(
  f^{{\rm adj}}_{ij}(x^n)e^{-in(\alpha_i-\alpha_j)}+
  \tilde{f}^{{\rm adj}}_{ij}(x^n)e^{-in(\tilde\alpha_i-\tilde\alpha_j)}
  \right)  \right]\ .
\end{equation}
For the trivial vacuum with no fluxes, all $f^{\rm adj}_{ij}$ and
$\tilde{f}^{\rm adj}_{ij}$ are zero that the adjoint determinant is
simply 1. This is consistent with the result in
\cite{Bhattacharya:2008bja}, where the vector multiplets including
gauge fields played no roles. See appendix B.2 for the derivation.

When evaluating the determinant in appendix B, one encounters an
overall factor
\begin{equation}
  \exp\left[-\beta\epsilon_0\right]\ ,\ \ \ {\rm where}\ \ \ \
  \epsilon_0\equiv\frac{1}{2}{\rm
  tr}\left[(-1)^F(\epsilon+j_3)\right]\ .
\end{equation}
This is similar to the ground state energy traced over all modes,
twisted by $j_3$ basically because we are only considering charges
commuting with $Q,S$ in our index. Although this is not just energy
due to $j_3$, we slightly abuse terminology and call this quantity
Casimir energy. With appropriate regulator respecting supersymmetry,
this is given by
\begin{equation}\label{casimir}
  \epsilon_0=\sum_{i,j=1}^N|n_i\!-\!\tilde{n}_j|-\sum_{i<j}|n_i\!-\!n_j|
  -\sum_{i<j}|\tilde{n}_i\!-\!\tilde{n}_j|\ .
\end{equation}
See appendix B.3 for the derivation and some properties of
$\epsilon_0$. In particular, $\epsilon_0$ is non-negative and
becomes zero if and only if the two sets of flux distributions
$\{n_i\}$, $\{\tilde{n}_i\}$ are identical. Some features of this
energy shift related to AdS/CFT is discussed in the next section.

Finally we integrate over the modes $\alpha_i$ and $\tilde\alpha_j$
with all factors explained above in the measure. The result for a
given saddle point labeled by $\{n_i\},\{\tilde{n}_i\}$ is
\begin{equation}\label{exact-index}
\boxed{\begin{gathered}
  \hspace*{-0.8cm}
  I(x,y_1,y_2)=
  x^{\epsilon_0}\int\frac{1}{\rm (symmetry)}
  \left[\frac{d\alpha_id\tilde\alpha_i}{(2\pi)^2}\right]
  \prod_{\substack{i<j;\\ n_i=n_j}}
  \left[2\sin\left(\frac{\alpha_i\!-\!\alpha_j}{2}\right)\right]^2
  \prod_{\substack{i<j;\\ \tilde{n}_i=\tilde{n}_j}}
  \left[2\sin\left(\frac{\tilde\alpha_i\!-\!\tilde\alpha_j}{2}\right)
  \right]^2\\
  \hspace{2cm}\times e^{ik\sum_{i=1}^N(n_i\alpha_i\!-\!
  \tilde{n}_i\tilde\alpha_i)}\!
  \prod_{i,j=1}^N\!\exp\left[\sum_{n=1}^\infty\frac{1}{n}\left(
  f^+_{ij}(x^n,y_1^n,y_2^n)e^{in(\tilde\alpha_j\!-\!\alpha_i)}+
  f^-_{ij}(x^n,y_1^n,y_2^n)e^{in(\alpha_i\!-\!\tilde\alpha_j)}\right)
  \right]\\
  \hspace{-2.3cm}
  \times\!\prod_{i,j=1}^N\!\exp\left[\sum_{n=1}^\infty\frac{1}{n}
  \left(f^{{\rm adj}}_{ij}(x^n)e^{-in(\alpha_i\!-\!\alpha_j)}+
  \tilde{f}^{{\rm adj}}_{ij}(x^n)e^{-in(\tilde\alpha_i\!-\!\tilde\alpha_j)}
  \right)\right]
\end{gathered}}
\end{equation}
with (\ref{letter-bif}), (\ref{letter-antibif}),
(\ref{letter-adj1}), (\ref{letter-adj2}), (\ref{casimir}) for the
definitions of various functions. The symmetry factor on the first
line divides by the factor of identical variables among
$\{\alpha_i\},\{\tilde\alpha_i\}$ according to `unbroken' gauge
group. For the $U(5)\!\rightarrow\!U(1)\times U(2)\times U(2)$
example above, this factor is $\frac{1}{1!\times 2!\times 2!}$. The
full index is the sum of (\ref{exact-index}) for all flux
distributions $\{n_i\}$, $\{\tilde{n}_i\}$.

Apart from the first phase factor on the second line,  the integrand
is invariant under the overall translation of
$\alpha_i,\tilde\alpha_i$. Therefore, the integral vanishes unless
$\sum_i n_i=\sum_i\tilde{n}_i$. This is of course a consequence of a
decoupled $U(1)$ as explained in \cite{Aharony:2008ug}. The
KK-momentum, or $\frac{k}{2}$ times the baryon-like charge, for
states counted by the above index is given by
(\ref{kk-momentum}). From the structure of this integral and the
letter indices, it is also easy to infer that the energy of the
states contributing to this index is bounded from below by
\begin{equation}\label{energy-bound}
  \epsilon\geq\frac{k}{2}\sum_{i=1}^N n_i=\frac{k}{2}
  \sum_{i=1}^N \tilde{n}_i
\end{equation}
if the two flux distributions $\{n_i\}$, $\{\tilde{n}_i\}$ are
identical. If the two distributions are different, the energy is
strictly larger than this bound. We discuss this in the next
section.

There is a unifying structure in the integrand of the above index if
one combines the matrix integral measure on the first line to the
last line. Note that the measure can be written as
\begin{equation}
  \prod_{\substack{i<j;\\ n_i=n_j}}\!
  \left[2\sin\!\left(\!\frac{\alpha_i\!-\!\alpha_j}{2}\!\right)
  \right]^2\!\!\prod_{\substack{i<j;\\ \tilde{n}_i=\tilde{n}_j}}\!
  \left[2\sin\!\left(\!\frac{\tilde\alpha_i\!-\!\tilde\alpha_j}{2}\!
  \right)\right]^2\!=\prod_{i\neq j}\exp\!
  \left[-\sum_{n=1}^\infty\frac{1}{n}\left(
  \delta_{n_in_j}e^{-in(\alpha_i\!-\!\alpha_j)}+
  \delta_{\tilde{n}_i\tilde{n}_j}e^{-in(\tilde\alpha_i\!-\!\tilde\alpha_j)}
  \right)\right]\ .\nonumber
\end{equation}
As in \cite{Sundborg:1999ue,Aharony:2003sx,Kinney:2005ej}, this
provides a 2-body repulsive effective potentials between
$\alpha_i$'s and $\tilde\alpha_i$'s in the same unbroken gauge
group. From the form of adjoint letter indices in
(\ref{letter-adj1}), (\ref{letter-adj2}), the above measure combines
with the last line and become
\begin{equation}
  \prod_{i\neq j}\exp
  \left[-\sum_{n=1}^\infty\frac{1}{n}\left(
  x^{n|n_i\!-\!n_j|}e^{-in(\alpha_i\!-\!\alpha_j)}+
  x^{n|\tilde{n}_i\!-\!\tilde{n}_j|}e^{-in(\tilde\alpha_i\!-\!\tilde\alpha_j)}
  \frac{}{}\!\right)\right]\ .
\end{equation}
Therefore, in the presence of fluxes, there are repulsive 2-body potentials
between all pairs with the strength $\frac{1}{n}$ weakened to
$\frac{1}{n}x^{n|n_i\!-\!n_j|}$ and
$\frac{1}{n}x^{n|\tilde{n}_i\!-\!\tilde{n}_j|}$, by factors of $x$.

\section{Large $N$ limit and index over gravitons}

We further analyze the gauge theory index we obtained in the
previous section in the large $N$ limit. The limit we take is
$N\rightarrow\infty$ while keeping the chemical potentials at order
1. Among other motivations, this setting lets us study the low energy
spectrum which can be compared to that from supergravity.

In this limit, only $\mathcal{O}(1)$, namely $\mathcal{O}(N^0)$,
numbers of $U(1)^N$'s in each $U(N)$ have nonzero magnetic flux.
This is because states with more than $\mathcal{O}(1)$ U(1)'s filled
with nonzero fluxes have energies bigger than $\mathcal{O}(1)$ from
(\ref{energy-bound}) and are suppressed in the large $N$ limit we
take. This implies that there always exist an
$U(N\!-\!\mathcal{O}(1))\times U(N\!-\!\mathcal{O}(1))$ part in the
integral over holonomies $\{\alpha_i,\tilde\alpha_i\}$. Let us call
this unbroken gauge group $U(N_1)\times U(N_2)$, where $N_1$ and
$N_2$ denote numbers of $U(1)$ with zero fluxes. In the large $N$
limit, there is a well-known way of calculating this part of the
integral \cite{Sundborg:1999ue,Aharony:2003sx}. We first introduce
\begin{equation}
  \rho_n=\frac{1}{N_1}\sum_i e^{-in\alpha_i}\ ,\ \
  \chi_n=\frac{1}{N_2}\sum_i e^{-in\tilde\alpha_i}\
\end{equation}
for nonzero integers $n$, where the summations are over $N_1$ and
$N_2$ $U(1)$ indices, respectively. In the large $N_1,N_2$ limit,
the integration over $\alpha_i,\tilde\alpha_i$ belonging to
$U(N_1)\times U(N_2)$ becomes
\begin{equation}
  \prod_{n=1}^\infty\left[N_1^2d\rho_nd\rho_{-n}\right]
  \left[N_2^2d\chi_nd\chi_{-n}\right]\ .
\end{equation}
The integrand containing $\rho_n,\chi_n$ is given by
\begin{eqnarray}
  &&\hspace{-1.5cm}\exp\left[-\sum_{n=1}^\infty\frac{1}{n}
  \left(N_1^2\rho_n\rho_{-n}\!+\!N_2^2\chi_n\chi_{-n}\!-\!
  N_1N_2f^+(x^n,y_1^n,y_2^n)
  \rho_n\chi_{-n}\!-\!N_1N_2f^-(x^n,y_1^n,y_2^n)\rho_{-n}\chi_n\right)\right]
  \\
  &&\hspace{-1.5cm}\times\exp\left[N_1\!\sum_{n=1}^\infty\frac{1}{n}
  \rho_n\left(\sum_{i=1}^{M_2}x^{n|\tilde{n}_i|}f^+(\cdot^n)
  e^{in\tilde\alpha_i}\!-\!\sum_{i=1}^{M_1}x^{n|n_i|}e^{in\alpha_i}
  \!\right)\!+\!\frac{1}{n}
  \rho_{-n}\!\left(\sum_{i=1}^{M_2}x^{n|\tilde{n}_i|}f^-(\cdot^n)
  e^{-in\tilde\alpha_i}\!-\!\sum_{i=1}^{M_1}x^{n|n_i|}e^{-in\alpha_i}
  \!\right)\!\right]\nonumber\\
  &&\hspace{-1.5cm}\times\exp\left[N_2\!\sum_{n=1}^\infty\frac{1}{n}
  \chi_n\!\left(\sum_{i=1}^{M_1}x^{n|n_i|}f^-(\cdot^n)
  e^{in\alpha_i}\!-\!\sum_{i=1}^{M_2}x^{n|\tilde{n}_i|}
  e^{in\tilde\alpha_i}\!\right)\!+\!
  \frac{1}{n}
  \chi_{-n}\!\left(\sum_{i=1}^{M_1}x^{n|n_i|}f^+(\cdot^n)
  e^{-in\alpha_i}\!-\!\sum_{i=1}^{M_2}x^{n|\tilde{n}_i|}
  e^{-in\tilde\alpha_i}\!\right)\!\right]
  \nonumber
\end{eqnarray}
where $M_1\equiv N-N_1$ and $M_2\equiv N-N_2$ are numbers (of order
$1$) of $U(1)$'s in two gauge groups with nonzero fluxes, and
$\cdot\ {}^n$ denotes taking $n$'th powers of all arguments
$x,y_1,y_2$. The integral of $\rho_n,\chi_n$ is Gaussian, where the
first line (with $N_1=N_2=N$) is the one encountered in
\cite{Bhattacharya:2008bja}. After this integration, one obtains
\begin{equation}\label{large-N-gaussian}
  I^{(0)}\exp\left[\frac{1}{2}\sum_{n=1}^\infty\frac{1}{n}
  V^T(\cdot\ {}^n)M(\cdot\ {}^n)V(\cdot\ {}^n)\right]
\end{equation}
where
\begin{equation}
  V=\left(\begin{array}{c}\displaystyle{
  \sum_{i=1}^{M_2}x^{|\tilde{n}_i|}f^+
  e^{i\tilde\alpha_i}\!-\!\sum_{i=1}^{M_1}x^{|n_i|}e^{i\alpha_i}}\\
  \displaystyle{\sum_{i=1}^{M_1}x^{|n_i|}f^-e^{i\alpha_i}\!-\!
  \sum_{i=1}^{M_2}x^{|\tilde{n}_i|}e^{i\tilde\alpha_i}}\\
  \displaystyle{\sum_{i=1}^{M_2}x^{|\tilde{n}_i|}f^-
  e^{-i\tilde\alpha_i}\!-\!\sum_{i=1}^{M_1}x^{|n_i|}e^{-i\alpha_i}}\\
  \displaystyle{\sum_{i=1}^{M_1}x^{|n_i|}f^+e^{-i\alpha_i}\!-\!
  \sum_{i=1}^{M_2}x^{|\tilde{n}_i|}e^{-i\tilde\alpha_i}}
  \end{array}\right)\ \ ,\ \
  M=\frac{1}{1-f^+f^-}\left(\begin{array}{cccc}
  &&1&f^-\\&&f^+&1\\1&f^+&&\\f^-&1&&\end{array}\right)
\end{equation}
and
\begin{equation}\label{IIA-index}
  I^{(0)}=\prod_{n=1}^\infty\det\left[M(x^n,y_1^n,y_2^n)\frac{}{}\!
  \right]^{\frac{1}{2}}=\prod_{n=1}^\infty\frac{(1-x^{2n})^2}
  {(1-x^ny_1^n)(1-x^ny_1^{-n})(1-x^ny_2^n)(1-x^ny_2^{-n})}\ .
\end{equation}
The factor $I^{(0)}$ was computed in
\cite{Bhattacharya:2008bja}. Since the second factor becomes $1$
(from $V=0$) if there are no fluxes in the saddle point, this is a
generalization of the large $N$ result of
\cite{Bhattacharya:2008bja}.

We now turn to the remaining part of the holonomy integral in
(\ref{exact-index}) apart from $I^{(0)}$. The integral over
$M_1\!+\!M_2$ variables $\alpha_i,\tilde\alpha_i$ including the
second factor in (\ref{large-N-gaussian}) can be written as
\begin{eqnarray}
  &&\hspace{-0.7cm}
  x^{\epsilon_0}\!\int_0^{2\pi}\!\!\!
  \frac{1}{{\rm (symmetry)}}\left[\frac{d\alpha}{2\pi}\right]\!
  \left[\frac{d\tilde\alpha}{2\pi}\right]\!\!
  \prod_{\substack{i,j;\\n_i\!=\!n_j}}\!
  \left(2\sin\frac{\alpha_i\!-\!\alpha_j}{2}\right)^2\!
  \prod_{\substack{i,j;\\ \tilde{n}\!=\!\tilde{n}_j}}\!
  \left(2\sin\frac{\tilde\alpha_i\!-\!\tilde\alpha_j}{2}\right)^2\!
  e^{ik(\sum n_i\alpha_i-\sum\tilde{n}_i\tilde\alpha_i)}\\
  &&\hspace{-0.7cm}\times
  \exp\left[\sum_{i=1}^{M_1}\sum_{j=1}^{M_2}\frac{1}{n}
  {\bf f}^{\rm bif}_{ij}(x^n,y_1^n,y_2^n,e^{in\alpha},e^{in\tilde\alpha})\!
  +\!\sum_{i,j=1}^{M_1}\frac{1}{n}
  {\bf f}^{\rm adj}_{ij}(x^n,y_1^n,y_2^n,e^{in\alpha})\!+\!
  \sum_{i,j=1}^{M_2}\frac{1}{n}\tilde{\bf f}^{\rm adj}_{ij}(x^n,y_1^n,y_2^n,
  e^{in\tilde\alpha})\right]\nonumber
\end{eqnarray}
where
\begin{eqnarray}
  {\bf f}^{\rm bif}_{ij}&=&
  \left(x^{|n_i\!-\!\tilde{n}_j|}-x^{|n_i|\!+\!|\tilde{n}_j|}\right)
  \left(f^+e^{i(\tilde\alpha_j\!-\!\alpha_i)}+
  f^-e^{i(\alpha_i\!-\!\tilde\alpha_j)}\right)\nonumber\\
  {\bf f}^{\rm adj}_{ij}&=&\left[-(1-\delta_{n_in_j})x^{|n_i\!-\!n_j|}+
  x^{|n_i|\!+\!|n_j|}\right]e^{-i(\alpha_i\!-\!\alpha_j)}\nonumber\\
  \tilde{\bf f}^{\rm adj}_{ij}&=&
  \left[-(1-\delta_{\tilde{n}_i\tilde{n}_j})
  x^{|\tilde{n}_i\!-\!\tilde{n}_j|}+x^{|\tilde{n}_i|\!+\!|\tilde{n}_j|}\right]
  e^{-i(\tilde\alpha_i\!-\!\tilde\alpha_j)}\
  ,\label{eff-letter-index}
\end{eqnarray}
and the symmetry factor again divides by the permutation symmetry of
identical variables $\alpha_i,\tilde\alpha_i$, depending on the
gauge symmetry unbroken by fluxes. Recall that $\epsilon_0$ is
Casimir energy like quantity which can be nonzero in the background
with nonzero flux.

The above integral can be factorized as follows. To explain this, we
decompose nonzero fluxes $\{n_i\}$, $\{\tilde{n}_i\}$ into positive
and negative ones $\{n^+_i: n^+_i>0, i=1,2,\cdots,M^+_1\}$,
$\{n^-_i:n^-_i<0,i=1,2,\cdots,M^-_1\}$ and similarly
$\{\tilde{n}^+_i: i=1,2,\cdots,M^+_2\}$, $\{\tilde{n}^-_i:
i=1,2,\cdots,M^-_2\}$. Having a look at the indices in
(\ref{eff-letter-index}), one can observe that none of these
functions get contribution from modes connecting two $U(1)$'s with
one positive and one negative flux. This simply follows from
$x^{|n_i\!-\!\tilde{n}_j|}\!=\!x^{|n_i|\!+\!|\tilde{n}_j|}$,
$x^{|n_i\!-\!n_j|}\!=\!x^{|n_i|\!+\!|n_j|}$ and
$x^{|\tilde{n}_i\!-\!\tilde{n}_j|}\!=\!x^{|\tilde{n}_i|\!+\!|\tilde{n}_j|}$
for pairs of fluxes with different signs. Furthermore, as explained
in appendix B.3, the Casimir energy also factorizes into
contributions coming from modes connecting positive fluxes or
negative fluxes only. This proves a complete factorization of the
integrand and the pre-factor into two pieces, each of which
depending only on fluxes $\{n_i^+\},\{\tilde{n}^+_i\}$ and
$\{n^-_i\},\{\tilde{n}^-_i\}$, respectively. Due to the overall
translational invariance of $\alpha_i,\tilde\alpha_i$ and
factorization, the integral is nonzero only if
\begin{equation}
  \sum_{i=1}^{M^+_1}n^+_i=\sum_{i=1}^{M^+_2}\tilde{n}^+_i\ \ ,\ \
  \sum_{i=1}^{M^-_1}n^-_i=\sum_{i=1}^{M^-_2}\tilde{n}^-_i\ ,
\end{equation}
namely the total positive and negative fluxes over two gauge groups
match separately.

We now write the expression for the full large $N$ index, summing
over all saddle points. Since $\frac{k}{2}$ times the total number of fluxes
is the Kaluza-Klein momentum along the Hopf fiber circle of
$S^7/\mathbb{Z}_k$, we grade the summation with the chemical
potential $y_3$ as $y_3^{\frac{k}{2}\sum_{i=1}^{M_1} n_i}$ (or
$y_3^{\frac{k}{2}\sum_{i=1}^{M_2} \tilde{n}_i}$). The large $N$
index is
\begin{equation}
  I_{N=\infty}(x,y_1,y_2,y_3)=I^{(0)}(x,y_1,y_2,y_3)
  I^{(+)}(x,y_1,y_2,y_3)I^{(-)}(x,y_1,y_2,y_3)\ ,
\end{equation}
where $I^{(0)}$ is given by (\ref{IIA-index}), and
\begin{equation}\label{gauge-pos-index}
\boxed{\begin{gathered}
  \hspace*{-3.5cm}I^{(+)}(x,y_1,y_2,y_3)=\sum_{M_1,M_2=0}^\infty\
  \sum_{\substack{n_1\geq\cdots\geq n_{M_1}>0\\ \tilde{n}_1\geq\cdots
  \geq \tilde{n}_{M_2}>0}}
  y_3^{\frac{k}{2}\sum n_i}x^{\sum|n_i\!-\!\tilde{n}_j|
  -\sum_{i<j}|n_i\!-\!n_j|-\sum_{i<j}|\tilde{n}_i\!-\!\tilde{n}_j|}\\
  \hspace*{-1.2cm}\times\int_0^{2\pi}\frac{1}{{\rm (symmetry)}}
  \left[\frac{d\alpha}{2\pi}\right]\left[\frac{d\tilde\alpha}{2\pi}\right]
  e^{ik(\sum n_i\alpha_i\!-\!\sum\tilde{n}_i\tilde\alpha_i)}
  \prod_{\substack{i,j;\\n_i\!=\!n_j}}\left[2\sin\frac{\alpha_i\!-\!\alpha_j}{2}
  \right]^2\prod_{\substack{i,j;\\ \tilde{n}_i\!=\!\tilde{n}_j}}
  \left[2\sin\frac{\tilde\alpha_i\!-\!\tilde\alpha_j}{2}\right]^2\\
  \hspace*{0.1cm}\times\exp\left[\sum_{i=1}^{M_1}\sum_{j=1}^{M_2}\frac{1}{n}
  {\bf f}^{\rm bif}
  _{ij}(x^n,y_1^n,y_2^n,e^{in\alpha},e^{in\tilde\alpha})+
  \sum_{i,j=1}^{M_1}{\bf f}^{\rm adj}_{ij}(x^n,y_1^n,y_2^n,e^{in\alpha})+
  \sum_{i,j=1}^{M_2}\tilde{\bf f}^{\rm adj}_{ij}(x^n,y_1^n,y_2^n,
  e^{in\tilde\alpha})\right]
\end{gathered}}
\end{equation}
with definitions for various functions given by
(\ref{eff-letter-index}). The last factor $I^{(-)}(x,y_1,y_2,y_3)$,
which is a summation of saddle points with negative fluxes only,
takes a form similar to $I^{(+)}$ with signs of $n_i,\tilde{n}_i$
flipped. The signs of these integers appear only in
$y_3^{\frac{k}{2}\sum n_i}$ and
$e^{ik(\sum n_i\alpha_i-\sum\tilde{n}_i\tilde\alpha_i)}$. The sign
flip in the first factor can be undone by replacing $y_3$ by
$1/y_3$, and that in the second factor can be undone by changing
integration variables from $\alpha,\tilde\alpha$ to
$-\alpha,-\tilde\alpha$. The latter change affects ${\bf f}^{\rm
bif}_{ij}$ by the exchange $f^+\leftrightarrow f^-$, which can be
achieved by changing
$\sqrt{\frac{y_1}{y_2}}\!+\!\sqrt{\frac{y_2}{y_1}}$ and
$\sqrt{y_1y_2}\!+\!\frac{1}{\sqrt{y_1y_2}}$. Collecting all, one
finds that
\begin{equation}
  I^{(-)}(x,y_1,y_2,y_3)=I^{(+)}(x,y_1,1/y_2,1/y_3)=
  I^{(+)}(x,1/y_1,y_2,1/y_3)\ .
\end{equation}
Since the knowledge of $I^{(+)}$ would be enough to obtain the full index,
we will mainly consider this function in the rest of this section.

We want to compare the above result with the index over
supersymmetric gravitons in $AdS_4\times S^7/\mathbb{Z}_k$. As shown
in appendix C, the index of multiple gravitons also split into three
parts, $I_{\rm mp}=I^{(0)}_{\rm mp} I^{(+)}_{\rm mp}I^{(-)}_{\rm
mp}$, essentially because gravitons with positive and negative
momenta do not mutually interact, even without the `statistical
interaction' for identical particles. It was shown in
\cite{Bhattacharya:2008bja} that $I^{(0)}_{\rm mp}=I^{(0)}$. For the
gauge theory and gravity indices to agree, one has to show
$I^{(+)}I^{(-)}=I^{(+)}_{\rm mp}I^{(-)}_{\rm mp}$, or
$\frac{I^{(+)}}{I^{(+)}_{\rm mp}}=\frac{I^{(-)}_{\rm mp}}{I^{(-)}}$.
Left hand side and right hand side can be Taylor-expanded in
$y_3^{\frac{1}{2}}$ and $y_3^{-\frac{1}{2}}$, respectively, together
with positive power expansions in $x$. The only way this equation
can hold is both sides being a constant, which is actually $1$.
Thus, one only has to show
\begin{equation}
  I^{(+)}(x,y_1,y_2,y_3)=I^{(+)}_{\rm mp}(x,y_1,y_2,y_3)
\end{equation}
to check the agreement of the indices in gauge theory and gravity.

By definition, saddle points for $I^{(+)}$ carry positive fluxes
only. Since a sequence of non-decreasing positive integers can be
represented by a Young diagram, we will sometimes represent positive
$\{n_i\},\{\tilde{n}_i\}$ by a pair of Young diagrams $Y$ and
$\tilde{Y}$, where the lengths of $i$'th rows are $n_i$ and
$\tilde{n}_i$. We denote by $d(Y)=d(\tilde{Y})$ the total number of
boxes in the Young diagram. The summation in $I^{(+)}$ can be
written as
\begin{eqnarray}\label{young-expand}
  &&\hspace{-0.7cm}I^{(+)}(x,y_1,y_2,y_3)=
  \sum_{Y,\tilde{Y}: d(Y)=d(\tilde{Y})}
  y_3^{\frac{k}{2}d(Y)}I_{Y\tilde{Y}}(x,y_1,y_2)\\
  &&\hspace{-0.6cm}
  =1+y_3^{\frac{k}{2}}I_{\Yboxdim4pt\yng(1)~\yng(1)}+
  y_3^k\left(I_{\Yboxdim4pt\yng(2)~\yng(2)}\!+\!
  I_{\Yboxdim4pt\yng(1,1)~\yng(1,1)}\!+\!
  2I_{\Yboxdim4pt\yng(2)~\yng(1,1)}\right)\!+\!
  y_3^{\frac{3k}{2}}\left(I_{\Yboxdim4pt\yng(3)~\yng(3)}\!+\!
  I_{\Yboxdim4pt\yng(2,1)~\yng(2,1)}\!+\!
  I_{\Yboxdim4pt\yng(1,1,1)~\yng(1,1,1)}\!+\!
  2I_{\Yboxdim4pt\yng(3)~\yng(2,1)}\!+\!
  2I_{\Yboxdim4pt\yng(3)~\yng(1,1,1)}\!+\!
  2I_{\Yboxdim4pt\yng(2,1)~\yng(1,1,1)}\right)\!+\!\cdots\nonumber
\end{eqnarray}
where we used $I_{Y\tilde{Y}}=I_{\tilde{Y}Y}$, which we do not prove
here but can be checked by suitable redefinitions of integration
variables in (\ref{gauge-pos-index}).

We did not manage to analytically prove $I^{(+)}=I^{(+)}_{\rm mp}$
generally. Below we provide nontrivial analytic and numerical checks
of this claim in various sectors: we consider the sectors in which
the total number of positive fluxes
$\sum_{i}n_i^+=\sum_i\tilde{n}^+$ is $1$, $2$ and $3$.

\subsection{One KK-momentum: analytic tests}

We analytically prove the agreement between gauge theory and gravity
indices in the sector with unit KK-momentum. This amounts to
comparing the coefficients of $y_3^{\frac{k}{2}}$ in $I^{(+)}$ and
$I^{(+)}_{\rm mp}$. The gravity result is simply
\begin{equation}\label{grav-1-flux}
  I_{k}^{\rm sp}(x,y_1,y_2)=\oint\frac{d\sqrt{y_3}}{2\pi i\sqrt{y_3}}
  y_3^{-\frac{k}{2}}I^{\rm sp}(x,y_1,y_2,y_3)\ ,
\end{equation}
namely index over single graviton with $k$ (i.e. minimal) units of
KK-momentum. The contour for $\sqrt{y_3}$ integration is the unit
circle in the complex plane. On the gauge theory side, the result
comes from one saddle point with fluxes given by
$n\!=\!\tilde{n}\!=\!1$: from the general formula
(\ref{gauge-pos-index}) one obtains
\begin{equation}\label{gauge-1-flux}
  I_{\Yboxdim4pt\yng(1)~\yng(1)}=
  \int_0^{2\pi}\frac{d\alpha d\tilde\alpha}{(2\pi)^2}
  e^{ik(\alpha\!-\!\tilde\alpha)}\exp\left[
  \sum_{n=1}^\infty\frac{1}{n}\left((1-x^{2n})\left(f^+(\cdot^n)
  e^{in(\tilde\alpha\!-\!\alpha)}+f^-(\cdot^n)
  e^{in(\alpha\!-\!\tilde\alpha)}\!\frac{}{}\right)+2x^{2n}
  \right)\right]\ .
\end{equation}
The `effective letter index'
$(1-x^2)\left(f^+e^{i(\tilde\alpha\!-\!\alpha)}+f^-
e^{i(\alpha\!-\!\tilde\alpha)}\!\frac{}{}\right)+2x^{2}$ in the
exponential is
\begin{equation}\label{eff-1-flux}
  \hspace{-1cm}\left[x^{\frac{1}{2}}\left(\!\sqrt{\frac{y_1}{y_2}}\!+\!
  \sqrt{\frac{y_2}{y_1}}\right)\!-\!x^{\frac{3}{2}}\left(\!\sqrt{y_1y_2}\!+\!
  \frac{1}{\sqrt{y_1y_2}}\right)\right]e^{i(\tilde\alpha\!-\!\alpha)}
  \!+\!\left[x^{\frac{1}{2}}\left(\!\sqrt{y_1y_2}\!+\!
  \frac{1}{\sqrt{y_1y_2}}\right)\!-\!x^{\frac{3}{2}}
  \left(\!\sqrt{\frac{y_1}{y_2}}\!+\!\sqrt{\frac{y_2}{y_1}}
  \right)\right]e^{i(\alpha\!-\!\tilde\alpha)}+2x^2\ .
\end{equation}
Note that $f^{\pm}$ have $\frac{1}{1-x^2}$ factors, coming from many
derivatives acting on fields, and take the form of infinite series
in $x$. The factor $(1\!-\!x^2)$ cancels these derivative factors
and lets the effective index be a finite series. Defining an
integration variable $z\equiv e^{i(\tilde\alpha\!-\!\alpha)}$ in
(\ref{gauge-1-flux}), and after exponentiating (\ref{eff-1-flux}),
one obtains
\begin{equation}\label{1-flux-gauge}
  I_{\Yboxdim4pt\yng(1)~\yng(1)}=
  \oint\frac{dz}{(2\pi i)z}z^{-k}
  \frac{\left(1-x\sqrt{xy_1y_2}z\right)\left(1-x\sqrt{\frac{x}{y_1y_2}}z\right)
  \left(1-x\sqrt{\frac{xy_1}{y_2}}z^{-1}\right)
  \left(1-x\sqrt{\frac{xy_2}{y_2}}z^{-1}\right)}
  {\left(1-\sqrt{\frac{xy_1}{y_2}}z\right)\left(1-\sqrt{\frac{xy_2}{y_1}}z\right)
  \left(1-\sqrt{xy_1y_2}z^{-1}\right)
  \left(1-\sqrt{\frac{x}{y_1y_2}}z^{-1}\right)(1-x^2)^2}\ .
\end{equation}
Using the relation (\ref{relation}), and identifying the integration variable
as $z=\sqrt{y_3}$, the above result can be rewritten as
\begin{equation}
  I_{\Yboxdim4pt\yng(1)~\yng(1)}=
  \oint\frac{d\sqrt{y_3}}{(2\pi i)\sqrt{y_3}}y_3^{-\frac{k}{2}}
  \left(I^{\rm sp}(x,y_1,y_2,y_3)+\frac{1-x^2+x^4}{(1-x^2)^2}\right)\ .
\end{equation}
Since the second term in the integrand does not survive the contour
integral, this is exactly the gravity expression
(\ref{grav-1-flux}), proving the agreement in this sector.

Before proceeding to more nontrivial examples, let us explain a bit
more on the above index. As stated in the previous section, the flux
provides a lower bound to the energy of states. In the sector with
unit flux, we can actually find from the integrand of
(\ref{1-flux-gauge}) that \#($\sqrt{x}$) in a term is always larger
than or equal to \#($z$). We can actually arrange the terms in
Taylor expansion of the integrand so that the number
\#($\sqrt{x}$)$-$\#($z$) ascends. The lowest order terms come from
the first two factors in the denominator containing $\sqrt{x}z$, for
which this number is $0$. The index for these states is
\begin{equation}\label{1-flux-gauge-low}
  \oint\frac{dz}{(2\pi i)z}z^{-k}
  \frac{1}{\left(1-\sqrt{\frac{xy_1}{y_2}}z\right)
  \left(1-\sqrt{\frac{xy_2}{y_1}}z\right)}=
  x^{\frac{k}{2}}\left(y_1^{\frac{k}{2}}y_2^{-\frac{k}{2}}+
  y_1^{\frac{k}{2}\!-\!1}y_2^{-\frac{k}{2}\!+\!1}+\cdots+
  y_1^{-\frac{k}{2}}y_2^{\frac{k}{2}}\right)\ .
\end{equation}
From Table 1, the two factors in the integrand originate from the
gauge theory letters $\bar{B}^{\dot{2}}$ and $\bar{B}^{\dot{1}}$ in
s-waves. The operators made of these letters form a subset of chiral
operators studied in \cite{Aharony:2008ug,Hanany:2008qc}. The
operator took the form of $k$'th product of $\bar{B}^{\dot{a}}$,
multiplied by a 't Hooft operator in the (Sym($\bar{\bf
N}^k$),Sym(${\bf N}^k$)) representation to make the whole operator
gauge invariant.

Since no fermionic letters can contribute in the lowest energy
sector due to their larger dimensions than scalars, the above index
equals to the partition function. This is not true any more
as one goes beyond lowest energy as fermionic letters start to
enter. This aspect in the lowest energy sector will continue to
appear with more fluxes below. The full spectrum of these chiral
operators, preserving specific $\mathcal{N}\!=\!2$ supersymmetry,
has been studied in \cite{Aharony:2008ug} by quantizing the moduli
space \cite{Kinney:2005ej,Mandal:2006tk,Benvenuti:2006qr}. We expect
our result to be identical to the result in \cite{Aharony:2008ug},
which we check explicitly for the case with two fluxes in the next
subsection.

\subsection{Two KK-momenta: analytic and numerical tests}

Monopole operators which have been studied in the context of
$\mathcal{N}\!=\!6$ Chern-Simons theory are in the conjugate
representations of the two $U(N)$ gauge groups, such as
(Sym($\bar{\bf N}^k$),Sym(${\bf N}^k$)) in the previous subsection
or more general examples studied in
\cite{Berenstein:2008dc,Klebanov:2008vq}. In our
analysis, these are related to the saddle points in which two flux
distributions $\{n_i\}$ and $\{\tilde{n}_i\}$ are the same. In the
sector with two fluxes, two of the four saddle points
$\Yboxdim8pt\yng(2)~\yng(2)~$ and $\Yboxdim8pt\yng(1,1)~\yng(1,1)$
are in this category, while the other two
$\Yboxdim8pt\yng(2)~\yng(1,1)$ and $\Yboxdim8pt\yng(1,1)~\yng(2)$
are not.Let us call the former kind of flux distributions as
`equal distributions.'

Incidently, the way of having equally distributing given amount of
fluxes to many $U(1)$ factors is the same as the way of distributing
same amount of momenta (in units of $\frac{k}{2}$) to multiple
gravitons, each carrying positive KK-momenta. For instance, the
first of the above two distributions maps to giving two units of
KK-momenta to a single graviton, while the second maps to picking
two gravitons and giving one unit of momentum to each. This might
let one suspect that there could be some relation between the index
from a saddle point with equal distribution and the multi-graviton
index with the corresponding momentum distribution. What we find
below empirically says that this is true up to a certain order in
the $x$ expansion. However, as we go beyond certain energy, it will
turn out that only the total sum over all saddle points with equal
distributions equals the total multi-graviton index. As one goes
beyond an even higher energy threshold, the saddle points with
unequal saddle points start to appear which add to the saddle points
with equal distributions to correctly reproduce the graviton index.

The saddle points with unequal flux distributions provide examples
in which monopole operators contribute to the energy, or the scaling
dimension, of the whole gauge invariant operator. Monopole operators
with vanishing scale dimensions are studied in
\cite{Aharony:2008ug,Berenstein:2008dc,Klebanov:2008vq}
in $\mathcal{N}\!=\!6$ Chern-Simons theory, while in general this
does not have to be the case. See \cite{Borokhov:2002ib,Klebanov:2008vq}
as well as recent works \cite{Imamura:2009ur,Gaiotto:2009tk} on monopole
operators in $\mathcal{N}\!=\!4$ and $\mathcal{N}\!=\!3$ theories.

Let us present various checks that we did in this sector. In this
subsection we write $p\equiv\sqrt{y_1y_2}$ and
$r=\sqrt{\frac{y_1}{y_2}}$ to simplify the formulae. Let us define
the following functions
\begin{eqnarray}
  f(x,p,r,z)
  &=&\sqrt{x}z(r+r^{-1})+\sqrt{x}z^{-1}(p+p^{-1})-x\sqrt{x}z(p+p^{-1})-
  x\sqrt{x}z^{-1}(r+r^{-1})\nonumber\\
  F(x,p,r,z)&=&\exp\left(\sum_{n=1}^\infty\frac{1}{n}f(x^n,p^n,r^n,z^n)\right)
  \nonumber\\
  &=&\frac{(1-x\sqrt{x}pz)(1-x\sqrt{x}p^{-1}z)(1-x\sqrt{x}rz^{-1})
  (1-x\sqrt{x}r^{-1}z^{-1})}
  {(1-\sqrt{x}rz)(1-\sqrt{x}r^{-1}z)(1-\sqrt{x}pz^{-1})
  (1-\sqrt{x}p^{-1}z^{-1})}\ ,\nonumber
\end{eqnarray}
which should be familiar from our analysis in the previous
subsection, and also define
\begin{eqnarray}
  f_m(x,p,r,z)&=&x^mf(x,p,r,z)\nonumber\\
  F_m(x,p,r,z)&=&\exp\left(\sum_{n=1}^\infty\frac{1}{n}f_m(x^n,p^n,r^n,z^n)\right)
  \nonumber\\
  &=&\frac{(1-x^{m\!+\!1}\sqrt{x}pz)(1-x^{m\!+\!1}\sqrt{x}p^{-1}z)
  (1-x^{m\!+\!1}\sqrt{x}rz^{-1})(1-x^{m\!+\!1}\sqrt{x}r^{-1}z^{-1})}
  {(1-x^m\sqrt{x}rz)(1-x^m\sqrt{x}r^{-1}z)(1-x^m\sqrt{x}pz^{-1})
  (1-x^m\sqrt{x}p^{-1}z^{-1})}\nonumber
\end{eqnarray}
Using these functions,
\begin{equation}
  f_{|n_i\!-\!\tilde{n}_j|}(x,p,r,e^{i(\alpha_i\!-\!\tilde\alpha_j)})+
  f_{|n_i\!-\!\tilde{n}_j|\!+\!2}(x,p,r,e^{i(\alpha_i\!-\!\tilde\alpha_j)})+
  \cdots+f_{n_i\!+\!\tilde{n}_j\!-\!2}(x,p,r,e^{i(\alpha_i\!-\!\tilde\alpha_j)})
\end{equation}
for $n_i\!\neq\!\tilde{n}_j$ is what we called ${\bf f}^{\rm
bif}_{ij}$ in the previous section.

To compare $I^{(+)}$ with $I^{(+)}_{\rm mp}$ in the sector with two
units of fluxes, i.e. at the order $y_3^{k}$, we write the integral
expressions for indices from four saddle points. Getting rid of a
trivial integral corresponding to the decoupled $U(1)$, one obtains
the following expressions.
\begin{eqnarray}\label{saddle-1}
  I_{\Yboxdim4pt\yng(2)~\yng(2)}&=&\frac{1}{2\pi i}\oint\frac{dz}{z}
  z^{-2k}\frac{F(x,p,r,z)F_2(x,p,r,z)}{(1-x^4)^2}
\end{eqnarray}
where $z\equiv e^{i(\tilde\alpha\!-\!\alpha)}$,
\begin{eqnarray}\label{saddle-2}
  I_{\Yboxdim4pt\yng(1,1)~\yng(1,1)}
  &=&\oint\frac{dz}{(2\pi i)z}z^{-2k}
  \oint\frac{dadb}{(2\pi i)^2ab}\left(1-\frac{a^2}{2}-\frac{1}{2a^2}\right)
  \left(1-\frac{b^2}{2}-\frac{1}{2b^2}\right)\nonumber\\
  &&\hspace{1cm}\times\frac{F(x,p,r,zab)F(x,p,r,xab^{-1})
  F(x,p,r,za^{-1}b)F(x,p,r,za^{-1}b^{-1})}
  {(1-x^2)^4(1-x^2a^2)(1-x^2a^{-2})(1-x^2b^2)(1-x^2b^{-2})}
\end{eqnarray}
where $e^{i\alpha_1}=z^{-1/2}a$, $e^{i\alpha_2}=z^{-1/2}a^{-1}$,
$e^{i\tilde\alpha_1}=z^{1/2}b$, $e^{i\tilde\alpha_2}=z^{1/2}b^{-1}$,
and
\begin{equation}\label{saddle-3}
  I_{\Yboxdim4pt\yng(2)~\yng(1,1)}=I_{\Yboxdim4pt\yng(1,1)~\yng(2)}
  =x^2\oint\frac{dz}{(2\pi i)z}z^{-2k}\oint\frac{da}{(2\pi i)a}
  \left(1-\frac{a^2}{2}-\frac{1}{2a^2}\right)\frac{F_1(x,p,r,za)
  F_1(x,p,r,za^{-1})}{(1-x^4)^2(1-x^2)^2(1-x^2a^2)(1-x^2a^{-2})}
\end{equation}
where $e^{i\alpha_1}\!=\!z^{-1/2}a$,
$e^{i\alpha_2}\!=\!z^{-1/2}a^{-1}$, $e^{i\tilde\alpha}\!=\!z^{1/2}$
for $I_{\Yboxdim4pt\yng(2)~\yng(1,1)}$ , and
$e^{i\alpha}\!=\!z^{-1/2}$, $e^{i\tilde\alpha_1}\!=\!z^{1/2}a$,
$e^{i\tilde\alpha_2}\!=\!z^{1/2}a^{-1}$ for
$I_{\Yboxdim4pt\yng(1,1)~\yng(2)}$ . All contour integrals here and
below are along the unit circle on the complex plane. The graviton
index with two units of momenta is given by
\begin{equation}\label{graviton-2-flux}
  \left.I^{(+)}_{\rm mp}(x,p,r)\frac{}{}\right|_{y_3^{k}}
  =I_{2k}(x,p,r)+\frac{1}{2}I_k(x^2,p^2,r^2)+\frac{1}{2}I_k(x,p,r)^2
\end{equation}
The first term is from single graviton with two units of momenta,
while the last two terms are from two identical gravitons, each with
unit momentum. We expect the sum of four contributions
(\ref{saddle-1}), (\ref{saddle-2}), (\ref{saddle-3}) to equal
(\ref{graviton-2-flux}).

We first study the lowest energy states in this sector analytically.
As in the previous section, we arrange the Taylor expansions of the
integrands in ascending orders in \#($\sqrt{x}$)$-$\#($z$). From the
behaviors of the functions $F,F_1,F_2$, one finds that unequal
distributions do not contribute to the lowest energy sector. The two
equal distributions contribute with $\epsilon\!=\!k$ as
\begin{equation}
  I_{\Yboxdim4pt\yng(2)~\yng(2)}\rightarrow
  \oint\frac{dz}{(2\pi i)z}z^{-2k}
  \frac{1}{\left(1-\sqrt{\frac{xy_1}{y_2}}z\right)
  \left(1-\sqrt{\frac{xy_2}{y_1}}z\right)}=
  x^{k}\left(y_1^{k}y_2^{-k}+y_1^{k\!-\!1}y_2^{-k\!+\!1}+\cdots+
  y_1^{-k}y_2^{k}\right)
\end{equation}
and
\begin{eqnarray}\label{2-low-gauge}
  I_{\Yboxdim4pt\yng(1,1)~\yng(1,1)}&\rightarrow&
  \oint\frac{dz}{(2\pi i)z}z^{-2k}
  \oint\frac{d\nu}{(2\pi i)\nu}\oint\frac{d\mu}{(2\pi i)\mu}
  (1-\mu\nu)(1-\mu\nu^{-1})\\
  &&\times\frac{1}{(1-tz\mu)(1-uz\mu)(1-tz\mu^{-1})(1-uz\mu^{-1})
  (1-tz\nu)(1-uz\nu)(1-tz\nu^{-1})(1-uz\nu^{-1})}\nonumber
\end{eqnarray}
where we redefined integration variables and the chemical potentials
as $a^2=\mu\nu$ and $b^2=\mu\nu^{-1}$, $x^{\frac{1}{2}}r=t$,
$x^{\frac{1}{2}}r^{-1}=u$. The two factors on the first line in the
integrand come from the $U(2)\times U(2)$ measure, which we can
effectively change to $(1\!-\!a^2)(1\!-\!b^2)$ since rest of the
integrand is invariant under $a\rightarrow a^{-1}$ and $b\rightarrow
b^{-1}$, separately.

From the results in appendix C, the two graviton indices at the
lowest energy are given by (upon identifying $z=\sqrt{y_3}$)
\begin{equation}
  I_{2k}\rightarrow\oint\frac{dz}{(2\pi i)z}
  z^{-2k}\frac{1}{(1-tz)(1-uz)}
  \stackrel{\rm lowest}{=}I_{\Yboxdim4pt\yng(2)~\yng(2)}
\end{equation}
and
\begin{eqnarray}\label{2-low-grav}
  \frac{1}{2}I_{k}(\cdot^2)+\frac{1}{2}I_k(\cdot)^2\!&\!\rightarrow\!&\!
  \frac{1}{2}\oint\frac{dz}{(2\pi i)z}z^{-2k}\frac{1}{(1-t^2z^2)(1-u^2z^2)}\\
  &&\hspace{1cm}
  +\frac{1}{2}\oint\frac{dz_1dz_2}{(2\pi i)^2z_1z_2}z_1^{-k}z_2^{-k}
  \frac{1}{(1-tz_1)(1-uz_1)(1-tz_2)(1-uz_2)}\nonumber\\
  &=&\oint\frac{dz}{(2\pi i)z}z^{-2k}\left[\!\frac{}{}\right.
  \frac{1}{2(1-t^2z^2)(1-u^2z^2)}\nonumber\\
  &&\hspace{1cm}+\frac{1}{2}\oint\frac{d\nu}{(2\pi i)\nu}\frac{1}
  {(1-tz\nu)(1-uz\nu)(1-tz\nu^{-1})(1-uz\nu^{-1})}
  \left.\frac{}{}\!\right]\nonumber
\end{eqnarray}
where we changed to variables $z_1=z\nu$, $z_2=z\nu^{-1}$ on the
last line. Keeping the $z$ integral, we contour-integrate $\mu$ and
$\nu$ in (\ref{2-low-gauge}) and (\ref{2-low-grav}) to compare them.
After some algebra, one finds
\begin{equation}
  \frac{1}{(1-t^2z^2)(1-u^2z^2)(1-tuz^2)}+{\rm\ const.}
\end{equation}
for both indices, where the last constant is $0$ and $-1$ for gauge
theory and gravity, respectively, which does not survive the
remaining $z$ integration. This shows the agreement of the indices from
gauge theory and gravity at the lowest energy with two fluxes. The
indices from two saddle points separately agree with two
corresponding gravity indices, as explained before. This result can
also be obtained by quantizing the moduli space
\cite{Aharony:2008ug}, restricted to the fields $\bar{B}^{\dot{1}}$
and $\bar{B}^{\dot{2}}$.

Since we are not aware of any further way of treating the integral
analytically, we compare the two indices order by order after
expanding in $x$. Firstly, for $k\!=\!1$, we find a perfect
agreement to $\mathcal{O}(x^{9})$ terms that we checked, as follows.
We find
\begin{eqnarray}
  I_{\Yboxdim4pt\yng(2)~\yng(2)}\!&=&\!
  x(r^2+1+r^{-2})+x^2(p+p^{-1})(r^3+r^{-3})
  +x^3\left[(p^2+p^{-2})(r^4\!+\!r^{-4})-2(r^2+r^{-2})\right]\nonumber\\
  &&\!+x^4\left[(p^3+p^{-3})(r^5+r^{-5})+(p+p^{-1})(r+r^{-1})\right]
  \nonumber\\
  &&\!+x^5\left[(p^4+p^{-4})(r^6+r^{-6})+(r^4-2r^2-4-2r^{-2}+r^{-4})\right]
  \nonumber\\
  &&\!+x^6\left[\left(p^5+p^{-5}\right)(r^7+r^{-7})+
  \left(p+p^{-1}\right)(-2r^3+2r\cdots)\right]\nonumber\\
  &&\!+x^7\left[(p^6+p^{-6})(r^8+r^{-8})+(p^2+p^{-2})(r^2+r^{-2})+
  (r^4-3+r^{-4})\right]\nonumber\\
  &&\!+x^8\left[(p^7+p^{-7})(r^9+r^{-9})+(p+p^{-1})(r^5-2r^3-2r\cdots)\right]
  \nonumber\\
  &&\!+x^9\left[(p^8\!+\!p^{-8})(r^{10}\!+\!r^{-10})+(p^2\!+\!p^{-2})
  (\!-2r^4\!+\!2r^2\!+\!3\cdots)
  +(\!-r^4\!+\!5r^2\!+\!6\cdots)\right]\!+\!\mathcal{O}(x^{10})\nonumber
\end{eqnarray}
\begin{eqnarray}
  I_{\Yboxdim4pt\yng(1,1)~\yng(1,1)}\!&=&\!
  x(r^2+1+r^{-2})+x^2(p+p^{-1})(r^3+r+r^{-1}+r^{-3})\nonumber\\
  &&\!+x^3\left[(p^2+p^{-2})(2r^4+r^2+1+r^{-2}+2r^{-4})
  +(r^4+r^{-4})\right]\nonumber\\
  &&\!+x^4\left[(p^3+p^{-3})(2r^5+r^3+r\cdots)+(p+p^{-1})(r^5-r^3-r\cdots)\right]
  \nonumber\\
  &&\!+x^5\left[(p^4+p^{-4})(3r^6+r^4+r^2+1\cdots)+(p^2+p^{-2})(r^6-r^4\cdots)
  +(r^6-r^4+3\cdots)\right]\nonumber\\
  &&\!+x^6\left[\left(p^5\!+\!p^{-5}\right)(3r^7\!+\!r^5\!+\!r^3\!+\!r\cdots)+
  \left(p^3\!+\!p^{-3}\right)(r^7\!-\!r^5\cdots)+
  (p\!+\!p^{-1})(r^7\!+\!r^3\cdots)\right]\nonumber\\
  &&\!+x^7\left[(p^6+p^{-6})(4r^8+r^6+r^4+r^2+1\cdots)+(p^4+p^{-4})
  (r^8-r^6\cdots)\right.\nonumber\\
  &&\!\left.\hspace{2cm}+(p^2+p^{-2})(r^8-r^4-r^2-2\cdots)+
  (r^8-r^4-3r^2-1\cdots)\right]\nonumber\\
  &&\!+x^8\left[(p^7+p^{-7})(4r^9+r^7+r^5+r^3+r\cdots)+(p^5+p^{-5})(r^9-r^7\cdots)
  \right.\nonumber\\
  &&\!\left.\hspace{2cm}+(p^3+p^{-3})(r^9-r^5\cdots)
  +(p+p^{-1})(r^9-r^5+r^3+4r\cdots)\right]\nonumber\\
  &&\!+x^9\left[(p^8+p^{-8})(5r^{10}+r^8+r^6+r^4+r^2+1\cdots)+(p^6+p^{-6})
  (r^{10}-r^8\cdots)\right.\nonumber\\
  &&\!\left.\hspace{0.5cm}+(p^4\!+\!p^{-4})(r^{10}\!-\!r^6\cdots)+
  (p^2\!+\!p^{-2})(r^{10}\!+\!r^4\!-\!1\cdots)+
  (r^{10}\!+\!r^4\!-\!2r^2\!-\!3\cdots)\right]\!+\!\mathcal{O}(x^{10})\nonumber
\end{eqnarray}
\begin{eqnarray}
  I_{\Yboxdim4pt\yng(2)~\yng(1,1)}=I_{\Yboxdim4pt\yng(1,1)
  ~\yng(2)}&=&x^5-x^6\left(p+p^{-1}\right)\left(r+r^{-1}\right)+
  x^7\left[(p^2+p^{-2})+(r^2+3+r^{-2})\right]\nonumber\\
  &&-x^8\left(p\!+\!p^{-1}\right)\left(r\!+\!r^{-1}\right)-x^9
  \left[(p^2\!+\!p^{-2})(r^2\!+\!1\!+\!r^{-2})\!+\!2(r^2\!+\!r^{-2})\right]
  \!+\!\mathcal{O}(x^{10})\nonumber
\end{eqnarray}
and on the gravity side we find
\begin{eqnarray}
  I_2(x,p,r)\!&=&\!x(r^2+1+r^{-2})+x^2(p+p^{-1})(r^3+r^{-3})+
  x^3\left[(p^2+p^{-2})(r^4+r^{-4})-(r^2+r^{-2})\right]\nonumber\\
  &&\!+x^4(p^3\!+\!p^{-3})(r^5\!+\!r^{-5})\!+\!
  x^5\left[(p^4\!+\!p^{-4})(r^6\!+\!r^{-6})\!-\!(r^2\!+\!r^{-2})
  \right]\!+\!x^6(p^5\!+\!p^{-5})(r^7\!+\!r^{-7})\nonumber\\
  &&\!+x^7\left[(p^6+p^{-6})(r^8+r^{-8})-(r^2+r^{-2})\right]+
  x^8(p^7+p^{-7})(r^9+r^{-9})\nonumber\\
  &&\!+x^9\left[(p^8+p^{-8})(r^{10}+r^{-10})
  -(r^2+r^{-2})\right]+\mathcal{O}(x^{10})\nonumber
\end{eqnarray}
\begin{eqnarray}
  &&\hspace{-1.5cm}
  \frac{1}{2}I_1(x^2,p^2,r^2)+\frac{1}{2}I_1(x,p,r)^2\nonumber\\
  &=&\!x(r^2\!+\!1\!+\!r^{-2})+x^2(p\!+\!p^{-1})(r^3\!+\!r\cdots)
  +x^3\left[(p^2\!+\!p^{-2})(2r^4\!+\!r^2\!+\!1\cdots)\!+\!
  (r^4\!-\!r^2\cdots)\right]\nonumber\\
  &&\!+x^4\left[(p^3+p^{-3})(2r^5+r^3+r\cdots)+(p+p^{-1})(r^5-r^3\cdots)\right]
  \nonumber\\
  &&\!+x^5\left[(p^4+p^{-4})(3r^6+r^4+r^2+1\cdots)+(p^2+p^{-2})(r^6-r^4\cdots)+
  (r^6-r^2+1\cdots)\right]\nonumber\\
  &&\!+x^6\left[(p^5\!+\!p^{-5})(3r^7\!+\!r^5\!+\!r^3\!+\!r\cdots)+
  (p^3\!+\!p^{-3})(r^7\!-\!r^5\cdots)
  +(p\!+\!p^{-1})(r^7\!-\!r^3\cdots)\right]\nonumber\\
  &&\!+x^7\left[(p^6+p^{-6})(4r^8+r^6+r^4+r^2+1\cdots)+
  (p^4+p^{-4})(r^8-r^6\cdots)\right.\nonumber\\
  &&\!\left.\hspace{2cm}
  +(p^2+p^{-2})(r^8-r^4\cdots)+(r^8+2+r^{-8})\right]\nonumber\\
  &&\!+x^8\left[(p^7+p^{-7})(4r^9+r^7+r^5+r^3+r\cdots)+
  (p^5+p^{-5})(r^9-r^7\cdots)\right.\nonumber\\
  &&\!\left.\hspace{2cm}
  +(p^3+p^{-3})(r^9-r^5\cdots)+(p+p^{-1})(r^9-r^3\cdots)\right]\nonumber\\
  &&\!+x^9\left[(p^8+p^{-8})(5r^{10}+r^8+r^6+r^4+r^2+1\cdots)+
  (p^6+p^{-6})(r^{10}-r^8\cdots)\right.\nonumber\\
  &&\!\left.\hspace{1cm}+(p^4\!+\!p^{-4})(r^{10}\!-\!r^6\cdots)+
  (p^2\!+\!p^{-2})(r^{10}\!-\!r^4\cdots)
  +(r^{10}\!+\!3\!+\!r^{-10})\right]+\mathcal{O}(x^{10})\ .\nonumber
\end{eqnarray}
The terms in `$\cdots$' take negative powers of $r$, and can be
completed from the fact that the expression in any parenthesis has
$r\rightarrow r^{-1}$ invariance. From this one can check
\begin{equation}
  I_{\Yboxdim4pt\yng(2)~\yng(2)}+
  I_{\Yboxdim4pt\yng(1,1)~\yng(1,1)}+
  I_{\Yboxdim4pt\yng(2)~\yng(1,1)}+
  I_{\Yboxdim4pt\yng(1,1)~\yng(2)}
  =I_2(x,p,r)+\frac{1}{2}I_1(x^2,p^2,r^2)+\frac{1}{2}I_1(x,p,r)^2
  +\mathcal{O}(x^{10})\nonumber
\end{equation}
for $k\!=\!1$.

We also found perfect agreement for $k\!=\!2$ for all terms that we
calculated. To reduce execution time for numerical calculation,
space of presenting our result and most importantly to reduce
possible typos, we set $r\!=\!p\!=\!1$ and keep $x$ only. We find
\begin{eqnarray}
  I_{\Yboxdim4pt\yng(2)~\yng(2)}\!&=&\!
  5x^2+4x^3+0x^4+8x^5-2x^6+4x^7+4x^8+4x^9-2x^{10}+
  8x^{11}+5x^{12}+\mathcal{O}(x^{13})\nonumber\\
  I_{\Yboxdim4pt\yng(1,1)~\yng(1,1)}\!&=&\!
  6x^2\!+\!12x^3\!+\!18x^4\!+\!16x^5\!+\!29x^6\!+\!28x^7\!+\!32x^8\!+\!
  44x^9\!+\!29x^{10}\!+\!72x^{11}\!+\!31x^{12}\!+\!\mathcal{O}(x^{13})
  \nonumber\\
  I_{\Yboxdim4pt\yng(2)~\yng(1,1)}\!=\!
  I_{\Yboxdim4pt\yng(1,1)~\yng(2)}\!&=&\!
  x^8-4x^9+10x^{10}-16x^{11}+11x^{12}+\mathcal{O}(x^{13})\nonumber
\end{eqnarray}
and
\begin{eqnarray}
  I_4(x)\!&\!=\!&\!5x^2+4x^3+2x^4+4x^5+2x^6+4x^7+2x^8+4x^9+2x^{10}
  +4x^{11}+2x^{12}+\mathcal{O}(x^{13})\nonumber\\
  \frac{I_2(x^2)\!+\!I_2(x)^2}{2}\!&\!=\!&\!
  6x^2\!+\!12x^3\!+\!16x^4\!+\!20x^5\!+\!25x^6\!+\!28x^7\!+\!36x^8\!+\!36x^9
  \!+\!45x^{10}\!+\!44x^{11}\!+\!56x^{12}\!+\!\mathcal{O}(x^{13})\nonumber
\end{eqnarray}
which proves
\begin{equation}
  I_{\Yboxdim4pt\yng(2)~\yng(2)}+
  I_{\Yboxdim4pt\yng(1,1)~\yng(1,1)}+
  I_{\Yboxdim4pt\yng(2)~\yng(1,1)}+
  I_{\Yboxdim4pt\yng(1,1)~\yng(2)}
  =I_4(x)+\frac{1}{2}I_2(x^2)+\frac{1}{2}I_2(x)^2
  +\mathcal{O}(x^{13})\nonumber
\end{equation}
for $k=2$.

One might think that theories with $k\!=\!1,2$ are somewhat special since
we expect enhancement of supersymmetry to $\mathcal{N}\!=\!8$
\cite{Aharony:2008ug}.
To ensure that the agreement has nothing to do with this property,
we also check the case with $k=3$. We find
\begin{eqnarray}
  \hspace*{-0cm}I_{\Yboxdim4pt\yng(2)~\yng(2)}\!&\!=\!&\!
  7x^3+4x^4+0x^5+8x^6-2x^7+4x^8+4x^9+4x^{10}-2x^{11}+8x^{12}+0x^{13}+
  \mathcal{O}(x^{14})\nonumber\\
  \hspace*{-0cm}I_{\Yboxdim4pt\yng(1,1)~\yng(1,1)}\!&\!=\!&\!
  10x^3\!+\!16x^4\!+\!20x^5\!+\!20x^6\!+\!31x^7\!+\!32x^8\!+\!36x^9\!+\!
  40x^{10}\!+\!49x^{11}\!+\!52x^{12}\!+\!40x^{13}\!+\!\mathcal{O}(x^{14})
  \nonumber\\
  \hspace*{-0cm}I_{\Yboxdim4pt\yng(2)~\yng(1,1)}\!=\!
  I_{\Yboxdim4pt\yng(1,1)~\yng(2)}\!&\!=\!&\!
  x^{11}-4x^{12}+10x^{13}+\mathcal{O}(x^{14})
  \nonumber
\end{eqnarray}
and
\begin{eqnarray}
  I_6(x)\!&\!=\!&\!
  7x^3+4x^4+2x^5+4x^6+2x^7+4x^8+2x^9+4x^{10}+2x^{11}+4x^{12}+2x^{13}+
  \mathcal{O}(x^{14})\nonumber\\
  \frac{I_3(x^2)\!+\!I_3(x)^2}{2}\!&\!=\!&\!
  10x^3\!+\!16x^4\!+\!18x^5\!+\!24x^6\!+\!27x^7\!+\!32x^8\!+\!38x^9\!+\!
  40x^{10}\!+\!47x^{11}\!+\!48x^{12}\!+\!58x^{13}\!+\!\mathcal{O}(x^{14})
  \nonumber
\end{eqnarray}
proving
\begin{equation}
  I_{\Yboxdim4pt\yng(2)~\yng(2)}+
  I_{\Yboxdim4pt\yng(1,1)~\yng(1,1)}+
  I_{\Yboxdim4pt\yng(2)~\yng(1,1)}+
  I_{\Yboxdim4pt\yng(1,1)~\yng(2)}
  =I_6(x)+\frac{1}{2}I_3(x^2)+\frac{1}{2}I_3(x)^2
  +\mathcal{O}(x^{14})\nonumber
\end{equation}
for $k=3$.

In these examples, the two saddle points with equal distributions
start to deviate from their `corresponding' graviton indices at two
orders higher than the lowest energy. The saddle points with unequal
distributions start to enter at $2k+2$ orders higher than the lowest
level. The $2k$ comes from the energy shifts in the letter indices
(\ref{eff-letter-index}), while $2$ comes from the Casimir energy
shift $\epsilon_0=2$ for the saddle points
$\Yboxdim8pt\yng(2)~\yng(1,1)$ and $\Yboxdim8pt\yng(1,1)~\yng(2)$.
See Table 2 in appendix B.3.

\subsection{Three KK-momenta: numerical tests}

We consider the case with $k\!=\!1$ only. The gauge theory indices
are given by
\begin{eqnarray}\label{3-equal-distribute}
  I_{\Yboxdim4pt\yng(3)~\yng(3)}&=&x^{\frac{3}{2}}
  \left(4+4x+0x^2+8x^3-4x^4+8x^5+2x^6+4x^7+0x^8+\mathcal{O}(x^9)\
  \right)\\
  I_{\Yboxdim4pt\yng(2,1)~\yng(2,1)}&=&x^{\frac{3}{2}}
  \left(6+20x+24x^2+28x^3+64x^4+34x^5+34x^6+166x^7-32x^8+\mathcal{O}(x^9)
  \ \right)\nonumber\\
  I_{\Yboxdim4pt\yng(1,1,1)~\yng(1,1,1)}&=&x^{\frac{3}{2}}
  \left(4+12x+30x^2+52x^3+52x^4+98x^5+170x^6+130x^7+106x^8+\mathcal{O}(x^9)
  \ \right)\nonumber
\end{eqnarray}
and
\begin{eqnarray}\label{3-unequal-distrib}
  I_{\Yboxdim4pt\yng(2,1)~\yng(1,1,1)}=I_{\Yboxdim4pt\yng(1,1,1)~\yng(2,1)}
  &=&x^{\frac{3}{2}}\left(0x^4+6x^5-10x^6-22x^7+88x^8+\mathcal{O}(x^9)
  \ \right)\nonumber\\
  I_{\Yboxdim4pt\yng(3)~\yng(2,1)}=
  I_{\Yboxdim4pt\yng(2,1)~\yng(3)}&=&x^{\frac{3}{2}}
  \left(2x^6-8x^7+16x^8+\mathcal{O}(x^9)\ \right)\\
  I_{\Yboxdim4pt\yng(3)~\yng(1,1,1)}=I_{\Yboxdim4pt\yng(1,1,1)~\yng(3)}
  &=&x^{\frac{3}{2}}\left(\ \mathcal{O}(x^{12})\ \right)\ .\nonumber
\end{eqnarray}
The last three pairs of saddle points with unequal distributions are
expected to enter from $(2k\!+\!2)$'th, $(4k\!+\!2)$'th and
$(6k\!+\!6)$'th level above the lowest level for general $k$,
respectively. Coefficients $0$ written above could be
accidental.\footnote{Unfortunately, we could not go to
$\mathcal{O}(x^{12})$ where one can start testing the last line of
(\ref{3-unequal-distrib}), due to the long execution time. The
bottleneck was at the third line of (\ref{3-equal-distribute}), in
which we had to integrate over the $U(3)\times U(3)$ holonomy with
nine factors of $F$ functions, etc., in the integrand. We hope we
can improve our calculation in the near future. We thank Sehun Chun
for his advice.}

From gravity, one can construct states carrying three units of
momenta in the following three ways: one graviton carrying three
momenta, one graviton with one momentum and another graviton with
two momenta, three identical particles where each carries one
momentum. Three contributions are
\begin{eqnarray}
  \hspace*{-1cm}I_3(x)\!&\!=\!&\!x^{\frac{3}{2}}\left(4+4x+2x^2+4x^3+2x^4+
  4x^5+2x^6+4x^7+2x^8\!+\!\mathcal{O}(x^9)\right)\nonumber\\
  \hspace*{-1cm}I_1(x)I_2(x)\!&\!=\!&\!x^{\frac{3}{2}}\left(6\!+\!20x\!+\!26x^2\!+\!
  36x^3\!+\!46x^4\!+\!52x^5\!+\!66x^6\!+\!68x^7\!+\!86x^8\!+\!
  \mathcal{O}(x^9)\right)
  \nonumber\\
  \hspace*{-1cm}
  \frac{1}{3}I_1(x^3)\!+\!\frac{1}{2}I_1(x)I_1(x^2)\!+\!\frac{1}{6}I_1(x)^3
  \!&\!=\!&\!
  x^{\frac{3}{2}}\left(4\!+\!12x\!+\!26x^2\!+\!48x^3\!+\!64x^4\!+\!
  96x^5\!+\!122x^6\!+\!168x^7\!+\!194x^8\!+\!\mathcal{O}(x^9)\right)\nonumber
\end{eqnarray}
From this we find
\begin{eqnarray}
  &&\hspace{-1cm}
  I_{\Yboxdim4pt\yng(3)~\yng(3)}+I_{\Yboxdim4pt\yng(2,1)~\yng(2,1)}
  +I_{\Yboxdim4pt\yng(1,1,1)~\yng(1,1,1)}+
  2I_{\Yboxdim4pt\yng(2,1)~\yng(1,1,1)}+
  2I_{\Yboxdim4pt\yng(3)~\yng(2,1)}+
  2I_{\Yboxdim4pt\yng(3)~\yng(1,1,1)}\nonumber\\
  &&=I_3(x)+I_1(x)I_2(x)+
  \frac{1}{3}I_1(x^3)+\frac{1}{2}I_1(x)I_1(x^2)+\frac{1}{6}I_1(x)^3
  +\mathcal{O}(x^{\frac{3}{2}+9})\ ,
\end{eqnarray}
which is again a perfect agreement.

\section{Conclusion}

In this paper we calculated the complete superconformal index for
$\mathcal{N}\!=\!6$ Chern-Simons-matter theory with gauge group
$U(N)_k\times U(N)_{-k}$ at level $k$. The low energy and large $N$
limit shows a perfect agreement with the index over supersymmetric
gravitons in M-theory on $AdS_4\times S^7/\mathbb{Z}_k$ in all
sample calculations we did.

Though we strongly suspect that the two large $N$ indices will
completely agree, it would definitely be desirable to develop a
general proof of this claim. Since the two indices assume very
different forms apparently, this would be a nontrivial check of the
AdS$_4$/CFT$_3$ proposal based on superconformal Chern-Simons
theories. Perhaps identities of unitary matrix integrals similar to those
explored in \cite{Dolan:2008qi} could be found to show this.

We expect our result to provide hints towards a better understanding
of 't Hooft operators and their roles in AdS$_4$/CFT$_3$. In
particular we found that monopole operators in nontrivial
representations, beyond those considered in
\cite{Aharony:2008ug,Berenstein:2008dc,Klebanov:2008vq}, played
important roles for the agreement of gauge/gravity indices. In our
calculation in the deformed theory, the degrees of freedom in the
vector multiplets turned out to be important, starting from
nontrivial contribution to the determinant and Casimir energy. It
would be interesting to understand it directly in the physical
Chern-Simons-matter theory, in which there are no propagating
degrees for the fields in vector multiplets. In a preliminary study,
we find that interaction between gauge fields and matters in
background fluxes transmutes some of the matter scalar degrees into
vector-like ones \cite{Kim:2009ia}.

We have kept the chemical potentials to be finite and order $1$ as
we take the large $N$ limit to obtain the low energy index. This
setting is also partly motivated from the physics that one expects
from the \textit{partition function}, namely a deconfinement phase
transition at order $1$ temperature. In the context of 4-dimensional
Yang-Mills theory, a first order deconfinement transition is found
in \cite{Sundborg:1999ue,Aharony:2003sx,Aharony:2005bq}. Although
the index does not see this either in the $d\!=\!4$
$\mathcal{N}\!=\!4$ Yang-Mills theory or here, possibly due to
cancelations from $(-1)^F$, it is not clear to us whether this means
that the trace of (supersymmetric) black holes and deconfined phase
is completely absent. It is a famous fact that the situation in
$d\!=\!3$ is more mysterious due to the replacement of $N^2$ by
$N^{\frac{3}{2}}$, or
$N^{\frac{3}{2}}\sqrt{k}=\frac{N^2}{\sqrt{\lambda}}$ in our case, in
the `deconfined' degrees of freedom. Another new aspect in $d\!=\!3$
compared to $d\!=\!4$ is the presence of sectors with topological
charges. It might be interesting to systematically investigate
finite $N$ effects in the flux distributions.

We think the finite $N$ and $k$ index that we obtained can be
straightforwardly extended to other superconformal Chern-Simons
theories. This can be used to solidly test many interesting ideas in
these theories. For example, a non-perturbative `parity duality' and
its generalization were proposed by \cite{Aharony:2008gk} for
\mbox{$\mathcal{N}\!=\!5,6$} Chern-Simons theories with gauge groups
$O(M)\times Sp(N)$ and $U(M)\times U(N)$, respectively, based on the
study of their gravity duals. The details of duality transformation
involves changes of parameters $M$, $N$ and the Chern-Simons level
$k$. The information on these parameters is of course wiped out in
the index in 't Hooft limit \cite{Choi:2008za}, and perhaps also in
the large $M$, $N$ limit keeping $k$ finite. The index with finite
$M$, $N$, $k$ should have a delicate structure for the duality to
hold. It would be interesting to explore this. An analysis of finite
$N$ indices for a class of 4 dimensional SCFT was reported in
\cite{Romelsberger:2007ec,Dolan:2008qi}.

For a class of superconformal Chern-Simons theories, the
superconformal index exhibits an interesting large $N$ phase
transition. For instance, it was explained that the index for
$\mathcal{N}\!=\!2,3$ $U(N)_k$ Chern-Simons theories with $N_f$
flavors (presented in \cite{Gaiotto:2007qi}) can undergo third order
phase transitions \cite{Bhattacharya:2008zy} in the so-called
Veneziano limit, which is very similar to that in the lattice gauge
theory explored by Gross, Witten and Wadia
\cite{Gross:1980he,Wadia:1979vk}. An interesting related issue is a
proposal by Giveon and Kutasov on Seiberg duality in these theories
\cite{Giveon:2008zn}, based on the study of brane constructions. See
also \cite{Niarchos:2008jb}. The $\mathcal{N}\!=\!3$ models were
suggested to be self strong-weak dual in that $\frac{N}{k}$ cannot
be small for both of the Seiberg-dual pair theories.
In some cases, it seems that the calculation of \cite{Bhattacharya:2008zy}
applies to only one of the two theories in the pair. We think that it could
be important to consider the sectors with magnetic fluxes to see the dual
phase transition. We hope to come back to this problem in a near future.

Finally, there have been active studies on superconformal
Chern-Simons theories which could be relevant to condensed matter
systems. For example, nonrelativistic versions of
Chern-Simons-matter theories were studied and some properties of the
superconformal indices were pointed out \cite{Nakayama:2008qm}. The
study of the last references in \cite{Nakayama:2008qm} suggests that
monopole operators are expected to play important roles. More recent
works include Chern-Simons-matter theories with fundamental matters
and their gravity duals \cite{Hohenegger:2009as,Gaiotto:2009tk}. It
should be interesting to apply our methods to these examples.

\vskip 0.5cm

\hspace*{-0.8cm} {\bf\large Acknowledgements}

\vskip 0.2cm

\hspace*{-0.75cm} I am grateful to Shiraz Minwalla for his early
collaboration and also for many helpful discussions and suggestions
throughout this work. I also thank Sayantani Bhattacharyya,
Kallingalthodi Madhu for collaborations, and Ofer Aharony, James
Bedford, Amihay Hanany, Ki-Myeong Lee, Sangmin Lee, Sungjay Lee,
Noppadol Mekareeya, Kyriakos Papadodimas, Mark van Raamsdonk,
Soo-Jong Rey, Riccardo Ricci, Ashoke Sen, Ho-Ung Yee, Piljin Yi and 
Shuichi Yokoyama for
helpful discussions. I acknowledge the support and hospitality of Tata
Institute of Fundamental Research and Korea Institute for Advanced
Study, respectively, where part of this work was done.

\appendix

\section{Notes on radial quantization}

In this appendix we summarize the conversion between conformal field
theories on $\mathbb{R}^{2+1}$ and $S^2\times\mathbb{R}$ via radial
quantization. In particular we would like to obtain the action on
the latter space starting from that on the former. The first
procedure is analytic continuation $x^0=-ix^3_E$ to $\mathbb{R}^3$
and obtaining the action in Euclidean $S^2\times\mathbb{R}$ by
setting $r=e^{\tau}$, where $\tau$ is the time of the latter space.
One may then continue back to Lorentzian $S^2\times\mathbb{R}$ with
Lorentzian time $t=-i\tau$, though in this paper we mainly consider
the Euclidean theory.

We first consider kinetic terms, involving derivatives acting on
fields. It suffices to consider free fields with ordinary
derivatives: covariantizing with gauge fields would take obvious
forms once the gauge fields are appropriately transformed into
fields in $S^2\times\mathbb{R}$ \cite{Grant:2008sk}. Let us start
from a (real) scalar field $\Phi$ on $\mathbb{R}^3$ with scale
dimension $\frac{1}{2}$. The field $\Phi_S$ on $S^2\times\mathbb{R}$
is related to $\Phi$ as
\begin{equation}
  \Phi=r^{-\frac{1}{2}}\Phi_S\ .
\end{equation}
The kinetic term on $\mathbb{R}^3$ can be rewritten as
\begin{eqnarray}
  \int d^3x\left(\nabla_{\mathbb{R}^3}\Phi\right)^2&=&
  \int {\rm vol}_{S^2}\ r^2dr\left[
  \partial_r\left(r^{-\frac{1}{2}}\Phi_S\right)^2+
  \frac{1}{r^2}\nabla_{S^2}\left(r^{-\frac{1}{2}}\Phi_S\right)^2\right]
  \nonumber\\
  &=&\int{\rm vol}_{S^2}\ d\tau\left[
  \left(\partial_\tau\Phi_S\right)^2+
  \left(\nabla_{S^2}\Phi_S\right)^2+\frac{1}{4}\left(\Phi_S\right)^2\right]
\end{eqnarray}
after eliminating surface terms. This is nothing but the action for
a scalar conformally coupled to the curvature. This result applies
to all eight real scalars $A_a,B_{\dot{a}}$.

The gauge fields $A_\mu,\tilde{A}_\mu$ and scalars
$\sigma,\tilde\sigma$ with dimensions $1$ can also be transformed
appropriately. We do not present the result here since we will
mostly work directly in $\mathbb{R}^3$ when we consider these
fields. One may see \cite{Grant:2008sk} for the details on
$\mathbb{R}^4$, which is essentially the same as our case. In
particular, $\sigma=r^{-1}\sigma_S$, $A_r=r^{-1}(A_S)_\tau$ and
$A_a=(A_S)_a$ for $a=\theta,\phi$. We note that the Chern-Simons
term takes the same form on $S^2\times\mathbb{R}$.

We also consider complex matter fermions $\Psi_\alpha$ with
dimension $1$ whose (Euclidean) kinetic term is given by
\begin{equation}
  -i\bar\Psi^\alpha(\gamma^m)_{\alpha\beta}\nabla_m\Psi^\beta
  =\bar\Psi_\alpha(\bar\sigma^m)^{\alpha\beta}
  \nabla_m\Psi_\beta\ \ \ (m=1,2,3\ {\rm for\ Cartesian\ coordi.})\ .
\end{equation}
In the representation with
$\gamma^m_E=(-\sigma^2,\sigma^1,\sigma^3)$, one obtains
$\bar\sigma^m=(1,i\sigma^3,-i\sigma^1)$ or an $SO(3)$ rotated one as
explained in section 2.1. Since we are considering fermions, we
should first change our dreibein frame on $\mathbb{R}^3$ from
Cartesian to spherical curvilinear one:
\begin{equation}
  e^r=dr\ ,\ \ e^\theta=rd\theta\ ,\ \ e^\phi=r\sin\theta d\phi\ .
\end{equation}
This frame is related to the Cartesian frame by a local $SO(3)$
transformation which we call $\Lambda$. One obtains
\begin{equation}
  \bar\Psi^\alpha(\gamma^m_E)_\alpha^{\ \beta}\nabla_m\Psi_\beta
  =\bar\Psi_{\rm cur}^\alpha
  \left[U(\Lambda^\dag)\gamma^m_EU(\Lambda)\right]_\alpha^{\ \ \beta}
  \Lambda_m^{\ \ n}
  E_n^\mu\nabla_\mu\Psi_{{\rm cur}\beta}=\bar\Psi_{\rm cur}^\alpha
  (\gamma^n_E)_\alpha^{\ \beta}E_n^\mu\nabla_\mu\Psi_{{\rm cur}\beta}\ ,
\end{equation}
where $U(\Lambda)$ is the spinor representation of $\Lambda$,
$\Psi=U(\Lambda)\Psi_{\rm cur}$, and $E_m$ is the inverse of the
above frame. At the last step we used the fact that action of any
local $SO(3)$ leaves $(\gamma^m)_\alpha^{\ \beta}$ invariant. Here
the final derivative $\nabla_\mu$ is covariantized with the
following spin connection (yet with zero curvature):
\begin{equation}
  \omega^{\theta r}=\frac{1}{r}e^\theta=d\theta\ ,\ \
  \omega^{\phi r}=\frac{1}{r}e^{\phi}=\sin\theta d\phi
  \ ,\ \ \omega^{\phi\theta}=\frac{\cot\theta}{r}e^\phi=
  \cos\theta d\phi\ :\ \
  \nabla_\mu=\partial_\mu+\frac{1}{4}\omega_\mu^{mn}\gamma_{mn}\ .
\end{equation}
Finally we change the field $\Psi_{\rm cur}$ to $\Psi_S$ living on
$S^2\times\mathbb{R}$, according to its dimension $1$, as
\begin{equation}
  \Psi_{\rm cur}=r^{-1}\Psi_S\ .
\end{equation}
This yields
\begin{equation}\label{fermion-redefine}
  \int {\rm vol}_{S^2}\ r^2 dr\ \bar\Psi^\alpha(\gamma^m_E)_\alpha^{\ \beta}
  \nabla_m\Psi_\beta=
  \int {\rm vol}_{S^2}\ dr\left[\bar\Psi_S^\alpha
  (\gamma^n_E)_\alpha^{\ \beta}E_n^\mu\nabla_\mu\Psi_{S\beta}
  +\frac{1}{r}\bar\Psi_S^\alpha(\sigma^2)_\alpha^{\ \beta}
  \Psi_{S\beta}\right]\ .
\end{equation}
The covariant derivative $\nabla_\mu$ is still that on
$\mathbb{R}^3$. Since we are trying to obtain an action on
$S^2\times\mathbb{R}$, with metric changing from
$ds^2_{\mathbb{R}^3}=dr^2+r^2ds^2_{S^2}=r^2ds^2_{S^2\times\mathbb{R}}$
to $ds^2_{S^2\times\mathbb{R}}$, we rewrite the spin connection of
$\mathbb{R}^3$ in terms of that of $S^2\times\mathbb{R}$. The
covariant derivatives are related as
\begin{equation}
  \nabla=\nabla_S+\frac{1}{2}\left(\frac{1}{r}e^\theta\gamma_{\theta r}+
  \frac{1}{r}e^\phi\gamma_{\phi r}\right)=
  \nabla_S-\frac{i}{2r}\left(e^\theta\sigma^3-e^\phi\sigma^1\right)\ .
\end{equation}
Thus one finds that
\begin{equation}\label{connection-redefine}
  \gamma^n_EE_n^\mu\nabla_\mu=\frac{1}{r}\gamma^n_E(E_S)_n^\mu
  (\nabla_S)_\mu-\frac{1}{r}\sigma^2\ ,
\end{equation}
where $E_S$ denotes the inverse frame for the metric
$ds^2_{S^2\times\mathbb{R}}$. The final Lagrangian for $\Psi_S$ is
\begin{equation}
  \int{\rm vol}_{S^2}d\tau\
  \bar\Psi_S\gamma^m_E(E_S)_m^\mu(\nabla_S)_\mu\Psi_S\ ,
\end{equation}
where the second term in (\ref{fermion-redefine}) is canceled by
that in (\ref{connection-redefine}). Note that there is no analogue
of conformal mass terms for scalars, which is well-known from
literatures on CFT in curved spaces \cite{Nicolai:1988ek}.

Finally we consider the kinetic term for the gaugino fields
$\lambda_\alpha$, $\tilde\lambda_\alpha$ in the $Q$-exact
deformation
\begin{equation}
  \int d^3x\ r\lambda\gamma^m_E\nabla_m\bar\lambda\ ,
\end{equation}
and a similar term for $\tilde\lambda$. These fields have dimensions
$\frac{3}{2}$. The analysis is almost the same as the previous
paragraph, apart from the fact that the fields on
$S^2\times\mathbb{R}$ are given by
\begin{equation}
  \lambda_{\rm cur}=r^{-\frac{3}{2}}\lambda_S\ ,\ \
  \tilde\lambda_{\rm cur}=r^{-\frac{3}{2}}\tilde\lambda_S\ .
\end{equation}
Therefore the step analogous to (\ref{fermion-redefine}) yields
\begin{equation}\label{gaugino-redefine}
  \int d^3x\ r\lambda^\alpha(\gamma^m_E)_\alpha^{\ \beta}\nabla_m
  \bar\lambda_\beta=\int{\rm vol}_{S^2}\ dr\left[\lambda_S^\alpha
  (\gamma^m_E)_\alpha^{\ \beta}E_m^\mu\nabla_\mu\bar\lambda_{S\beta}
  +\frac{3}{2r}\lambda_S^\alpha(\sigma^2)_\alpha^{\ \beta}
  \bar\lambda_{S\beta}\right]\ .
\end{equation}
As in the previous paragraph, rewriting the covariant derivative on
$\mathbb{R}^3$ in terms of that on $S^2\times\mathbb{R}$ provides a
term $-\frac{1}{r}\lambda_S^\alpha(\sigma^2)_\alpha^{\
\beta}\bar\lambda_{S\beta}$, which in this case \textit{does not}
completely cancel the second term in (\ref{gaugino-redefine}). The
final kinetic term for $\lambda_S$ is, in terms of
$(\sigma^\mu)_{\alpha\beta}=(1,i\vec\sigma)$,
\begin{equation}
  \int{\rm vol}_{S^2}\ d\tau\ \lambda_S^\alpha\left[
  (\sigma^m)_{\alpha\beta}(E_S)_m^\mu
  (\nabla_S)_\mu
  -\frac{1}{2}\delta_{\alpha\beta}\right]\bar\lambda_S^\beta\ ,
\end{equation}
where $\delta_{\alpha\beta}$ comes from
$(\bar\sigma^0)_{\alpha\beta}$. Action for $\tilde\lambda_S$ is
similar.

The terms in the action which do not involve derivatives, such as
potential, Yukawa interaction, etc., transform rather obviously.
Starting from $\int d^3x\mathcal{L}(\Phi)$ and transforming the
fields according to their dimensions $\Delta$ as
$\Phi=r^{-\Delta}\Phi_S$, one obtains
$\int_{S^2\times\mathbb{R}}\mathcal{L}(\Phi_S)$.

\section{Details of 1-loop calculation}

The 1-loop determinant comes from two contributions: firstly from
the quadratic fluctuations of the matter fields and secondly from
the fields in vector multiplets. We explain them in turn.

\subsection{Determinant from matter fields}

We first consider the matter scalar fields. The quadratic action for
the scalars (on $S^2\times S^1$) in the presence of nonzero
$A_\mu,\tilde{A}_\mu,\sigma,\tilde\sigma$ background takes the
following form:
\begin{eqnarray}\label{quad-scalar}
  \mathcal{L}_{s2}&=&{\rm tr}\left[\frac{}{}\right.\!
  -\bar{A}^aD^\mu D_\mu A_a -\bar{B}^{\dot{a}}D^\mu D_\mu B_{\dot{a}}
  +\frac{1}{4}\left(A_a\bar{A}^a+B_{\dot{a}}
  \bar{B}^{\dot{a}}\right)\nonumber\\
  &&\hspace{1cm}\left.+(\sigma A_a\!-\!A_a\tilde\sigma)
  (\bar{A}^a\sigma\!-\!\tilde\sigma\bar{A}^a)+
  (\tilde\sigma B_{\dot{a}}\!-\!\sigma B_{\dot{a}})
  (\bar{B}^{\dot{a}}\tilde\sigma\!-\!\bar{B}^{\dot{a}}\sigma)\frac{}{}\right]
\end{eqnarray}
where derivatives $D_\mu$ are covariantized with the background
gauge fields, external gauge field corresponding to the twisting and
$S^2$ spatial connection. The fields are regarded as those on
$S^2\times\mathbb{R}$, $A_{aS}$, etc. $\sigma,\tilde\sigma$ assume
their background values $(\sigma_S)_{ij}=-\frac{n_i}{2}\delta_{ij}$
and $(\tilde\sigma_S)_{ij}=-\frac{\tilde{n}_i}{2}\delta_{ij}$ where
$i,j=1,2,\cdots, N$. The third term in the trace is the conformal
mass term.

Since there are $U(1)^N\times U(1)^N$ background magnetic fields on
$S^2$, each component of the scalars has to be spanned by
appropriate monopole spherical harmonics on $S^2$ \cite{Wu:1976ge}.
Scalars are either in $(N,\bar{N})$ or $(\bar{N},N)$ representation
of $U(N)\times U(N)$. We consider the $ij$'th component of the
scalar, where $i$ ($j$) runs over $1,2,\cdots,N$ and labels the
components in the first (second) gauge group. A scalar $\Phi_{ij}$
couples to the background magnetic field whose strength is
\begin{equation}
  \frac{s(n_i\!-\!\tilde{n}_j)}{2}
  \sin\theta d\theta\wedge d\phi\ \ \ \ \
  (s=\pm 1\ {\rm for}\ (N,\bar{N}),\ (\bar{N},N))\ ,
\end{equation}
corresponding to a Dirac monopole carrying $s(n_i\!-\!\tilde{n}_j)$
units of minimal charge. The monopole spherical harmonics $Y_{jm}$
in this background, with angular momentum quantum numbers $j,m$
given by
\begin{equation}
  j=\frac{|n_i\!-\!\tilde{n}_j|}{2},\ \frac{|n_i\!-\!\tilde{n}_j|}{2}+1,\
  \cdots\ {\rm and}\ \ m=-j,-(j\!-\!1)\cdots,j\!-\!1,j\ ,
\end{equation}
diagonalize the spatial Laplacian on $S^2$ as
\begin{equation}\label{monopole-eigen}
  -D^aD_aY_{jm}=\left(j(j+1)-\frac{(n_i\!-\!\tilde{n}_j)^2}{4}\right)
  Y_{jm}
\end{equation}
where $a=1,2$ labels the coordinates of $S^2$.

Plugging this mode expansion into the quadratic action, one can
easily see that the second term on the right hand side of
(\ref{monopole-eigen}) is canceled by the second line of
(\ref{quad-scalar}). Collecting all, the $Y_{jm}$ mode
$\Phi_{ij}^{jm}$ has a quadratic term
\begin{equation}
  \bar\Phi_{ij}^{jm}\left[-(D_\tau)^2+\left(j+\frac{1}{2}\right)^2
  \right]\Phi_{ij}^{jm}\ .
\end{equation}
We hope our bad notation of using two kinds of $j$ is not too
confusing. The time derivative is
\begin{equation}
  D_\tau=\partial_\tau-is\frac{\alpha_i\!-\!\tilde\alpha_j}
  {\beta\!+\!\beta^\prime}
  -\frac{\beta-\beta^\prime}{\beta+\beta^\prime}m
  +\frac{\beta^\prime}{\beta+\beta^\prime}h_3-\frac{\gamma_1}
  {\beta+\beta^\prime}h_1-\frac{\gamma_2}{\beta+\beta^\prime}h_2\ ,
\end{equation}
where $h_{1,2,3}$ are eigenvalues of $SO(6)_R$ Cartans for the given
field $\Phi$. The determinant is evaluated for each conjugate pair
of scalar fields $(\Phi,\bar\Phi)$, where $\Phi$ may run over four
complex scalars, say, $A_a,B_{\dot{a}}$. The determinant from the
pair $\Phi,\bar\Phi$ is given by
\begin{eqnarray}
  &&\hspace{-0.5cm}
  \prod_{j=\frac{|n_i\!-\!\tilde{n}_j|}{2}}^\infty\prod_{j_3=-j}^j
  \det\left[-\left(\partial_\tau-is\frac{\alpha_i\!-\!\tilde\alpha_j}
  {\beta\!+\!\beta^\prime}-\frac{\beta-\beta^\prime}{\beta+\beta^\prime}j_3
  +\frac{\beta^\prime}{\beta+\beta^\prime}h_3-
  \frac{\gamma_1h_1+\gamma_2h_2}{\beta+\beta^\prime}
  \right)^2+\left(j+\frac{1}{2}\right)^2\right]\nonumber\\
  &&\hspace{-0.3cm}=\prod_{n=-\infty}^\infty\prod_{j,j_3}
  \left[\left(\frac{2\pi n}{\beta+\beta^\prime}
  +s\frac{\alpha_i\!-\!\tilde\alpha_j}{\beta\!+\!\beta^\prime}
  -i\frac{\beta-\beta^\prime}{\beta+\beta^\prime}j_3
  +i\frac{\beta^\prime}{\beta+\beta^\prime}h_3-
  i\frac{\gamma_1h_1+\gamma_2h_2}{\beta+\beta^\prime}
  \right)^2+\left(j+\frac{1}{2}\right)^2\right]\ .\nonumber
\end{eqnarray}
Following the prescription in \cite{Aharony:2003sx}\footnote{See
eqns. (4.23), (4.24) there and surrounding arguments.}, we factor
out a divergent constant, set it to unity, and obtain
\begin{eqnarray}\label{scalar-pair-det}
  &&\hspace{-0.5cm}\prod_{j,j_3}(-2i)
  \sin\left[\frac{1}{2}\left(\frac{}{}\!s
  (\tilde\alpha_j\!-\!\alpha_i)+i\beta(\epsilon_j\!+\!j_3)+
  i\beta^\prime(\epsilon_j\!-\!h_3\!-\!j_3)+i(\gamma_1h_1\!+\!\gamma_2h_2)
  \right)\right]\nonumber\\
  &&\times(-2i)
  \sin\left[\frac{1}{2}\left(\!\!\frac{}{}\!-\!s
  (\tilde\alpha_j\!-\!\alpha_i)+i\beta(\epsilon_j\!-\!j_3)+
  i\beta^\prime(\epsilon_j\!+\!h_3\!+\!j_3)-i(\gamma_1h_1\!+\!\gamma_2h_2)
  \right)\right],
\end{eqnarray}
where $\epsilon_j\equiv j\!+\!\frac{1}{2}$. Generalizing
\cite{Aharony:2003sx}, the two sine factors in the final form has an
obvious interpretation as contributions from a pair of particle and
anti-particle modes, since all charges except `energy' $\epsilon_j$
have different signs. Therefore, the determinant from the scalars
admits a simple form
\begin{eqnarray}\label{scalar-det}
  \hspace*{-0.7cm}\det\!{}_{\rm scalar}&=&\prod_{i,j}
  \prod_{8\ {\rm scalars}}\prod_{j,j_3}
  \sin\left[\frac{1}{2}\left(\frac{}{}\!s
  (\tilde\alpha_j\!-\!\alpha_i)+i\beta(\epsilon_j\!+\!j_3)+
  i\beta^\prime(\epsilon_j\!-\!h_3\!-\!j_3)+i(\gamma_1h_1\!+\!\gamma_2h_2)
  \right)\right]\\
  &=&\prod_{i,j}\prod_{8\ {\rm scalars}}\prod_{j,j_3}
  e^{\frac{is}{2}(\alpha_i\!-\!\tilde\alpha_j)+\frac{\beta}{2}(\epsilon_j\!+\!j_3)+
  \frac{\beta^\prime}{2}(\epsilon_j\!-\!h_3\!-\!j_3)+\gamma_1h_2+\gamma_2h_2}
  \left(1-e^{is(\tilde\alpha_j\!-\!\alpha_i)}
  x^{\epsilon_j\!+\!j_3}
  (x^\prime)^{\epsilon_j\!-\!h_3\!-\!j_3}y_1^{h_1}y_2^{h_2}\right)\nonumber
\end{eqnarray}
where $x^\prime\equiv e^{-\beta^\prime}$. The product is over $8$
scalars regarding conjugate pairs as independent fields.

We would like to write $(\det_{\rm scalar})^{-1}$, appearing in the
index, in terms of functions which we call the indices over
`letters,' or modes. We find that
\begin{eqnarray}\label{scalar-det-log}
  \hspace*{-0cm}\log(\det\!{}_{\rm scalar})^{-1}
  &\equiv&-\sum_{i,j}\sum_{{\rm scalars}}\sum_{j,j_3}
  \left[\frac{is}{2}(\alpha_i\!-\!\tilde\alpha_j)+\frac{\beta}{2}(\epsilon_j\!+\!j_3)+
  \frac{\beta^\prime}{2}(\epsilon_j\!-\!h_3\!-\!j_3)+\gamma_1h_2+\gamma_2h_2
  \right]\\
  &&+\sum_{i,j=1}^N\sum_{n=1}^\infty\frac{1}{n}\left[
  f^{+B}_{ij}(x^n,(x^\prime)^n,y_1^n,y_2^n)
  e^{in(\tilde\alpha_j-\alpha_i)}+f^{-B}_{ij}
  (x^n,(x^\prime)^n,y_1^n,y_2^n)e^{in(\alpha_i-\tilde\alpha_j)}
  \!\frac{}{}\right]\ .\nonumber
\end{eqnarray}
The first line provides a quantity analogous to the Casimir energy,
which will be computed in appendix B.3. The contribution from
scalars to the letter index is given by
\begin{equation}
  f^{\pm B}_{ij}(x,x^\prime,y_1,y_2)\equiv
  \sum_{\substack{4\ {\rm scalars}\\s=\pm 1}}
  \sum_{j=\frac{|n_i\!-\!\tilde{n}_j|}{2}}^\infty\sum_{j_3=-j}^j
  \left(\frac{}{}\!x^{\epsilon_j+j_3}(x^\prime)^{\epsilon_j-h_3-j_3}
  y_1^{h_2} y_2^{h_2}\right)
\end{equation}
where the first summation is restricted to fields with one of $s=\pm
1$. Explicitly summing over the scalar modes, one obtains
\begin{eqnarray}
  \hspace*{-1cm}f^{+B}_{ij}(x,x^\prime,y_1,y_2)&=&
  \left(\!\sqrt{\frac{y_1}{y_2}}\!+\!\sqrt{\frac{y_2}{y_1}}\right)
  \sum_{j=\frac{|n_i\!-\!\tilde{n}_j|}{2}}^\infty x^{\frac{1}{2}}
  \left[(x^\prime)^{2j}+(x^\prime)^{2j-1}x^{}+\cdots+
  x^\prime x^{2j\!-\!1}+x^{2j}\right]\\
  &&+\left(\!\sqrt{y_1y_2}\!+\!\frac{1}{\sqrt{y_1y_2}}\right)
  \sum_{j=\frac{|n_i\!-\!\tilde{n}_j|}{2}}^\infty x^\prime x^{\frac{1}{2}}
  \left[(x^\prime)^{2j}+(x^\prime)^{2j-1}x^{}+\cdots+
  x^\prime x^{2j\!-\!1}+x^{2j}\right]\nonumber
\end{eqnarray}
where the two lines come from $\bar{B}^{\dot{a}}$ and $A_a$,
respectively, and
\begin{eqnarray}
  \hspace*{-1cm}f^{-B}_{ij}(x,x^\prime,y_1,y_2)&=&
  \left(\!\sqrt{y_1y_2}\!+\!\frac{1}{\sqrt{y_1y_2}}\right)
  \sum_{j=\frac{|\tilde{n}_i\!-\!n_j|}{2}}^\infty x^{\frac{1}{2}}
  \left[(x^\prime)^{2j}+(x^\prime)^{2j-1}x+\cdots+
  x^\prime x^{2j\!-\!1}+x^{2j}\right]\\
  &&+\left(\!\sqrt{\frac{y_1}{y_2}}\!+\!\sqrt{\frac{y_2}{y_1}}\right)
  \sum_{j=\frac{|\tilde{n}_i\!-\!n_j|}{2}}^\infty x^\prime x^{\frac{1}{2}}
  \left[(x^\prime)^{2j}+(x^\prime)^{2j-1}x^{}+\cdots+
  x^\prime x^{2j\!-\!1}+x^{2j}\right]\ .\nonumber
\end{eqnarray}
where the two lines come from $\bar{A}^a$ and $B_{\dot{a}}$,
respectively. The dependence on $x^\prime$ is to be canceled against
the contribution from fermions.

We also consider the determinant from fermions. Fermionic quadratic
action is given by
\begin{equation}
  \mathcal{L}_{f2}=\bar\psi^a_\alpha(\bar\sigma^\mu)^{\alpha\beta}D_\mu
  \psi_a^\beta+\bar\psi^{\dot{a}}_\alpha(\bar\sigma^\mu)^{\alpha\beta}D_\mu
  \chi_{\dot{a}\beta}\ ,
\end{equation}
where $D_3\psi_a=-i\sigma\psi_a+i\psi_a\tilde\sigma$ and
$D_3\chi_{\dot{a}}=i\chi_{\dot{a}}\sigma-i\tilde\sigma\chi_{\dot{a}}$.
As explained in section 2.1,
$\bar\sigma^\mu=(1,i\sigma^3,-i\sigma^1,i\sigma^2)$ is changed to
$\bar\sigma^\mu=(1,-i\sigma^1,-i\sigma^2,-i\sigma^3)$ by an $SO(3)$
frame rotation. Since the latter basis is more convenient in that
spin operator on $S^2$ is diagonalized, we do our computation in
this basis.

Let us denote by $\Psi_{ij}$ the $i$'th and $j$'th component of
fermions $\psi_a$ or $\chi_{\dot{a}}$ in the first and second gauge
group, respectively. We want to compute the determinant of the
matrix differential operator
\begin{equation}\label{spinor-operator}
  \bar\sigma^\mu D_\mu=
  D_\tau-i\sigma^aD_a+s\sigma^3\frac{n_i\!-\!\tilde{n}_j}{2}\ ,
\end{equation}
where $a=1,2$, and the last term comes from the coupling with
background $\sigma,\tilde\sigma$. We would first like to obtain the
complete basis of spinor spherical harmonics diagonalizing
\begin{equation}\label{spinor-eigen}
  \left(i\sigma^aD_a-s\sigma^3\frac{n_i\!-\!\tilde{n}_j}{2}\right)
  \Psi=\lambda\Psi
\end{equation}
with eigenvalue $\lambda$. Acting the same operator again on the
above equation, one obtains
\begin{equation}\label{spinor-square}
  \left(-D^aD_a+\frac{1-s(n_i\!-\!\tilde{n}_j)\sigma^3}{2}+
  \frac{|n_i\!-\!\tilde{n}_j|^2}{4}\right)\Psi=\lambda^2\Psi
\end{equation}
where the second term comes from the commutator of two covariant
derivatives and is the sum of the spatial curvature and the field
strength. The first operator $-D^aD_a$ is a $2\times 2$ diagonal
matrix since the derivative involves
$+\frac{i}{2}\omega^{\theta\phi}\sigma^3$. The spectrum of this
operator is known and may be found, for instance, in
\cite{Kim:2001tw}. For the spinor component $\alpha=\pm$, its
eigenvalue is given by
\begin{equation}
  l_\pm\left(l_\pm+|s(n_i\!-\!\tilde{n}_j)\mp 1|+1\frac{}{}\!\right)+
  \frac{|s(n_i\!-\!\tilde{n}_j)\mp 1|}{2}
\end{equation}
where $l_\pm=0,1,2,\cdots$, and $\Psi_\pm$ is given by scalar
monopole harmonics with
$j_\pm=l_\pm+\frac{|s(n_i\!-\!\tilde{n}_j)\mp 1|}{2}$, coupled to
$s(n_i\!-\!\tilde{n}_j)\mp 1$ units of minimal Dirac monopoles.
Plugging this in (\ref{spinor-square}) and studying the upper/lower
components, one obtains
\begin{eqnarray}\label{spinor-eigenvalue}
  \lambda^2&=&\left(l_+ +\frac{|n_i\!-\!\tilde{n}_j|}{2}\right)^2=
  \left(l_- +\frac{|n_i\!-\!\tilde{n}_j|}{2}+1\right)^2\ \ \
  (l_+\!=\!l_-\!+\!1=1,2,3,\cdots)
  \ \ \ {\rm if}\ s(n_i\!-\!\tilde{n}_j)>0\nonumber\\
  \lambda^2&=&\left(l_+ +\frac{|n_i\!-\!\tilde{n}_j|}{2}+1\right)^2=
  \left(l_- +\frac{|n_i\!-\!\tilde{n}_j|}{2}\right)^2\ \ \
  (l_-\!=\!l_+\!+\!1=1,2,3,\cdots)
  \ \ \ {\rm if}\ s(n_i\!-\!\tilde{n}_j)<0\nonumber\\
  \lambda^2&=&(l_\pm+1)^2\ \ \ (l_+=l_-=0,1,2,\cdots)\ \ \ {\rm if}\
  n_i\!=\!\tilde{n}_j\ .
\end{eqnarray}
The eigenspinors are given as follows. In all three cases, one finds
a pair of eigenspinors corresponding to $\lambda\gtrless 0$,
\begin{equation}\label{eigen-paired}
  \left(\begin{array}{c}\Psi_+\\ \Psi_-\end{array}\right)=
  \left(\begin{array}{c}Y_{jm}\\ \pm Y_{jm}\end{array}\right)
\end{equation}
with $j\equiv j_+=j_-$, where the latter two are equal if one
relates $l_+$ and $l_-$ as explained in (\ref{spinor-eigenvalue}).
$j\geq\frac{|n_i\!-\!\tilde{n}_j|+1}{2}$ is the total angular
momentum of the mode.

Apart from the above modes, there is a set of exceptional modes in
the complete set if $n_i\neq\tilde{n}_j$. For the first and second
cases in (\ref{spinor-eigenvalue}), there exist nonzero modes
\begin{eqnarray}\label{eigen-unpaired}
  &&\left(\begin{array}{c}Y_{\frac{|n_i\!-\!\tilde{n}_j|-1}{2},m}\\0
  \end{array}\right)\ \ \ \ {\rm if}\ \
  \ s(n_i\!-\!\tilde{n}_j)>0\nonumber\\
  &&\left(\begin{array}{c}0\\Y_{\frac{|n_i\!-\!\tilde{n}_j|-1}{2},m}
  \end{array}\right)\ \ \ \ {\rm if}\ \
  \ s(n_i\!-\!\tilde{n}_j)<0\ .
\end{eqnarray}
These modes corresponds to $l_\pm=0$ on the first/second line of
(\ref{spinor-eigenvalue}), respectively. By directly studying
(\ref{spinor-eigen}), one finds that the eigenvalue is always
negative for both cases, i.e.
$\lambda=-\frac{|n_i\!-\!\tilde{n}_j|}{2}=-(j\!+\!\frac{1}{2})$.

Expanding the operator (\ref{spinor-operator}) in the above basis,
and following steps similar to those for the scalar determinant, one
obtains
\begin{eqnarray}
  \hspace*{-1cm}\det\!{}_{\rm f}\!&\!=\!&\!\prod_{i,j}
  \prod_{8\ \rm fermions}
  \prod_{j\geq\frac{|n_i\!-\!\tilde{n}_j|\!+\!1}{2}}\prod_{j_3}
  (-2i)\sin\left[\frac{1}{2}\left(\frac{}{}\!s
  (\tilde\alpha_j\!-\!\alpha_i)+i\beta(\epsilon_j\!+\!j_3)+
  i\beta^\prime(\epsilon_j\!-\!h_3\!-\!j_3)+i(\gamma_1h_1\!+\!\gamma_2h_2)
  \right)\right]\nonumber\\
  &&\times\prod_{i,j}\prod_{\bar\psi^a,\bar\chi^{\dot{a}}}
  \prod_{j_3=-\frac{|n_i\!-\!\tilde{n}_j|\!-\!1}{2}}^{
  \frac{|n_i\!-\!\tilde{n}_j|\!-\!1}{2}}(-2i)\sin\left[\frac{1}{2}\left(\frac{}{}\!s
  (\tilde\alpha_j\!-\!\alpha_i)+i\beta(\epsilon_j\!+\!j_3)+
  i\beta^\prime(\epsilon_j\!-\!h_3\!-\!j_3)+i(\gamma_1h_1\!+\!\gamma_2h_2)
  \right)\right]\nonumber
\end{eqnarray}
where $\epsilon_j=j+\frac{1}{2}$ for fermions as well. Let us
explain how each term is derived. In the first line, 8 fermions in
the product denote
$\psi_{a},\bar\psi^a,\chi_{\dot{a}},\bar\chi^{\dot{a}}$. This comes
from the paired eigenmodes (\ref{eigen-paired}) as one evaluates the
determinant of the operator (\ref{spinor-operator}). The second line
is multiplied over four fields only since the modes in
(\ref{eigen-unpaired}) do not appear in a paired form. From the fact
that $\lambda$ is negative when the differential operator acts on
the chiral spinors $\psi_a,\chi_{\dot{a}}$, one can easily check
that only the charges of $\bar\psi^a,\bar\chi^{\dot{a}}$ have to be
inserted on the second line.

One can also write this determinant in terms of indices over letters
as follows:
\begin{eqnarray}\label{ferm-det-log}
  \hspace*{-0.3cm}\log(\det\!{}_{\rm fermion})\!&\!=\!&\!
  +\sum_{i,j}\sum_{{\rm fermions}}\sum_{j,j_3}
  \left[\frac{is}{2}(\alpha_i\!-\!\tilde\alpha_j)+
  \frac{\beta}{2}(\epsilon_j\!+\!j_3)+\frac{\beta^\prime}{2}
  (\epsilon_j\!-\!h_3\!-\!j_3)+\gamma_1h_2+\gamma_2h_2\right]\\
  \hspace*{-0.3cm}&&+\sum_{i,j=1}^N\sum_{n=1}^\infty\frac{1}{n}\left[
  f^{+B}_{ij}(x^n,(x^\prime)^n,y_1^n,y_2^n)
  e^{in(\tilde\alpha_j-\alpha_i)}+f^{-B}_{ij}
  (x^n,(x^\prime)^n,y_1^n,y_2^n)e^{in(\alpha_i-\tilde\alpha_j)}
  \!\frac{}{}\right]\ ,\nonumber
\end{eqnarray}
where
\begin{eqnarray}
  \hspace*{-1.5cm}f^{+F}_{ij}(x,x^\prime,y_1,y_2)\!&=&\!
  -\left(\!\sqrt{y_1y_2}\!+\!\frac{1}{\sqrt{y_1y_2}}\right)
  \sum_{j=\frac{|n_i\!-\!\tilde{n}_j|+1}{2}}^\infty x^{\frac{1}{2}}
  \left[(x^\prime)^{2j}+(x^\prime)^{2j-1}x^{}+\cdots+
  x^\prime x^{2j\!-\!1}+x^{2j}\right]\nonumber\\
  &&\!-\left(\!\sqrt{\frac{y_1}{y_2}}\!+\!\sqrt{\frac{y_2}{y_1}}\right)
  \sum_{j=\frac{|n_i\!-\!\tilde{n}_j|-1}{2}}^\infty x^\prime x^{\frac{1}{2}}
  \left[(x^\prime)^{2j}+(x^\prime)^{2j-1}x^{}+\cdots+
  x^\prime x^{2j\!-\!1}+x^{2j}\right]\nonumber
\end{eqnarray}
from $\psi_{a\alpha}$ and $\bar\chi^{\dot{a}}_\alpha$, and
\begin{eqnarray}
  \hspace*{-1.5cm}f^{-F}_{ij}(x,x^\prime,y_1,y_2)\!&=&\!
  -\left(\!\sqrt{\frac{y_1}{y_2}}\!+\!\sqrt{\frac{y_2}{y_1}}\right)
  \sum_{j=\frac{|n_i\!-\!\tilde{n}_j|+1}{2}}^\infty x^{\frac{1}{2}}
  \left[(x^\prime)^{2j}+(x^\prime)^{2j-1}x^{}+\cdots+
  x^\prime x^{2j\!-\!1}+x^{2j}\right]\nonumber\\
  &&\!-\left(\!\sqrt{y_1y_2}\!+\!\frac{1}{\sqrt{y_1y_2}}\right)
  \sum_{j=\frac{|n_i\!-\!\tilde{n}_j|-1}{2}}^\infty x^\prime x^{\frac{1}{2}}
  \left[(x^\prime)^{2j}+(x^\prime)^{2j-1}x^{}+\cdots+
  x^\prime x^{2j\!-\!1}+x^{2j}\right]\nonumber
\end{eqnarray}
from $\chi_{\dot{a}\alpha}$ and $\bar\psi^a_\alpha$.

We combine the bosonic and fermionic determinants and obtain
\begin{eqnarray}\label{matter-det-log}
  \hspace*{-0cm}\log\left(\frac{\det\!{}_{\rm fermion}}
  {\det\!{}_{\rm scalar}}\right)&=&
  -\sum_{i,j}\sum_{{\rm matter}}\sum_{j,j_3}(-1)^F
  \left[\frac{is}{2}(\alpha_i\!-\!\tilde\alpha_j)+
  \frac{\beta}{2}(\epsilon_j\!+\!j_3)+\frac{\beta^\prime}{2}
  (\epsilon_j\!-\!h_3\!-\!j_3)+\gamma_1h_2+\gamma_2h_2\right]\nonumber\\
  &&+\sum_{i,j=1}^N\sum_{n=1}^\infty\frac{1}{n}\left[
  f^{+}_{ij}(x^n,(x^\prime)^n,y_1^n,y_2^n)
  e^{in(\tilde\alpha_j-\alpha_i)}+f^{-}_{ij}
  (x^n,(x^\prime)^n,y_1^n,y_2^n)e^{in(\alpha_i-\tilde\alpha_j)}
  \!\frac{}{}\right]\ ,\nonumber
\end{eqnarray}
where
\begin{equation}
  f^+_{ij}(x,y_1,y_2)=f^{+B}_{ij}+f^{+F}_{ij}=
  x^{|n_i\!-\!\tilde{n}_j|}\left[\frac{x^{\frac{1}{2}}}{1-x^2}
  \left(\!\sqrt{\frac{y_1}{y_2}}\!+\!\sqrt{\frac{y_2}{y_1}}\right)
  -\frac{x^{\frac{3}{2}}}{1-x^2}
  \left(\!\sqrt{y_1y_2}\!+\!\frac{1}{\sqrt{y_1y_2}}\right)\right]
\end{equation}
and similarly
\begin{equation}
  f^-_{ij}(x,y_1,y_2)=
  x^{|n_i\!-\!\tilde{n}_j|}\left[\frac{x^{\frac{1}{2}}}{1-x^2}
  \left(\!\sqrt{y_1y_2}\!+\!\frac{1}{\sqrt{y_1y_2}}\right)
  -\frac{x^{\frac{3}{2}}}{1-x^2}
  \left(\!\sqrt{\frac{y_1}{y_2}}\!+\!\sqrt{\frac{y_2}{y_1}}\right)
  \right]\ .
\end{equation}
This proves the assertion in section 2.3 on determinant from matter
fields.

\subsection{Determinant from fields in vector multiplets}

We also consider the 1-loop determinant from fields in vector
multiplets. We consider the multiplet $A_\mu,\sigma,\lambda_\alpha$:
the other vector multiplet can be treated in a completely same way.

We start from the bosonic part. We expand the quadratic fluctuation
in the $Q$-exact deformation, which is dominant in the limit
$g\rightarrow 0$. Denoting the fluctuation by $\delta
A_\mu,\delta\sigma$, one finds the following quadratic term:
\begin{equation}\label{vec-bos-quadrataic}
  \left|\vec{D}\times\delta\vec{A}-\vec{D}\delta\sigma
  -i[\sigma,\delta\vec{A}]\right|^2\ .
\end{equation}
We are directly working in $\mathbb{R}^3$ with $1\leq r\leq e^\beta$
rather than going to $S^2\times S^1$. The boundary conditions are
\begin{equation}
  \delta\vec{A}(r\!=\!e^{\beta})=e^{-\beta}\delta\vec{A}(r\!=\!1)\ ,\ \
  \delta\sigma(r\!=\!e^\beta)=e^{-\beta}\delta\sigma(r\!=\!1)\ ,
\end{equation}
associated with their scale dimensions $1$.

$\delta\sigma_{ij}$ is expanded with monopole spherical harmonics
with $n_i\!-\!n_j$ units of magnetic charge. We can also expand
$\delta\vec{A}_{ij}$ using monopole vector spherical harmonics,
which is nicely presented in \cite{Weinberg:1993sg}. For $j\geq q+1$
where $q\equiv\frac{n_i\!-\!n_j}{2}\geq 0$, it has three components
$\vec{C}^{\lambda}_{qjm}$ (with $\lambda=+1,0,-1$) and are related
to the scalar harmonics as
\begin{eqnarray}
  \vec{C}^{+1}_{qjm}&=&\frac{1}{\sqrt{2(\mathcal{J}^2+q)}}
  \left(\vec{D}+i\hat{r}\times\vec{D}\right)Y_{qjm}\ \ \ \ \
  (j\geq q>0)\\
  \vec{C}^0_{qjm}&=&\frac{\hat{r}}{r}Y_{qjm}\ \ \ \ \ (j\geq q\geq 0)\\
  \vec{C}^{-1}_{qjm}&=&\frac{1}{\sqrt{2(\mathcal{J}^2-q)}}
  \left(\vec{D}-i\hat{r}\times\vec{D}\right)Y_{qjm}\ \ \ \ \
  (j>q\geq 0)
\end{eqnarray}
where $\mathcal{J}^2\equiv j(j+1)-q^2$. Knowledge on vector spherical harmonics
for $q\geq 0$ would turn out to be enough to calculate the 1-loop determinant.
For $j=q$,
$\vec{C}^{-1}_{qjm}$ is absent instead of the above. For $j=q-1$,
both $\vec{C}^{-1}$ and $\vec{C}^0$ are absent. We expand the fields
as
\begin{equation}
  \delta\vec{A}=\sum_{n=-\infty}^\infty\sum_{j,m}\sum_{\lambda=0,\pm
  1}a^\lambda_{njm}r^{-i\frac{2\pi n}{\beta+\beta^\prime}}
  \vec{C}^\lambda_{jm}\ ,\ \
  \delta\sigma=\sum_{n,j,m}b_{njm}r^{-i\frac{2\pi n}{\beta+\beta^\prime}}
  \frac{Y_{jm}}{r}\ .
\end{equation}
One can expand (\ref{vec-bos-quadrataic}) by inserting these
expansions, using the following properties of vector
harmonics\footnote{We correct sign typos of \cite{Weinberg:1993sg}
in some of these formulae.},
\begin{eqnarray}
  &&\vec{D}\cdot\vec{C}^0_{qjm}=\frac{1}{r^2}Y_{qjm}\ ,\ \
  \vec{D}\cdot\vec{C}^{\pm 1}_{qjm}=-\frac{1}{r^2}
  \sqrt{\frac{\mathcal{J}^2\pm q}{2}}Y_{qjm}\\
  &&\vec{D}\times\vec{C}^0_{qjm}=\frac{i}{r}\left(
  \sqrt{\frac{\mathcal{J}^2+q}{2}}\ \vec{C}^{+1}_{qjm}-
  \sqrt{\frac{\mathcal{J}^2-q}{2}}\ \vec{C}^{-1}_{qjm}\right)\\
  &&\vec{D}\times\vec{C}^{\pm 1}_{qjm}=
  \frac{i}{r}\left(\mp\sqrt{\frac{\mathcal{J}^2\pm q}{2}}
  \ \vec{C}^0_{qjm}\right)
\end{eqnarray}
and
\begin{equation}
  \vec{D}\left(\frac{1}{r}Y_{qjm}\right)=
  \frac{1}{r}\left(-\vec{C}^0_{qjm}+\sqrt{\frac{\mathcal{J}^2+q}{2}}
  \ \vec{C}^{+1}_{qjm}+\sqrt{\frac{\mathcal{J}^2-q}{2}}\
  \vec{C}^{-1}_{qjm}\right)\ .
\end{equation}
From (\ref{vec-bos-quadrataic}) one finds
\begin{equation}\label{quad-generic}
  \sum_{i,j=1}^N\sum_{n=-\infty}^\infty\sum_{j,m}
  v_{-n,j,-m}^T\left(\mathcal{M}^{\frac{n_j\!-\!n_i}{2}}_{-n,j,-m}\right)^T
  \mathcal{M}^{\frac{n_i\!-\!n_j}{2}}_{njm}v_{njm}
\end{equation}
where
\begin{equation}\label{matrix-generic}
  \mathcal{M}^q_n=\left(\begin{array}{cccc}
  -\lambda+iq&0&is_+&-s_+\\0&\lambda+iq&-is_-&-s_-\\
  -is_+&is_-&iq&i\lambda+1
  \end{array}\right)\ ,\ \
  v_n=\left(\begin{array}{c}a_+\\a_-\\a_0\\b\end{array}\right)
\end{equation}
for the modes with $j\geq q+1$, $\lambda=\frac{2\pi
n}{\beta+\beta^\prime}-i\frac{\beta-\beta^\prime}{\beta+\beta^\prime}m
+(\alpha_i\!-\!\alpha_j)$, and this result is for
$q=\frac{n_i\!-\!n_j}{2}\geq 0$.\footnote{The matrix $\mathcal{M}$
with $n_i<n_j$ can be easily obtained as follows. In
(\ref{quad-generic}), one of $\mathcal{M}^T$ and $\mathcal{M}$
satisfy $n_i\geq n_j$. Suppose $\mathcal{M}$ satisfies this
condition. Firstly complex conjugate all $i$ which are explicit in
(\ref{matrix-generic}). Then change the sign of $\lambda$, which is
due to the sign changes of $n,j_3$ and $\alpha_i\!-\!\alpha_j$.
Transposing it gives the pair $\mathcal{M}^T$.} We also defined
$s_\pm\equiv\sqrt{\frac{\mathcal{J}^2\pm q}{2}}$. For $j=q$, there
is no $a_-$ modes and one finds
\begin{equation}
  \mathcal{M}^q_n=\left(\begin{array}{ccc}
  -\lambda+iq&is_+&-s_+\\-is_+&iq&i\lambda+1
  \end{array}\right)\ ,\ \
  v_n=\left(\begin{array}{c}a_+\\a_0\\b\end{array}\right)
\end{equation}
where $s_+=\sqrt{q}$. Finally for $j=q-1$ (possible only when $q\geq
1$), both $a_-$ and $a_0$ modes are absent. The scalar monopole
harmonics mode $b$ is also absent. One simply finds
\begin{equation}
  \mathcal{M}^q_n=\left(\begin{array}{c}
  -\lambda+iq\end{array}\right)\ ,\ \
  v_n=\left(\begin{array}{c}a_+\end{array}\right)
\end{equation}
where we used $s_+=0$ in this case.

Before evaluating the determinant we fix the gauge for these
fluctuations. For the $ij$'th mode for which $n_i\!=\!n_j$, we
choose the Coulomb gauge $s_+a_++s_-a_-=0$ following
\cite{Aharony:2003sx}. Since they are not coupled to magnetic
fields, the corresponding infinitesimal gauge transformation, call
it $\epsilon$, is expanded by ordinary spherical harmonics. The
Coulomb gauge condition requires $\partial^a\partial_a\epsilon=0$ on
$S^2$, which leaves the s-wave component of $\epsilon$ unfixed. We
impose a residual gauge condition
$\frac{d}{d\tau}\int_{S^2}A_\tau=0$ to fix this. The corresponding
Faddeev-Popov determinant can be calculated following
\cite{Aharony:2003sx}. For the Coulomb gauge, The Faddeev-Popov
determinant is that of the operator $D^a\partial_a\approx
\partial^a\partial_a$ over nonzero modes. For the residual gauge,
the determinant is given by
\begin{equation}\label{residual-fp}
  \prod_{\substack{i<j;\\n_i\!=\!n_j}}\left[2\sin
  \frac{\alpha_i\!-\!\alpha_j}{2}\right]^2\ .
\end{equation}
We also fix the gauge for the modes for which $n_i\neq n_j$.
Analogous to the previous case, we choose the `background Coulomb
gauge' $s_+a_++s_-a_-=0$. The infinitesimal gauge transformation
$\epsilon$ acquires the condition $D^aD_a\epsilon=0$. The
corresponding Faddeev-Popov determinant is $\det D^aD_a$. This, and
$\det\partial^a\partial_a$ above, will be canceled by a factor in
the 1-loop determinant to be calculated below. See \cite{Aharony:2003sx}
for the similar results. Contrary to the operator
$\partial^a\partial_a$, $D^2$ has no zero modes due to the absence
of s-waves in monopole spherical harmonics. So we do not have a
residual gauge fixing or a corresponding measure like
(\ref{residual-fp}). For $j=q$, our gauge implies $a_+\!=\!0$. For
$j=q\!-\!1$, there is no need to fix the gauge.

In the Coulomb gauge, we may write $a_+=s_- a$ and $a_-=-s_+ a$. Now
the quadratic terms for modes with $j\geq q\!+\!1$ takes the form
(\ref{quad-generic}) with $\mathcal{M}_n$ and $v_n$ given by
\begin{equation}
  \mathcal{M}^q_n=\left(\begin{array}{ccc}-s_-(\lambda-iq)&is_+&-s_+\\
  -s_+(\lambda+iq)&-is_-&-s_-\\-2is_+s_-&iq&i(\lambda-i)
  \end{array}\right)\ ,\ \
  v_n=\left(\begin{array}{c}a\\a_0\\b\end{array}\right)\ .
\end{equation}
The determinant of this matrix is
$\det(\mathcal{M}^q_n)=-\mathcal{J}^2\left[\left(\lambda-\frac{i}{2}\right)^2\!+\!
\left(j+\frac{1}{2}\right)^2\right]$. $-\mathcal{J}^2$ is nothing but the
eigenvalue of $D^aD_a$, whose determinant partly cancels with the Faddeev-Popov
measure as claimed. The remaining determinant of
bosonic fields with $j\geq q\!+\!1$ is
\begin{equation}
  \prod_{i,j=1}^N\prod_{n=-\infty}^\infty\prod_{j,j_3}\det\left(
  \mathcal{M}_{njj_3}^{\frac{|n_i\!-\!n_j|}{2}}\right)
  =\prod_{i,j}\prod_{n=-\infty}^\infty
  \prod_{j,j_3}\left[\left(j+\frac{1}{2}\right)^2+\left(
  \lambda-\frac{i}{2}\right)^2\right]\ .
\end{equation}
We can arrange the product over $n$ to sine functions:
\begin{equation}
  \hspace*{-1cm}\prod_{i,j}\prod_{j=\frac{|n_i\!-\!n_j|}{2}+1}^\infty
  \prod_{j_3}\sin\left[\frac{1}{2}\left(\beta(j\!-\!j_3)+
  \beta^\prime(j\!+\!j_3)
  -i(\alpha_i\!-\!\alpha_j)\!\frac{}{}\right)\right]
  \sin\left[\frac{1}{2}\left(\beta(j\!+1\!+\!j_3)+\beta^\prime
  (j\!+1\!-\!j_3)+i(\alpha_i\!-\!\alpha_j)\!\frac{}{}\right)\right]\ .
\end{equation}
Note that in each of the two sine factors, the role of energy is
played by the quantities $j$ and $j+1$, respectively. Signs of $j_3$
are not important since the product is symmetric under
$j_3\rightarrow-j_3$.

We also consider the modes with $j=q$ and $j=q\!-\!1$ (for $q\geq
1$). For $j=q$, with the gauge choice $a_+=0$, one finds
\begin{equation}
  \mathcal{M}^q_n=\left(\begin{array}{cc}i\sqrt{q}&-\sqrt{q}\\
  iq&i\lambda+1\end{array}\right)\ ,\ \ v_n=\left(\begin{array}{c}a_0\\b
  \end{array}\right)\ .
\end{equation}
From $\det(\mathcal{M}^q_n)=-\sqrt{q}(\lambda-i-iq)=-\frac{\sqrt{q}}
{\beta+\beta^\prime}\left[2\pi
n\!-\!i\beta(q\!+\!1\!+\!j_3)-i\beta^\prime(q\!+\!1\!-\!j_3)+
(\alpha_i\!-\!\alpha_j)\right]$, one obtains
\begin{equation}
  \prod_{n=-\infty}^\infty\prod_{j_3=-q}^q\sin\left[
  \frac{1}{2}\left(\frac{}{}\!\beta(q\!+\!1\!+\!j_3)+\beta^\prime(q\!+\!1\!-\!
  j_3)+i(\alpha_i\!-\!\alpha_j)\right)\right]\ .
\end{equation}
Note that this is a contribution from modes with energy
$q\!+\!1(=j\!+\!1)$. For $j=q\!-\!1$, one similarly finds
\begin{equation}
  \prod_{n=-\infty}^\infty\prod_{j_3=-q}^q\sin\left[
  \frac{1}{2}\left(\frac{}{}\!\beta(q\!+\!j_3)+\beta^\prime(q\!-\!
  j_3)+i(\alpha_i\!-\!\alpha_j)\right)\right]\ .
\end{equation}
This is again a contribution from modes with energy $q$($=j\!+\!1$).

Collecting all, the determinant of bosonic modes can be casted in
the following form (after relabeling $j_3\rightarrow-j_3$ for some
terms)
\begin{eqnarray}
  \log(\det\!{}_{\rm boson})^{-1}&=&
  -\frac{1}{2}{\rm tr}_B\left[
  \left(\frac{}{}\!\beta(j\!+\!j_3)+\beta^\prime(j\!-\!j_3)\right)+
  \left(\frac{}{}\!\beta(j\!+1\!+\!j_3)+\beta^\prime(j+\!1\!-\!j_3)\right)
  \right]\nonumber\\
  &&+\sum_{i,j=1}^\infty\sum_{n=1}^\infty\frac{1}{n}f^{\rm
  adj,B}_{ij}(x)e^{-in(\alpha_i\!-\!\alpha_j)}
\end{eqnarray}
where ${\rm tr}_B$ is trace over all bosonic modes explained above.
Contribution to the adjoint letter index from modes with
$j\geq\frac{|n_i\!-\!n_j|}{2}\!+\!1$ is
\begin{eqnarray}
  \hspace*{-0.5cm}f^{\rm adj,B}_{ij}&\leftarrow&
  \sum_{j=\frac{|n_i\!-\!n_j|}{2}\!+\!1}^\infty
  \left[(x^\prime)^{2j}+(x^\prime)^{2j\!-\!1}x+\cdots x^{2j}\right]+
  \left[(x^\prime)^{2j\!+\!1}x+(x^\prime)^{2j}x^2+\cdots
  x^\prime x^{2j\!+\!1}\right].
\end{eqnarray}
Additional contribution from modes with $j=\frac{|n_i\!-\!n_j|}{2}$
is given by
\begin{equation}
  f^{\rm adj,B}_{ij}\leftarrow
  (x^\prime)^{|n_i\!-\!n_j|\!+\!1}x+(x^\prime)^{|n_1\!-\!n_j|}x^2+
  \cdots+x^\prime x^{|n_i\!-\!n_j|\!+\!1}\ .
\end{equation}
Finally, when $\frac{|n_i\!-\!n_j|}{2}\geq 1$,
\begin{equation}
  f^{\rm adj,B}_{ij}\leftarrow
  (x^\prime)^{|n_i\!-\!n_j|\!-\!1}x+(x^\prime)^{|n_1\!-\!n_j|\!-\!2}x^2+
  \cdots+x^\prime x^{|n_i\!-\!n_j|\!-\!1}
\end{equation}
from modes with $j=\frac{|n_i\!-\!n_j|}{2}\!-\!1$. A similar
determinant from $\tilde{A}_\mu,\tilde\sigma$ is obtained with
$\alpha_i,n_i$ replaced by $\tilde\alpha_i,\tilde{n}_i$.

To complete the computation we consider contribution from the
fermion $\lambda_\alpha$. The Lagrangian on $S^2\times S^1$ is given
in appendix A, with a novel mass-like term. The calculation is
similar to that in appendix B.1 for matter fermions except for the
addition of this term. The operator acting on
$(\bar\lambda^\alpha)_{ij}$ is
\begin{equation}
  D_\tau+i\sigma^a D_a-\frac{n_i\!-\!n_j}{2}\sigma^3-\frac{1}{2}\ .
\end{equation}
The eigenvalue problem for the operator consisting of second and
third terms is solved, replacing $s(n_i\!-\!\tilde{n}_j)$ by
$n_i\!-\!n_j$ here. Again there appears eigenspinors
(\ref{eigen-paired}) with eigenvalues $\lambda=\pm\frac{j+1}{2}$ for
$j\geq\frac{|n_i\!-\!n_j|\!+\!1}{2}$, as well as additional modes
only if $n_i\!\neq\!n_j$ with $j\!=\!\frac{|n_i\!-\!n_j|\!-\!1}{2}$
and $\lambda\!=\!-\frac{|n_i\!-\!n_j|}{2}$. The combination
appearing in the determinant gets shifted by $-\frac{1}{2}$:
\begin{equation}
  D_\tau+\lambda-\frac{1}{2}\rightarrow-\frac{i}{\beta+\beta^\prime}
  \left[2\pi n+i\beta\left(\lambda-j_3-\frac{1}{2}\right)+
  i\beta^\prime\left(\lambda-\frac{1}{2}+h_3+j_3\right)+
  (\alpha_i\!-\!\alpha_j)\right]\ .
\end{equation}
For the modes with $j\geq\frac{|n_i\!-\!n_j|\!+\!1}{2}$, since
$\lambda$ appears in both signs, the `energy' $|\lambda|$ appearing
in the determinant gets shifted in two ways
$|\lambda|\rightarrow|\lambda|\mp\frac{1}{2}$ where upper (lower)
sign is for the positive (negative) $\lambda$. However, since
$\lambda$ is always negative for modes with
$j\!=\!\frac{|n_i\!-\!n_j|\!-\!1}{2}$, the shifted energy is always
$|\lambda|+\frac{1}{2}$ in this case. For $\bar\lambda_\alpha$, one
inserts $h_3=1$. Collecting all, the fermionic determinant is
($j_3\rightarrow-j_3$ relabeled for some terms)
\begin{eqnarray}
  \hspace*{-1cm}\log(\det\!{}_{\rm fermion})&=&
  \frac{1}{2}{\rm tr}_F\left[
  \left(\frac{}{}\!\beta(j\!+\!j_3)+\beta^\prime(j\!+\!1\!-\!j_3)\right)+
  \left(\frac{}{}\!\beta(j\!+1\!+\!j_3)+\beta^\prime(j\!-\!j_3)\right)
  \right]\nonumber\\
  &&+\sum_{i,j=1}^\infty\sum_{n=1}^\infty\frac{1}{n}f^{\rm
  adj,F}_{ij}(x)e^{-in(\alpha_i\!-\!\alpha_j)}\ .
\end{eqnarray}
The modes with $j\geq\frac{|n_i\!-\!n_j|\!+\!1}{2}$ contribute to
the letter index as
\begin{equation}
  f^{\rm adj,F}_{ij}\leftarrow
  -\!\!\!\!\sum_{j=\frac{|n_i\!-\!n_j|\!+\!1}{2}}^\infty\!\!\!\!
  \left[(x^\prime)^{2j\!+\!1}+(x^\prime)^{2j\!-\!1}x+\cdots+
  x^\prime x^{2j}\right]+
  \left[(x^\prime)^{2j}x+(x^\prime)^{2j\!-\!1}x^2+\cdots+
  x^{2j\!+\!1}\right]\ .
\end{equation}
Additional contribution from modes with
$j\!=\!\frac{|n_i\!-\!n_j|\!-\!1}{2}$ is given by
\begin{equation}
  f^{\rm adj,F}_{ij}\leftarrow-
  \left[(x^\prime)^{|n_i\!-\!n_j|\!-\!1}x+(x^\prime)^{|n_1\!-\!n_j|\!-\!2}x^2+
  \cdots+x^{|n_i\!-\!n_j|}\right]
\end{equation}
if $n_i\!\neq\!n_j$.

Comparing the determinants from bosons and fermions, one can
immediately find a vast cancelation. In fact, contribution from
bosonic modes with $j\geq\frac{|n_i\!-\!n_j|}{2}$ completely cancels
with that from fermionic modes with
$j\geq\frac{|n_i\!-\!n_j|\!+\!1}{2}$. In particular, this means that
there is no net contribution from modes which do not feel the flux,
i.e. $q=0$. This is of course consistent with the result of
\cite{Bhattacharya:2008bja}, in which the authors use combinatoric
methods in the free theory where the gauge fields play no role. In
our case, there are exceptional modes when $n_i\neq n_j$.
Contributions from fermion modes with $j=q\!-\!\frac{1}{2}$ and
bosonic modes with $j=q\!-\!1$ (if they exist) do not perfectly
cancel and yield
\begin{equation}
  f^{\rm adj}_{ij}=-x^{|n_i\!-\!n_j|}\ \ \ \ ({\rm if}\ n_i\neq n_j)\ .
\end{equation}
Generally one can write $f^{\rm adj}_{ij}(x)=-(1-\delta_{n_in_j})
x^{|n_i\!-\!n_j|}$. The final result is simply
\begin{eqnarray}
  \frac{\det\!{}_{\rm fermion}}{\det\!{}_{\rm boson}}
  =\prod_{i,j=1}^N\exp\left[\sum_{n=1}^\infty\frac{1}{n}
  \left(f^{\rm adj}_{ij}(x^n)e^{-in(\alpha_i\!-\!\alpha_j)}+
  \tilde{f}^{\rm adj}_{ij}(x^n)e^{-in(\tilde\alpha_i\!-\!\tilde\alpha_j)}
  \!\frac{}{}\right)\right]\ ,
\end{eqnarray}
with similarly defined $\tilde{f}^{\rm adj}_{ij}(x)$.
The evaluation of the Casimir-like energy is relegated to
appendix B.3 below.

\subsection{Casimir energy}

We finally turn to the Casimir-energy like shift in the effective
action
\begin{equation}
  \beta\epsilon_0\equiv
  \frac{1}{2}{\rm tr}\left[(-1)^F\left(\frac{}{}\!\beta(\epsilon\!+\!j_3)+
  \beta^\prime(\epsilon\!-\!h_3\!-\!j_3)
  +\gamma_1h_1+\gamma_2h_2\right)\right]\ ,
\end{equation}
where we have dropped the holonomy variables inside the trace,
$is(\alpha_i\!-\!\tilde\alpha_j)$ for matters and
$i(\alpha_i\!-\!\alpha_j)$ etc., for adjoints, since their traces
are trivially zero. To compute this formally divergent quantity, one
has to correctly regularize it. A constraint is that it has to be
compatible with our special supersymmetry. The most general
regularization would be insertion of
\begin{equation}
  x^{\epsilon+j_3}(x^\prime)^{\epsilon-h_3-j_3}y_1^{h_1}y_2^{h_2}\ .
\end{equation}
inside the trace. The parameters $x,x^\prime,y_1,y_2$ are not to be
confused with the chemical potentials in the rest of this paper:
they are regulators and should be taken to $x,x^\prime\rightarrow
1^-$, $y_1,y_2\rightarrow 1$ after computing the trace. Anyway, the
trace is formally very similar to the total summation of all letter
indices we computed in the previous subsections. Actually the above
trace, regularized as above, is
\begin{equation}
  \beta\epsilon_0=\frac{1}{2}\lim_{x,x^\prime,y_1,y_2\rightarrow 1}
  \left(\beta\partial_x\!+\!
  \beta^\prime\partial_{x^\prime}\!+\!\gamma_1\partial_{y_1}\!+\!
  \gamma_2\partial_{y_2}\right)
  \sum_{i,j=1}^N
  \left[f^{+}_{ij}(x,y_1,y_2)+f^-_{ij}(x,y_1,y_2)+f^{\rm adj}_{ij}(x)
  +\tilde{f}^{\rm adj}_{ij}(x)\right]\ .
\end{equation}
Since $x^\prime$ disappears in the letter indices,
$\partial_{x^\prime}$ is zero. Also, it is easy to see from the
$y_1,y_2$ dependence of $f^{\pm}_{ij}$ that $\partial_{y_1}$ and
$\partial_{y_2}$ are zero at $y_1,y_2=1$. Thus we only need to
compute $\beta\partial_x$ acting on various functions. At
$y_1=y_2=1$, they are given by
\begin{eqnarray}
  &&\lim_{x\rightarrow 1}\partial_xf^+_{ij}=
  +|n_i\!-\!\tilde{n}_j|\ \ ,\ \
  \lim_{x\rightarrow 1}\partial_xf^-_{ij}=
  +|n_i\!-\!\tilde{n}_j|\nonumber\\
  &&\lim_{x\rightarrow 1}\partial_xf^{\rm adj}_{ij}(x)=-|n_i\!-\!n_j|
  \ \ ,\ \ \lim_{x\rightarrow 1}\partial_x\tilde{f}^{\rm adj}_{ij}(x)
  =-|\tilde{n}_i\!-\!\tilde{n}_j|\ .
\end{eqnarray}
Therefore one finds
\begin{equation}
  \epsilon_0=\sum_{i,j=1}^N|n_i\!-\!\tilde{n}_j|-\sum_{i<j}|n_i\!-\!n_j|
  -\sum_{i<j}|\tilde{n}_i\!-\!\tilde{n}_j|\ .
\end{equation}
We list a few nonzero values of $\epsilon_0$ for some positive flux
distributions in Table 2.
\begin{table}[t]
$$
\begin{array}{c|c|c|c|c|c}
  \hline
  {\rm flux}&~\Yboxdim7pt\yng(2)~~\yng(1,1)~&
  ~\Yboxdim7pt\yng(2,1)~~\yng(1,1,1)~&\Yboxdim7pt\yng(3)~\yng(2,1)&
  \Yboxdim7pt\yng(3)~~\yng(1,1,1)~&\Yboxdim5pt\yng(6,4,3,2)~\yng(5,5,2,2,1)\\
  \hline
  \epsilon_0&2&2&2&6&39\!-\!13\!-\!22\!=\!4\\
  \hline
\end{array}
$$
\caption{Casimir energy for some positive flux distributions}
\end{table}

We explain a few useful properties of $\epsilon_0$. The fluxes may
involve positive, negative integers and zero. We first show that
contributions to $\epsilon_0$ from modes carrying $U(1)$ indices
with zero fluxes cancel to zero. Then we show that contributions
from modes ending on one $U(1)$ with positive flux and another
$U(1)$ with negative flux also cancel.

To show the first, since modes ending on two $U(1)$'s both with zero
flux is trivial, we restrict to the modes whose one end has zero
flux and another nonzero. Then contribution of these modes to
$\epsilon_0$ is
\begin{equation}
  \left(\!N_2\!\sum_{n_i\neq 0}|n_i|\!+\!N_1\!
  \sum_{\tilde{n}_i\neq 0}|\tilde{n}_i|\!\right)-
  N_1\sum_{n_i\neq 0}|n_i|-N_2\sum_{\tilde{n}_i\neq 0}|\tilde{n}_i|=
  (N_1\!-\!N_2)\left(\!\sum_{\tilde{n}_i\neq 0}|\tilde{n}_i|\!-\!
  \sum_{n_i\neq 0}|n_i|\!\right)\ .
\end{equation}
Here we use the fact that, for the index to be nonzero, total sum of positive
(negative) fluxes on both gauge groups should be equal. This implies that
expression in the second parenthesis  is zero, proving our claim.
This result implies that, to compute $\epsilon_0$,
one only has to consider contribution from modes connecting
nonzero fluxes.

To show the second, note that for such modes
$|n_i\!-\!\tilde{n}_j|=|n_i|+|\tilde{n}_j|$,
$|n_i\!-\!n_j|=|n_i|+|n_j|$ and
$|\tilde{n}_i\!-\!\tilde{n}_j|=|\tilde{n}_i|+|\tilde{n}_j|$ due to
the opposite sign of the two fluxes. After an analysis similar to the previous
parenthesis, their contribution to $\epsilon_0$ is
\begin{equation}
  (M^-_1\!-\!M^-_2)\left(\sum|\tilde{n}^+_i|\!-\!
  \sum|n^+_i|\right)+(M^+_1\!-\!M^+_2)\left(\sum|\tilde{n}^-_i|\!-\!
  \sum|n^-_i|\right)\ .
\end{equation}
Again from the equality of total positive/negative fluxes on two
gauge groups, this quantity is zero. This result implies that one
can separate $\epsilon_0=\epsilon_0^++\epsilon_0^-$, first one
coming from modes connecting positive fluxes only and second from
modes connecting negative fluxes only. This property will be
important when we discuss the large $N$ factorization in section 3.

Finally, we show that $\epsilon_0$ is always non-negative, and
becomes zero if and only if the two sets $\{n_i\}$ and
$\{\tilde{n}_i\}$ are the same. We shall actually prove a slightly
more general claim. Suppose we have two decreasing functions $f(x)$
and $g(x)$ defined on $0\leq x\leq\ell$. Then we claim that the
functional $\mathcal{E}[f,g]$ defined by
\begin{equation}
  \mathcal{E}[f,g]\equiv\int dxdy\left(|f(x)-g(y)|-\frac{1}{2}|f(x)-f(y)|
  -\frac{1}{2}|g(x)-g(y)|\right)
\end{equation}
is always non-negative, and assumes its minimum at $0$ if and only
if $f(x)=g(x)$ everywhere.\footnote{Requiring these functions to
assume integral values, admitting decreasing step function like
singularities, brings us back to our original problem. We think
discontinuity would not cause any problem, but if one prefers, one
may slightly regularize them to smooth decreasing functions while
staying arbitrarily close to our problem.} To prove our claim, we
vary the functional by $\delta f(x)$. Note that
\begin{eqnarray}
  &&\delta|f(x)-g(y)|=\delta f(x)\left[2\theta(f(x)\!-\!g(y))-1\right]
  =\delta f(x)\left[2\theta(y\!-\!g^{-1}\!f(x))-1\right]\
  ,\nonumber\\
  &&\delta|f(x)\!-\!f(y)|=\delta f(x)\left[2\theta(y\!-\!x)-1\right]+
  \delta f(y)\left[2\theta(x\!-\!y)-1\right]
\end{eqnarray}
under this variation, where $\theta(x)$ is the step function
(assuming $1$ for $x>0$, and $0$ for $x<0$). From these one finds
\begin{equation}\label{variation}
  \delta\mathcal{E}[f,g]=\int dx\ 2\delta f(x)
  \left[x\!-\!g^{-1}\!f(x)\right]\ .
\end{equation}
The condition for the extremal points is $x=g^{-1}\!f(x)$, or
$f(x)=g(x)$. Same result is obtained by the variation $\delta g(x)$.
To show this is a minima, we compute the Hessian. From
(\ref{variation}) and analogous result for $\delta g(x)$, one finds
\begin{equation}
  \left(\begin{array}{cc}\frac{\delta^2 \mathcal{E}}{\delta f(x)\delta f(y)}&
  \frac{\delta^2 \mathcal{E}}{\delta f(x)\delta g(y)}\\
  \frac{\delta^2 \mathcal{E}}{\delta g(x)\delta f(y)}&
  \frac{\delta^2 \mathcal{E}}{\delta g(x)\delta
  g(y)}\end{array}\right)=2
  \left(\begin{array}{cc}-\frac{\delta(x\!-\!y)}{g^\prime(g^{-1}\!f(x))}&
  \frac{\delta(g^{-1}\!f(x)\!-\!y)}{g^\prime(g^{-1}\!f(x))}
  \\\frac{\delta(f^{-1}\!g(x)\!-\!y)}{f^\prime(f^{-1}\!g(x))}&
  -\frac{\delta(x\!-\!y)}{f^\prime(f^{-1}\!g(x))}\end{array}\right)
\end{equation}
where we used $\delta f^{-1}(x)=-\frac{\delta
f(f^{-1}(x))}{f^\prime(f^{-1}(x))}$ and similar formula for $\delta
g^{-1}$. At the extrema $f=g$, the last matrix becomes
\begin{equation}
  2\delta(x\!-\!y)\left(\begin{array}{cc}-\frac{1}{f^\prime(x)}&
  \frac{1}{f^\prime(x)}\\\frac{1}{f^\prime(x)}&
  -\frac{1}{f^\prime(x)}\end{array}\right)=
  -\frac{2\delta(x\!-\!y)}{f^\prime(x)}
  \left(\begin{array}{cc}1&-1\\-1&1\end{array}\right)\ .
\end{equation}
Since $f$ is decreasing, the coefficient in front of the matrix is
positive. The last $2\times 2$ matrix has eigenvalue $0$ for $\delta
f(x)=\delta g(x)$ and $+2$ for $\delta f(x)=-\delta g(x)$. The first
is the expected zero direction since the variation leaves the
relation $f=g$ unchanged. The second shows that the extrema $f=g$ is
actually a minima, proving our claim.

\section{Index over gravitons in $AdS_4\times S^7/\mathbb{Z}_k$}

Index of many gravitons in $AdS_4\times S^7/\mathbb{Z}_k$ can be
obtained from the index of single graviton in $AdS_4\times S^7$, as
explained in \cite{Bhattacharya:2008bja}. The index of single
graviton in $AdS_4\times S^7$ is given by
\begin{equation}
  I^{\rm sp}(x,y_1,y_2,y_3)=
  \frac{\rm (numerator)}{\rm (denominator)}
\end{equation}
where
\begin{eqnarray}
  {\rm numerator}&=&\sqrt{y_1y_2y_3}\left(1+y_1y_2+y_2y_3+y_3y_1\right)
  x^{\frac{1}{2}}-\sqrt{y_1y_2y_3}
  (y_1\!+\!y_2\!+\!y_3\!+\!y_1y_2y_3)x^{\frac{7}{2}}
  \nonumber\\
  &&+(y_1y_2\!+\!y_2y_3\!+\!y_3y_1\!+\!y_1y_2y_3(y_1\!+\!y_2\!+\!y_2))
  (x^3-x)\\
  {\rm denominator}&=&(1-x^2)(\sqrt{y_3}-\sqrt{xy_1y_2})
  (\sqrt{y_1}-\sqrt{xy_2y_3})(\sqrt{y_2}-\sqrt{xy_3y_1})
  (\sqrt{y_1y_2y_3}-\sqrt{x})\ .\nonumber
\end{eqnarray}
A very useful property of this function is
\begin{eqnarray}\label{relation}
  I^{\rm sp}&=&\frac{\left(1-x\sqrt{xy_1y_2y_3}\right)
  \left(1-x\sqrt{\frac{xy_3}{y_1y_2}}\right)
  \left(1-x\sqrt{\frac{xy_1}{y_2y_3}}\right)
  \left(1-x\sqrt{\frac{xy_2}{y_1y_3}}\right)}
  {\left(1-\sqrt{\frac{xy_1y_3}{y_2}}\right)
  \left(1-\sqrt{\frac{xy_2y_3}{y_1}}\right)
  \left(1-\sqrt{\frac{xy_1y_2}{y_3}}\right)\left(1-\sqrt{\frac{x}{y_1y_2y_3}}\right)
  (1-x^2)^2}-\frac{1-x^2+x^4}{(1-x^2)^2}\nonumber\\
  &\equiv&\frac{F(x,y_1,y_2,y_3)}{(1-x^2)^2}
  -\frac{1-x^2+x^4}{(1-x^2)^2}
\end{eqnarray}
where the function $F(x,y_i)$ is defined in section 3.2. The index
of single gravitons in $AdS_4\times S^7/\mathbb{Z}_k$ is obtained by
expanding $I^{\rm sp}$ in $y_3$ as
\begin{equation}
  I^{\rm sp}=\sum_{n=-\infty}^\infty y_3^{\frac{n}{2}}I^{\rm sp}_n(x,y_1,y_2)
\end{equation}
and keeping terms in which $n$ is a multiplet of $k$:
\begin{equation}
  I^{\rm sp}_{\mathbb{Z}_k}(x,y_1,y_2,y_3)\equiv\sum_{n=-\infty}^\infty
  y_3^{\frac{kn}{2}}I^{\rm sp}_{kn}(x,y_1,y_2)\ .
\end{equation}
Each term $I_{kn}$ represents a single particle index of gravitons
carrying Kaluza-Klein momentum $kn$. Finally, the index of multiplet
gravitons in $AdS_4\times S^7/\mathbb{Z}_k$ is given by
\begin{equation}
  I_{\rm mp}(x,y_1,y_2,y_3)=\exp\left[\sum_{n=1}^\infty\frac{1}{n}
  I^{\rm sp}_{\mathbb{Z}_k}(x^n,y_1^n,y_2^n,y_3^n)\right]\ .
\end{equation}
One can decompose this index into three factors, each coming from
gravitons with positive/negative/zero KK-momenta, respectively, as
\begin{equation}
  I_{\rm mp}(x,y_1,y_2,y_3)=I^{(0)}_{\rm mp}I^{(+)}_{\rm mp}
  I^{(-)}_{\rm mp}\ ,
\end{equation}
where
\begin{equation}
  I^{(0)}_{\rm mp}=\exp\left[\sum_{n=1}^\infty\frac{1}{n}
  I^{\rm sp}_0(\cdot^n)\right]\ ,\ \ I^{(\pm)}_{\rm mp}=\exp\left[\sum_{n=1}^\infty
  \frac{1}{n}I^{\rm sp(\pm)}_{\mathbb{Z}_k}(\cdot^n)\right]
\end{equation}
and
\begin{equation}
  I^{\rm sp(\pm)}_{\mathbb{Z}_k}=\sum_{n=1}^{\infty}
  y_3^{\pm\frac{kn}{2}}I^{\rm sp}_{\pm kn}(x,y_1,y_2)\ .
\end{equation}
$I^{(\pm)}_{\rm mp}$ satisfies the property $I^{(-)}_{\rm
mp}(x,y_1,y_2,y_3)= I^{(+)}_{\rm mp}(x,y_1,1/y_2,1/y_3)$, similar to
the relation between $I^{(-)}$ and $I^{(+)}$ for the gauge theory
index defined in section 3.

\end{document}